\documentclass[fleqn,usenatbib]{mnras}

\usepackage{newtxtext,newtxmath}

\usepackage[T1]{fontenc}
\usepackage{ae,aecompl}
\usepackage{color}

\usepackage{graphicx}	
\usepackage{amsmath}	
\usepackage{amssymb}	

\usepackage{breqn}

\usepackage{siunitx}

\usepackage{xcolor}

\newcommand{\hlm}[1]{#1}
\newcommand{\hlt}[1]{#1}
\newcommand{\hlcom}[1]{}

\DeclareSIUnit \h {\ensuremath{\mathit{h}}}
\DeclareSIUnit \parsec {pc}

\definecolor{myorange}{RGB}{220,100,40}
\definecolor{mydarkblue}{RGB}{50,100,200}
\definecolor{mydarkgreen}{RGB}{50,180,50}
\definecolor{mypurple}{RGB}{180,50,180}

\newcommand{\myvec}[1]{{ \mathbf{#1} }}

\newcommand{\vecnorm}[1]{{ \lVert \myvec{#1} \rVert }}
\newcommand{\norm}[1]{{ \lVert #1 \rVert }}

\newcommand{\Dxq}[0]{\mathbfss{D}_\text{xq}}
\newcommand{\Dvq}[0]{\mathbfss{D}_\text{vq}}

\newcommand{\erfc}[0]{\text{erfc}}

\newcommand{\Dxqdot}[0]{\dot{\mathbfss{D}}_\text{xq}}
\newcommand{\Dvqdot}[0]{\dot{\mathbfss{D}}_\text{vq}}

\newcommand{\rhos}[0]{\hlm{\rho_{\text{s}}}}

\newcommand{\phis}[0]{\hlm{\phi_{\text{s}}}}
\newcommand{\phil}[0]{\hlm{\phi_{\text{l}}}}
\newcommand{\Gs}[0]{\hlm{G_{\phi,\text{s}}}}

\newcommand{\rs}[0]{\hlm{r_{\text{s}}}}

\newcommand{\Lmin}[0]{L_{\text{min}}}

\newcommand{\T}[0]{\mathbfss{T}}

\graphicspath{{img/}{./}}

\title[The Complexity of the Dark Matter Sheet]{Simulating the
Complexity of the Dark Matter Sheet I: Numerical Algorithms}

\author[J. St\"ucker et al.]{
Jens St\"ucker,$^{1,2}$\thanks{\hlt{E-mail: jstuecker@dipc.org}}
Oliver Hahn,$^{3}$
Raul E. Angulo$^{2,4}$
and Simon D.M. White$^{1}$
\\
$^{1}$Max-Planck-Institut f\"ur Astrophysik, Postfach 1317, D-85741 Garching, Germany.\\
\hlt{$^{2}$Donostia International Physics Centre (DIPC), Paseo Manuel de Lardizabal 4, 20018 Donostia-San Sebastian, Spain.}\\
$^{3}$Laboratoire Lagrange, Universit\'e C\^ote d'Azur, Observatoire de la C\^ote d'Azur, CNRS,
      Blvd de l'Observatoire,\\\hskip0.15in CS 34229, 06304 Nice cedex 4, France.\\
$^{4}$IKERBASQUE, Basque Foundation for Science, E-48013, Bilbao, Spain.
}

\date{Accepted XXX. Received YYY; in original form ZZZ}

\pubyear{2019}

\begin{document}
\label{firstpage}
\pagerange{\pageref{firstpage}--\pageref{lastpage}}
\maketitle

\begin{abstract}
At early times dark matter has a thermal velocity dispersion of unknown amplitude which, for warm dark matter models, can influence the formation of nonlinear structure on observable scales.  
We propose a new scheme to simulate cosmologies with a small-scale suppression of perturbations that combines two previous methods in a way that avoids the numerical artefacts which have so far prevented either from producing fully reliable results. At low densities and throughout most of the cosmological volume, we represent the dark matter phase-sheet directly using high-accuracy interpolation, thereby avoiding the artificial fragmentation which afflicts particle-based methods in this regime. Such phase-sheet methods are, however, unable to follow the rapidly increasing complexity of the denser regions of dark matter haloes, so for these we switch to an N-body scheme  which uses the geodesic deviation equation to track phase-sheet properties local to each particle. In addition, we present  a novel high-resolution force calculation scheme based on an oct-tree of cubic force resolution elements which is well suited to approximate the force-field of our combined sheet+particle distribution. Our hybrid simulation scheme enables the first reliable simulations of the internal structure of low-mass haloes in a warm dark matter cosmology.
\end{abstract}





\section{Introduction}

While the $\Lambda$CDM model explains most key observations of our Universe remarkably well, the physical nature of its main ingredients -- dark energy and dark matter -- remains unknown. So far dark matter appears to be cold and collisionless on those scales which have been reliably probed. However, in all dark matter models which are well motivated from particle physics, dark matter has a residual thermal (or non-thermal) velocity dispersion at early times. In the case of weakly interacting massive particles (WIMPs) the thermal velocity dispersion is so small that it only affects structure formation on very small scales ($\sim$ Earth mass). However, for other dark matter candidates - for example, sterile neutrinos - the thermal velocity dispersion might be large enough to cause effects on observable (or soon observable) scales. In this case we speak of warm dark matter. If we were to measure such effects, we could constrain the nature of dark matter. It is therefore of fundamental interest to understand the implications of the ``warmth'' of dark matter, and to search for deviations from the perfectly cold scenario. It is the aim of this paper to improve upon existing methods for simulating cosmologies with a small scale cut-off in the power spectrum, sothat reliable predictions for warm dark matter cosmologies become possible. 

Cosmological N-body simulations have been very successful at predicting the large-scale structure of the universe for cold dark matter scenarios \citep[see e.g.][for reviews]{frenk_white_2012,kuhlen_2012}. In N-body simulations, the (statistically) well known linear density field of the early universe is discretized to a finite number of macroscopic particles. These particles are then evolved under their self-gravity to infer an approximation to the late-time non-linear density field. The particles are often interpreted as a \hlt{discrete representation} of the continuous non-linear density field of the dark matter fluid.

However, in the case of warm dark matter cosmologies, N-body simulations give rise to numerical artefacts during the first phases of nonlinear structure formation. In practice the difference between cold and warm dark matter simulations lies merely in the choice of initial conditions. The initial density field is smoothed by the free streaming motion of the particles and therefore its power spectrum has a small scale cut-off, which is on larger scales for warmer dark matter. Since at later times the thermal velocity dispersion is relatively small when compared to the bulk velocities of the dark matter fluid (the first is a decaying, the second a growing mode), the thermal velocity dispersion can be neglected once the cut-off in power on small scales is established \citep{bode_halo_2001}. Therefore, to excellent approximation one simulates a perfectly cold fluid in all cases, however either with (WDM) or without (CDM) an additional truncation scale in the initial perturbation spectrum. N-body simulations of warm dark matter form a large number of small haloes - most prominently found regularly spaced in filaments - aligning like beads on a string \citep{bode_halo_2001}. \citet{wang_white_2007} showed that these haloes are not of physical nature, but are merely numerical artefacts. They found such fragments even in the case of N-body simulations of the collapse of a perfectly homogeneous filament.

The fragmentation is a natural consequence of the anisotropic collapse with incomplete thermalisation in cosmology. This anisotropy of collapse means that, as structure forms, it collapses first to a one-dimensional sheet, or ``pancake'' \citep{zeldovich_1970}, followed by collapse to a filamentary strucuture, and only then a halo \citep[e.g.][]{bond_kofman_pogosyan_1996}. In each case, the structures are supported by velocity dispersion only along the already collapsed directions, while the temperature is still effectively zero in the uncollapsed dimensions \citep[cf.][]{buehlmann_hahn_2018}, making them unstable to spurious collapse seeded by numerical noise. The underlying reason is of course that in a collisionless fluid no thermalisation (and therefore isotropisation of the temperature) takes place.

In recent years, a new set of simulation schemes has been designed which are unaffected by this artificial fragmentation \citep{hahn_2013,hahn_angulo_2016,sousbie_colombi_2016}. These employ a density estimate that is much closer to the continuum limit than that obtained from the particles in standard N-body simulations. This density estimate is obtained by interpolating between the positions of tracer particles in phase space.  This is possible since in the limit of a cold distribution function, these tracers occupy only a three-dimensional (Lagrangian) submanifold of phase space, also known as the dark matter sheet \citep{arnold_1982,shandarin_zeldovich_1989,shandarin_2012,abel_2012}. While this approach has successfully been used in \citet{angulo_2013} to measure the WDM halo mass function below the cut-off scale, there are still major limitations to the range of applications of the schemes. Inside haloes, the dark matter sheet grows rapidly in complexity making it hard to reconstruct the sheet accurately \citep{vogelsberger_white_2011,sousbie_colombi_2016}. Therefore schemes which do not refine the resolution of the interpolated mass elements give biased densities inside haloes, and schemes which use refinement \citep{hahn_angulo_2016,sousbie_colombi_2016} quickly become unfeasibly complex. This is so since the detailed fine-grained evolution of the distribution function has to be followed at any instance in time, so that new tracers need to be inserted in order to not lose information about the dynamics.

This is an important difference between the two approaches. In the $N$-body method, one benefits from ergodicity, i.e. in a time-averaged sense one obtains an accurate representation of the underlying distribution function, even if at any moment in time, the particular realisation might not be perfect. This is also the underlying reason, why the $N$-body method has problems with anisotropic collapse from cold initial conditions: ergodicity has not been established in the uncollapsed subspace, where the mean field dynamics is now very noisy. This is circumvented by following the distribution function explicitly, which can be done as long as its structure is not yet too complex. In this case, there is no noise and the cold uncollapsed subspaces can be followed accurately. Ultimately however, rapid phase and chaotic mixing inside of haloes lead to close to ergodic dynamics, rendering it increasingly complex and ultimately impossible to follow the evolution of the sheet, but making $N$-body attractive, since it relies on exactly that assumption.

Finally, another short-coming of the previous implementations of the sheet method is that they have so far only worked at very low force-resolution. So far these have used only a single mesh for the force calculation which smooths the density field on scales much larger than what is necessary to resolve the centres of haloes. An accurate treatment requires an adaptive scheme for the force-calculation and the time-stepping.

\begin{figure}
	\includegraphics[width=\columnwidth]{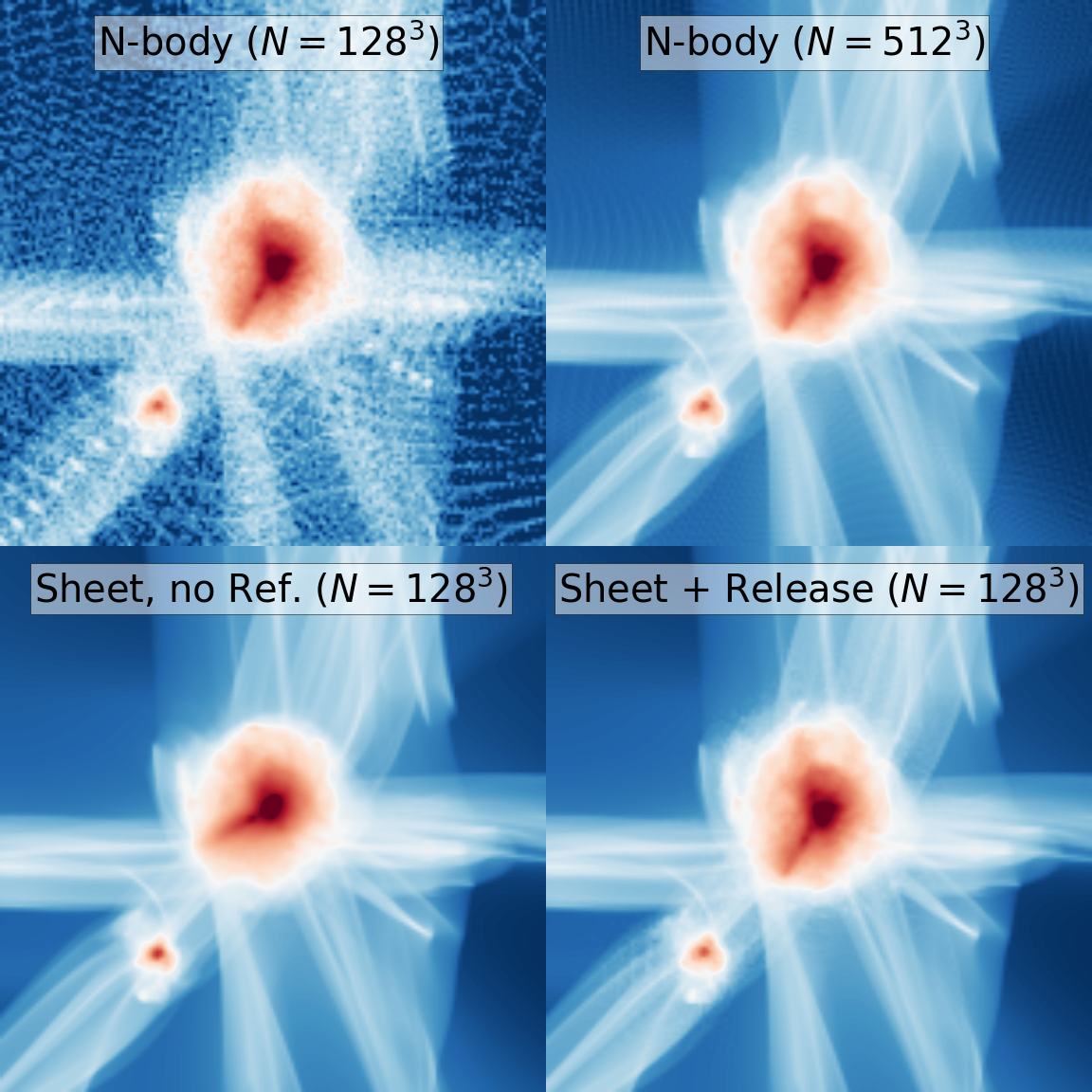}
    \caption{A projection of the density field in and around a halo for different simulation setups. Top left: N-body with $N=128^3$ particles. Top right: N-body with $N=512^3$ as a reference case. Bottom left: sheet (no refinement) with $N=128^3$. Bottom right: sheet + release (no refinement) with $N=128^3$. ``Sheet + release'' means that most of the mass is traced by the sheet interpolation, but mass elements which were detected to become too complex are traced by an N-body approach instead. The low resolution N-body scheme appears to get the shape of the centre of the halo correct, but fragments in the low density regions. The pure sheet scheme captures the low density regions very well, but creates a biased overly-round halo, since its phase space structure is too complex for reconstruction by interpolation. The sheet + release case inherits the best of both worlds and avoids the problems with fragmentation or biased halo structure. It comes closest to the much higher resolution reference case in the top right panel. Note that the $N=512^3$ N-body case would also fragment if the force-resolution were increased significantly, while the sheet cases can also avoid fragmentation in that scenario. It is the subject of section \ref{sec:chapterrelease} to elaborate the details of the sheet + release scheme. 
    }
    \label{fig:halo_projection}
\end{figure}

 In this paper, we propose solutions to these various short-comings of sheet-based dark matter simulations. We employ a hybrid scheme which uses sheet-based simulation techniques wherever the interpolation is reliable, and switches to N-body based simulation techniques where the sheet becomes untraceable, but where we are reasonably confident that in a time-averaged sense the particles reproduce the correct mean field dynamics. We illustrate this in Figure~\ref{fig:halo_projection} by a projection of the density field in and around a halo. We present - going from top left to bottom right - a low resolution $N=128^3$ N-body simulation, a high resolution $N=512^3$ N-body simulation as reference, a low resolution $N=128^3$ sheet simulation (without using refinement techniques) and a low resolution $N=128^3$ ``sheet + release'' simulation (without refinement) that switches to an N-body scheme when the sheet becomes too complex. The N-body case produces the correct halo shape, but fragments in low density regions. The pure sheet case captures the low density regions with stunning accuracy, but produces a deformed halo. However, the sheet + release case inherits the best of both worlds and seems authentic everywhere - thereby coming closest to the much higher resolution reference simulation at a much reduced number of degrees of freedom. It is the subject of section \ref{sec:chapterrelease} to guide the reader to an understanding of how and how well the sheet + release scheme works.

In section \ref{sec:force}, we further develop a new tree-based discretization of the force-field which is compatible with both N-body and sheet approaches. This makes it possible to use the sheet + release simulation approach all the way down to the high force resolution scale that is needed to resolve the centres of haloes.

Thus, in this paper, we present for the first time a full scheme which makes possible non-fragmenting and unbiased warm dark matter simulations with high force-resolution. In a subsequent companion paper, we will present its predictions for the case of one of the smallest haloes in a warm dark matter universe.




\section{A fragmentation-free and unbiased scheme for cosmological warm dark matter simulations} \label{sec:simulationscheme} \label{sec:chapterrelease}

\subsection{The Dark Matter Sheet}

The artificial fragmentation of N-body simulations (as for example in Figure \ref{fig:halo_projection}, top left panel) appears to
be  a  major  shortcoming  of  the  N-body  method.  The  difference  between  simulations  of  warm  dark  matter  (which
tend to fragment) and cold dark matter (where artificial fragments are subdominant to real small-scale clumps)  lies  merely  in  the
choice  of  the  initial  conditions.  In  principle,  simulations  of
warm  dark  matter  should  be  simpler  than  cold  dark  matter ones, since there is much less structure and complexity.
Solving the problem of fragmentation of N-body simulations
is thus not only important for testing warm dark matter models, but also as a test, beyond the customary numerical  convergence tests, of the validity of the N-body scheme as a whole in the cold dark matter case.

\begin{figure}
	\includegraphics[width=0.49 \columnwidth]{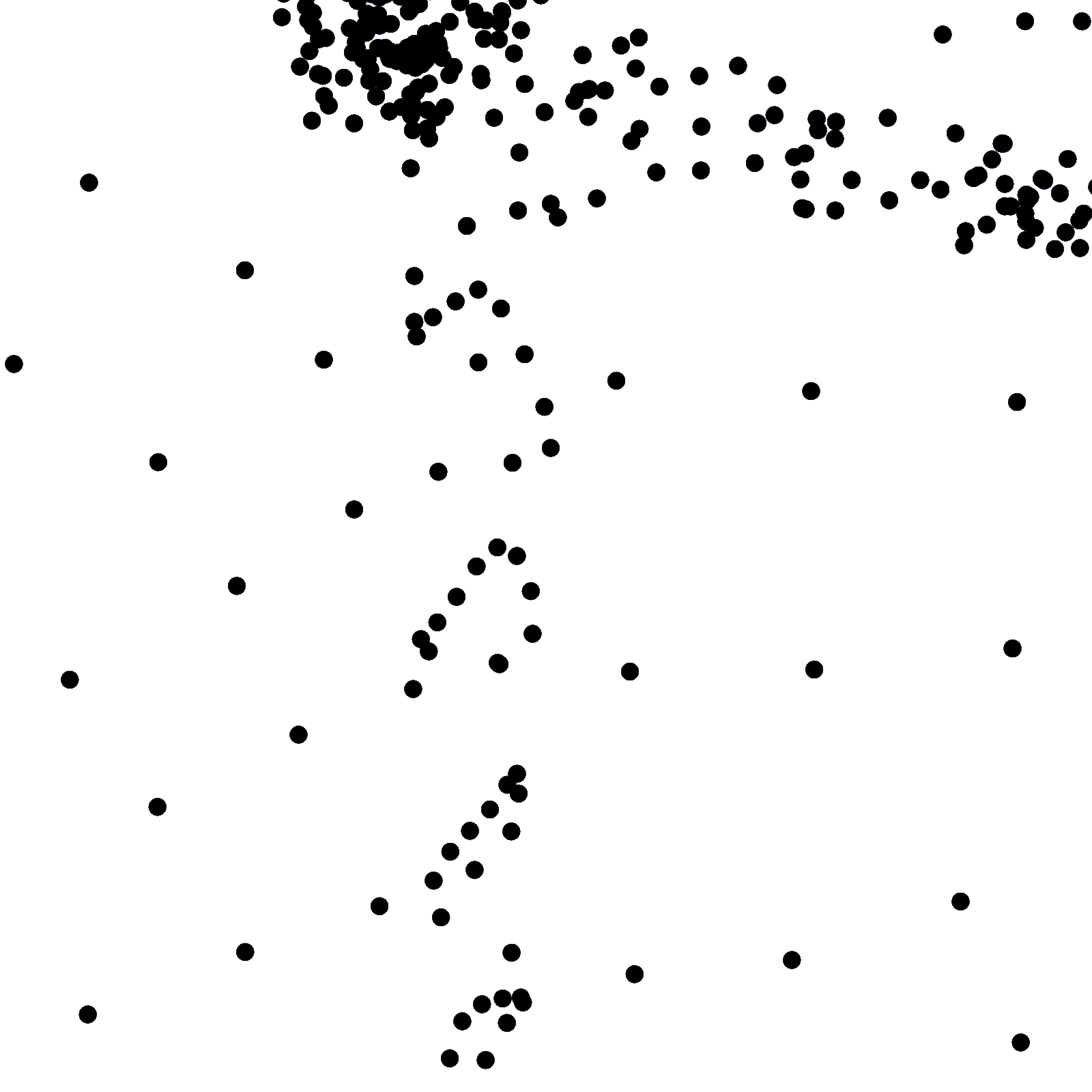}
    \includegraphics[width=0.49 \columnwidth]{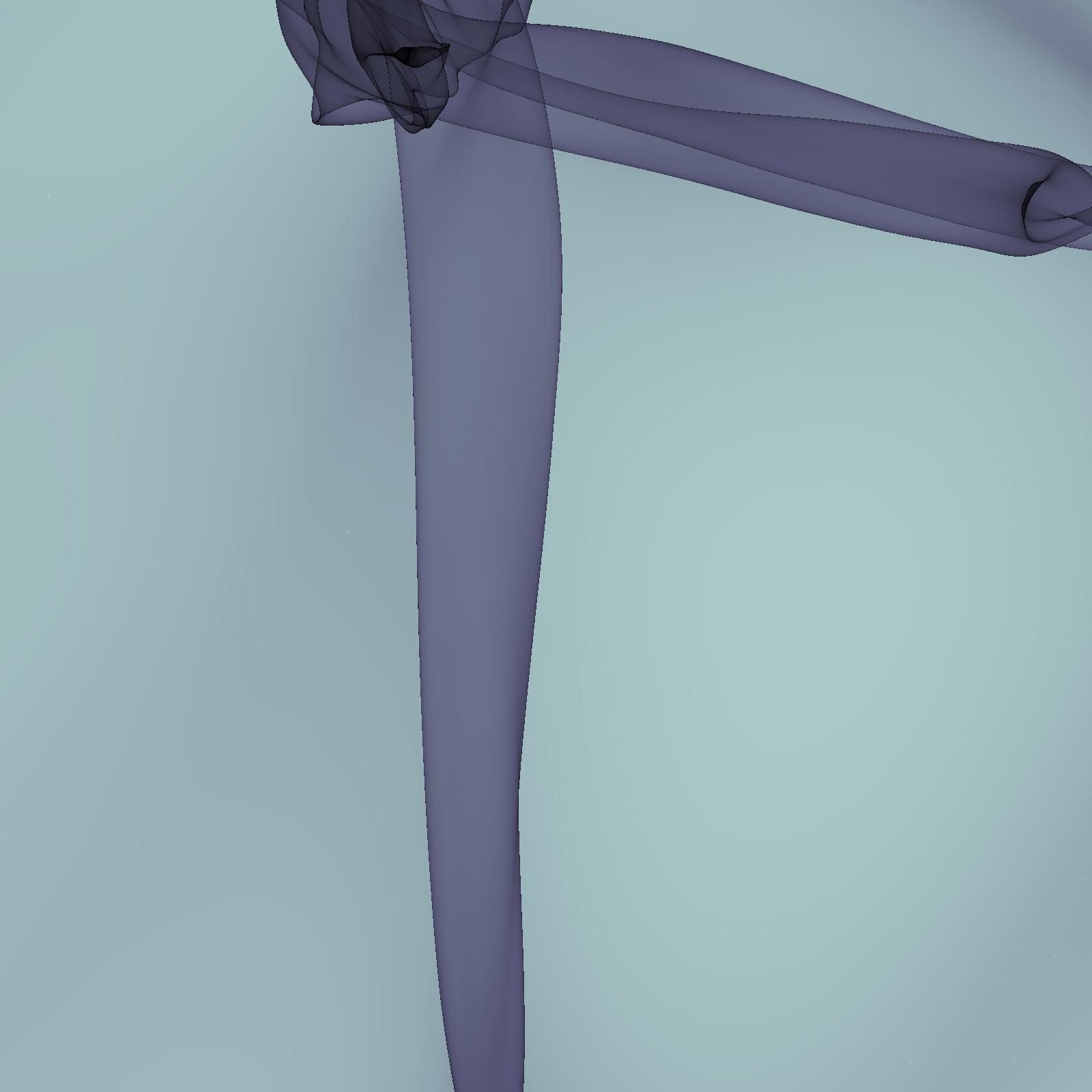}
    \caption{N-body density estimate versus a density estimate inferred by interpolation of the dark matter sheet in phase space. The N-body density estimate shows regular lumps which will grow into fragmented artificial haloes. However, the continuum density estimate shows no such artefacts. }
    \label{fig:density-estimates}
\end{figure}

The fragmentation of N-body simulations can be overcome by using a smoother density estimate than the granular ``N-body density estimate'' in the Poisson-solver in warm dark matter simulations. Such a smooth density estimate can be obtained by considering the phase space structure of the dark matter ``fluid'' \citep{abel_2012,shandarin_2012,hahn_angulo_2016,sousbie_colombi_2016}. We show a visualization of such an estimate in comparison to an N-body density estimate for a two dimensional simulation in Figure \ref{fig:density-estimates}. What appear as artificial lumps in the N-body density estimate do not appear at all in the continuum density estimate. We will briefly review here how such high quality density estimates can be constructed, and what are the limitations of such an approach.

\begin{figure}
  \includegraphics[width=\columnwidth]{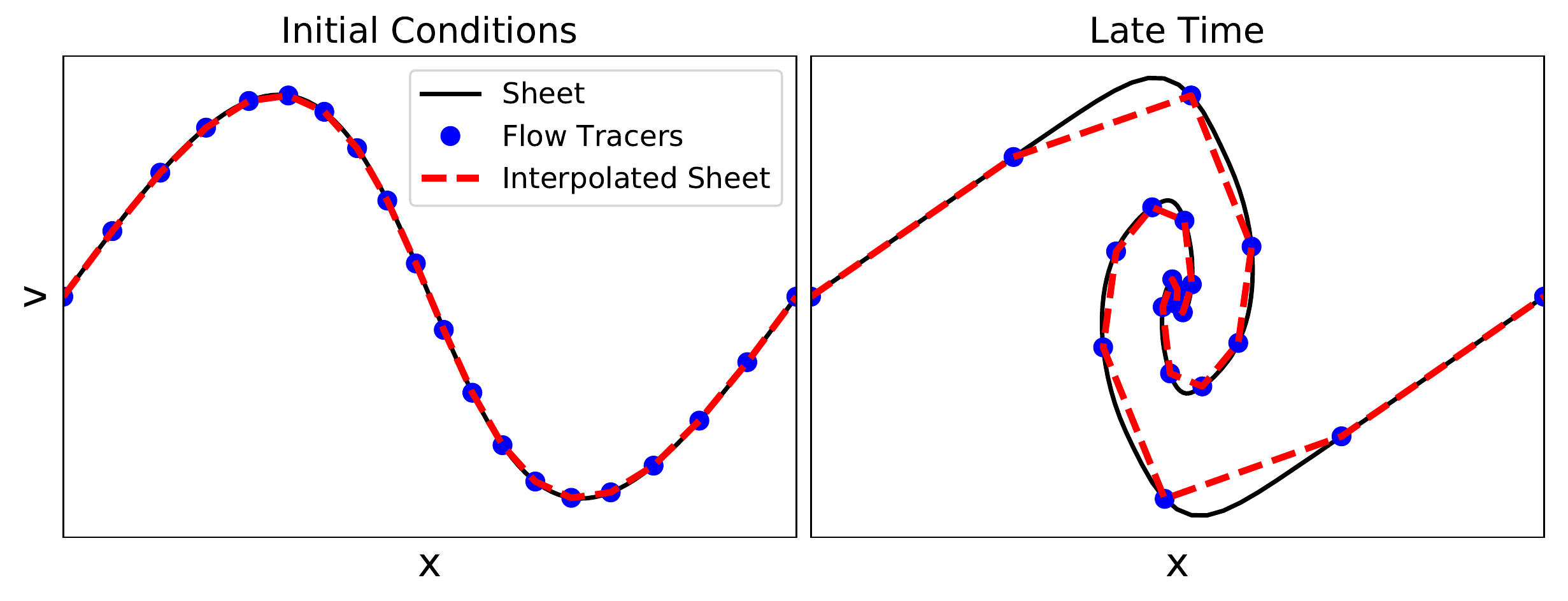}
  \begin{center}
    \includegraphics[width=0.7 \columnwidth]{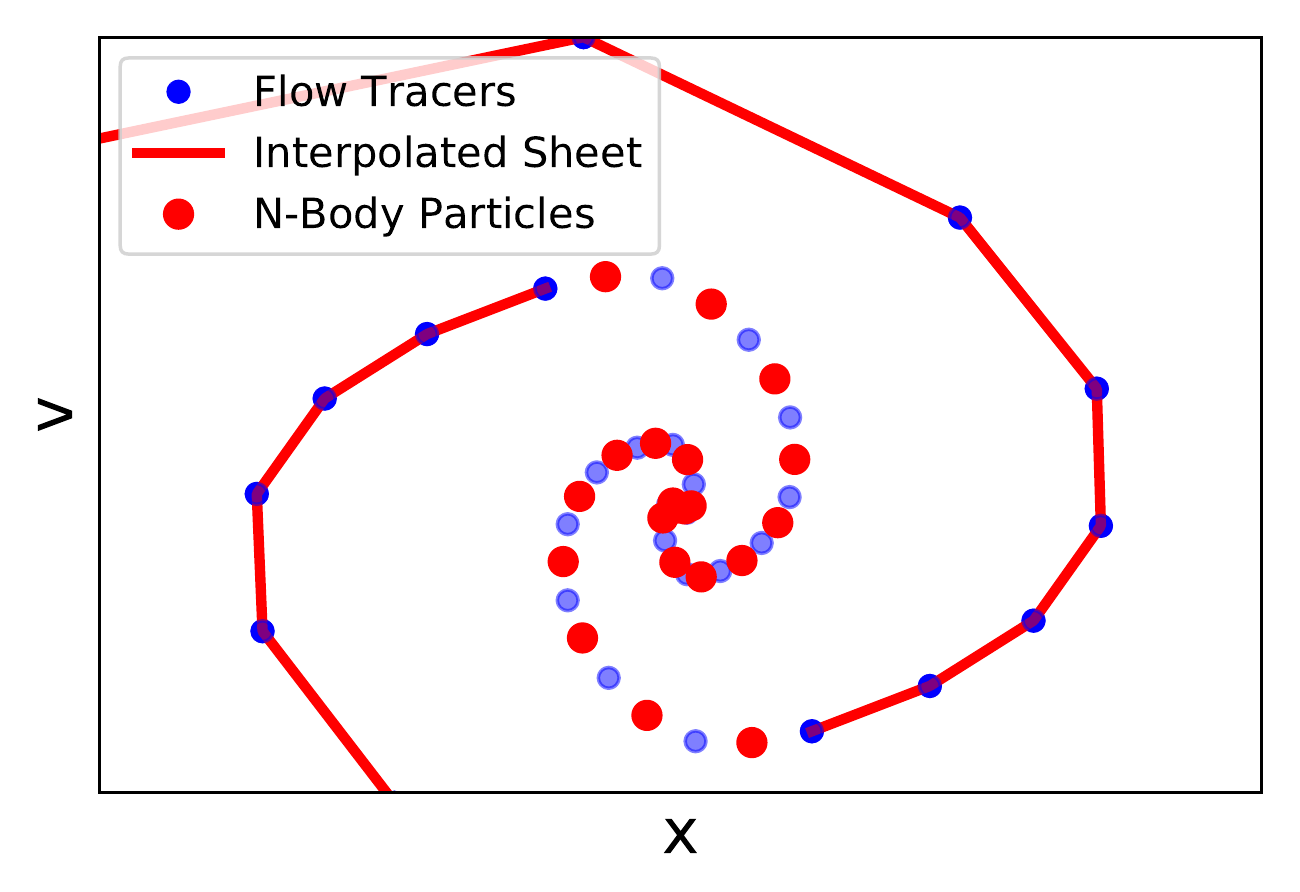}
  \end{center}
  \caption{Top: Illustration of the dark matter sheet in phase space\hlt{, where dark matter occupies a submanifold}. This submanifold can be reconstructed by interpolation from a finite number of tracers. Note \hlt{that} in this image we purposely use only linear interpolation to emphasize the difference between the true dark matter sheet and the one reconstructed from interpolation. 
  Bottom:  Illustration of the release. Originally the mass (in red) in all Lagrangian volume elements was traced by the interpolated sheet. However, in the course of the simulation some mass elements have been flagged for release and are now represented by N-body particles instead. }
  \label{fig:sheet1d}\label{fig:sheet1d_release}
\end{figure}

In comparison to its bulk velocities, dark matter has a tiny thermal velocity dispersion at low redshift. Its bulk velocities and the gravitationally induced dispersion velocities within nonlinear objects typically reach hundreds of $\SI{}{\kilo\metre\per\second}$ at the present day, whereas the thermal velocity dispersion in unstructured regions is well below $\SI{100}{\metre\per\second}$, even for the hottest non-excluded warm dark matter models. Therefore dark matter forms a thin three-dimensional submanifold in six-dimensional phase space. Since it is collisionless, its dynamics are governed by smooth long-range gravitational forces, but are unaffected by short-range scattering depending on the local phase space distribution. \hlt{The} equations of motion are symplectic, phase space densities are conserved, and the dark matter fluid continues to occupy a three-dimensional submanifold at all times. We refer to this submanifold as the \emph{dark matter sheet} \citep{arnold_1982,shandarin_zeldovich_1989,white_vogelsberger_2008,abel_2012,shandarin_2012}. 

This is most easily visualized in the two-dimensional phase space of a one-dimensional universe as shown in the top panel of Figure \ref{fig:sheet1d}. While the exact dark matter sheet can only be inferred by simulating an infinite number of particles, a very good approximation to it can be obtained by using a finite number of tracer points and interpolating between them. The quality of the interpolation will depend on the complexity of the dark matter sheet, on the number of sampling points (particles) and on the order of interpolation.

As can already be suspected from the top panel of Figure \ref{fig:sheet1d}, the complexity of the dark matter sheet grows over time. While the sheet is very well represented by a linear interpolation between the particles in the initial conditions (top left panel), the representation by interpolating between the particles at later times is already considerably worse. The function that is to be captured by the interpolation has grown in complexity. A better representation of it can be obtained by using a higher interpolation order, or by using a higher number of particles. \citet{hahn_angulo_2016} and \citet{sousbie_colombi_2016} have explored both of these approaches by using a higher order triquadratic interpolation, and by implementing refinement schemes in Lagrangian space. Those schemes try to detect when and where the sheet is growing in complexity and create additional flow tracers to trace the interpolation more accurately in those regions.

With these schemes it is, in principle, possible to make a simulation that exactly traces the dark matter sheet. However, this turns out to be impossible in practice, since the dark matter sheet grows in complexity very rapidly. Tracing it requires an extraordinary amount of computational resources. \citet{sousbie_colombi_2016} managed to carry out a $m_X = \SI{250}{\electronvolt}$ WDM simulation in a $\SI{28}{\mega\parsec}$ box until $a = 0.31$ and found the number of simplexes required to scale with the twelfth power of time. Assuming that this scaling remains valid until $a=1$, running that simulation until the present time would require roughly $10^{6}$ times as much memory and probably also computational time. However, it is already an optimistic assumption that this scaling can be extrapolated so easily. So soon after their formation, their haloes had probably had no mergers yet and so had maintained a relatively simple phase space structure. It cannot be excluded that chaotic orbits with an exponentially growing complexity arise from merging haloes. Further the simulation described in \citet{sousbie_colombi_2016} has a relatively low force-resolution of $\SI{28}{\mega\parsec} / 1024 \approx \SI{27}{\kilo\parsec}$. Therefore, the central structure of haloes is resolved poorly. The complexity of the sheet is expected to grow most rapidly in the centres of haloes. \citet{vogelsberger_white_2011} find that the number of streams at a single point in the centre of the halo of a Milky-Way-type dark matter halo in a cold dark matter universe might already get as high as $10^{16}$. If that is true, a dark matter sheet plus refinement based simulation scheme for such a halo would require far more than $10^{16}$ resolution elements.

Even in the most optimistic case it seems unlikely that a simulation like that in \citet{sousbie_colombi_2016} can be run until the present day, $a = 1$.  We will demonstrate in this paper how to deal with this with affordable computational costs. We propose a simulation scheme with a ``release'' mechanism that uses a sheet-interpolation scheme where it is well converged, and switches to a particle based N-body approach in regions where the sheet becomes too complex. This allows us to perform the first warm dark matter simulations that do not fragment in low density regions while remaining accurate in the inner regions of haloes. While thus giving up the qualities of a density field estimated from the smooth sheet, we pay particular attention that N-body particle noise remains unimportant at all times (see discussion in Section \ref{sec:timevolgde}) so that no disadvantage, apart from the loss of the possibility to keep tracking the full fine-grained distribution function, arises in the evolution by resorting to N-body. We illustrate qualitatively in the bottom panel of Figure \ref{fig:sheet1d_release} how the release could look in the phase space of a one dimensional world. Note that in the case of a three dimensional simulation, the complexity in the released region would be much higher - for example it could have $\sim 10^{16}$ foldings in the same region \citep{vogelsberger_white_2011}.

The next sections will explain how we identify regions where the interpolation scheme is no longer reliable without refinement.

\subsection{Quantifying Complexity - The Geodesic Deviation Equation}
The dynamics of the dark matter sheet is best described by considering the Lagrangian map $\myvec{q}\mapsto\left(\myvec{x}(\myvec{q}),\myvec{v}(\myvec{q})\right)$ between the three-dimensional Lagrangian coordinates $\myvec{q}$ that give the location on the sheet, to the Eulerian coordinate $\myvec{x}$ and velocity $\myvec{v}$. $N$-body methods correspond then to a finite and discrete set of $\left\{\myvec{q}_i\right\}$, in interpolation schemes, such a finite set is used to approximate the submanifold itself. While interpolation schemes trace the deformations of the dark matter sheet $\myvec{x} (\myvec{q})$ well as long as $\myvec{x} (\myvec{q})$ varies slowly, they break down when the function varies rapidly on the separation scale $\Delta q$ of the sampling points $(\myvec{q}_i, \myvec{x}_i)$ (i.e. particles).

In this case it is no longer possible to trace the whole fine-grained phase space sheet exactly, but its deformation local to each tracer can still be traced by the Geodesic Deviation Equation \hlt{(GDE)} \citep{vogelsberger_white_2008,white_vogelsberger_2008,vogelsberger_white_2011}. The deformations of the dark matter sheet can be characterised by the Jacobian of the mapping from Lagrangian to Eulerian space $\frac{\partial \myvec{x}}{\partial \myvec{q}}$, also known as the real-space distortion tensor, which simply represents the space tangent to the dark matter sheet. It quantifies how an infinitesimal Lagrangian volume element gets stretched and rotated during the course of the simulation \citep{vogelsberger_white_2008,vogelsberger_white_2011}. 

\begin{figure*}
	\includegraphics[width=0.9\textwidth]{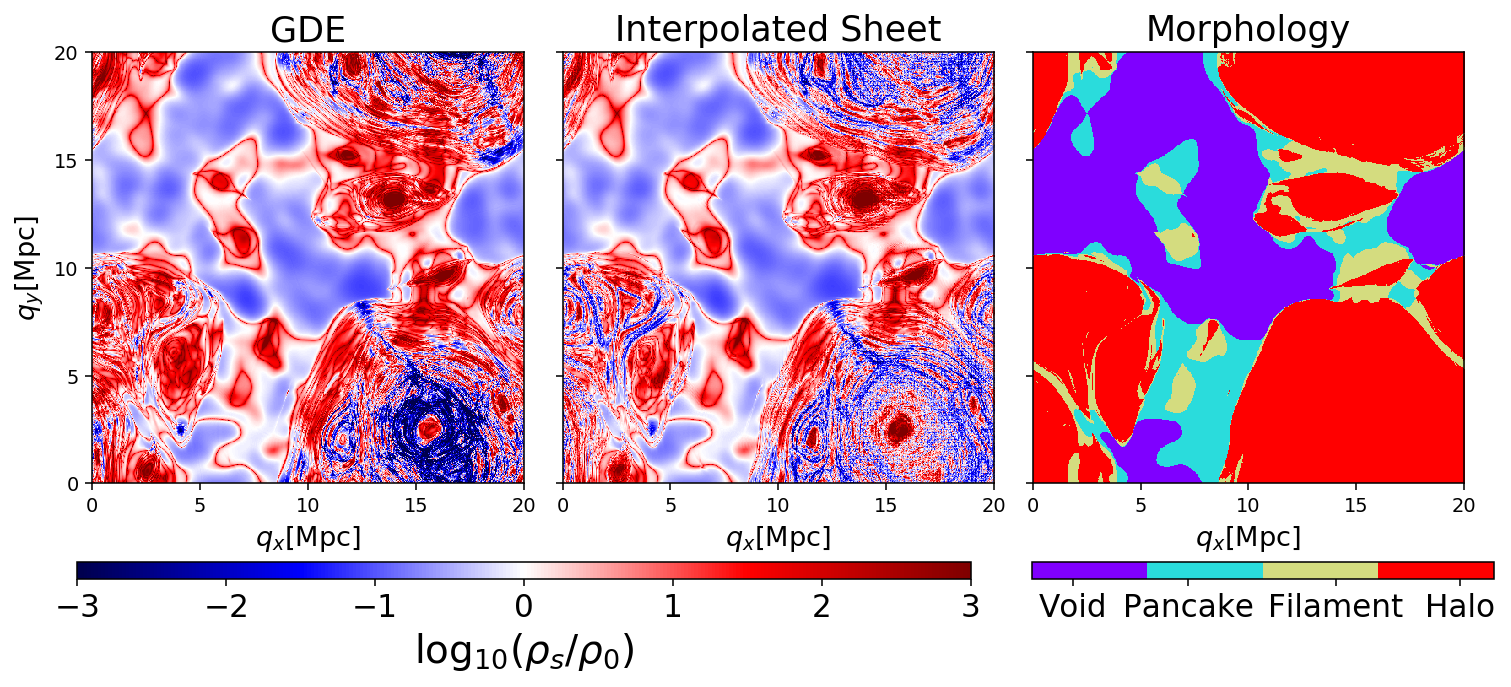}
    \caption{Stream-Densities on an infinitesimally thin plane in Lagrangian Space. Left: \hlt{Stream densities defined at} \hlt{tracer positions as given by the GDE distortion tensor.} Centre: Finite differences approximation, representing the derivatives of the interpolated sheet. Right: Morphology classification as described in section \ref{sec:structure_classification}. The stream densities agree extraordinarily well in regions where they vary slowly with the Lagrangian coordinates, but get into complete disagreement in regions where they vary rapidly -- \hlt{this is the case in haloes}. The sheet is too complex here for accurate reconstruction.}
    \label{fig:lagrangian_stream_densities}
\end{figure*}

In comoving coordinates the equations of motion of arbitrary particles in an expanding universe can be written as
\begin{align}
  \dot{\myvec{x}} &= a^{-2} \myvec{v} \\
  \dot{\myvec{v}} &= - a^{-1}  \nabla \phi(\myvec{x})\label{eqn:equations_of_motion}
\end{align}
\citep[e.g.][]{peebles_1980}\hlt{,} where $\myvec{v} / a$ are the peculiar velocities, $\myvec{x} a$ are physical coordinates and $\phi$ is the comoving potential which obeys the Poisson equation
\begin{align}
  \nabla^2 \phi &= \frac{3}{2}H_0^2 \Omega_0 a^{-1} \delta \\
  \delta            &= \frac{\rho - \rho_{0}}{\rho_{0}},
\qquad \rho_{0}=\Omega_0 \rho_{c,0}\hlt{\,,}
\end{align}
where all spatial derivatives are taken with respect to comoving coordinates, $\delta$ is the over-density, $\rho_0$ the mean matter density of the universe today, $\rho_{c,0}$ the critical density today, $H_0$ the Hubble constant and $\Omega_0$ the matter density parameter\hlcom{(declared more variables)}. One can infer the equations of motion of the distortion tensor by differentiating \eqref{eqn:equations_of_motion} with respect to the Lagrangian coordinates $\myvec{q}$. Writing $\Dxq := \frac{\partial \myvec{x}}{\partial \myvec{q}}$ and $\Dvq := \frac{\partial \myvec{v}}{\partial \myvec{q}}$ and defining the tidal tensor $\mathbfss{T} = - \nabla\otimes\nabla \phi$ these can be written as
\begin{align}
  \Dxqdot &= a^{-2} \Dvq  \nonumber \\
  \Dvqdot &= a^{-1} \T \Dxq \label{eqn:equations_of_motion_dxq} \hlt{\,.}
\end{align}
While it is also possible to investigate how small velocity displacements in the initial conditions affect the final positions by using the full six-dimensional distortion tensor \citep{vogelsberger_white_2008,vogelsberger_white_2011}, we limit our discussion to the real-space part here. In the case of the dark matter sheet, the dynamics of $\Dxq$ describe the evolution of the space tangent to the sheet, while those of $\Dvq$ describe the space normal to it.

The geodesic deviation equation \eqref{eqn:equations_of_motion_dxq} can be integrated along with the other equations of motion for each particle in a cosmological simulation to determine the real-space distortion tensor as an additional property of each particle. The distortion tensor can be initialized by using finite differencing methods on the initial conditions, so that at 2nd order

\begin{align}
  \frac{\partial{x_i}}{\partial{q_j}} &\approx \frac{{x}_i(\myvec{q} + \Delta q \myvec{e}_j) - {x}_i(\myvec{q} - \Delta q \myvec{e}_j)}{2 \Delta q} \label{eqn:fd_distortion_tensor} \\
  \frac{\partial{v_i}}{\partial{q_j}} &\approx \frac{{v}_i(\myvec{q} + \Delta q \myvec{e}_j) - {v}_i(\myvec{q} - \Delta q \myvec{e}_j)}{2 \Delta q}\hlt{\,,}
\end{align}
where $\myvec{e}_j$ is the unit vector along the $j$-th coordinate axis. If the simulation particles are initially located on a grid, the evaluation points of the finite differencing can be chosen so that the distortion tensor can be approximated purely by taking differences of particle positions. We note that these tensors can also be explicitly calculated when generating cosmological initial conditions. We found however that finite differences are accurate enough, and can be conveniently calculated from just particle positions and velocities.

In the initial conditions of a typical cosmological simulation this will always be a reasonably good approximation, since initial conditions are typically set at a time where the displacement field varies only moderately between Lagrangian neighbors. However, the finite-differencing scheme can also be used to obtain an approximation for the distortion tensor at any later time. If the Lagrangian map $\myvec{x} (\myvec{q})$ varies slowly with the Lagrangian coordinate this will be a good approximation, but if it varies rapidly, the approximation will break down. These are the cases where the dark matter sheet becomes too complex to be reconstructed from particle positions.

To illustrate this, we show in Figure \ref{fig:lagrangian_stream_densities} a comparison of the stream densities\hlt{,}
\begin{align}
  \rhos = \frac{\rho_0}{| \det (\Dxq) |}\hlt{\,,}
\end{align}
that can be obtained from the finite difference distortion tensor as in \eqref{eqn:fd_distortion_tensor} and the infinitesimal distortion tensor that has been evaluated by the GDE. Additionally, we show the result of a morphology classification based on the distortion tensor which we will describe in section \ref{sec:structure_classification}. The Figure shows a slice through Lagrangian space, where each particle is plotted at its initial comoving location $\myvec{q}$ (for $a \rightarrow 0$), but is coloured by the stream density that it has at a later \hlt{time} in the simulation. In Lagrangian space, the volume is proportional to the mass, therefore for example haloes appear as large regions in Lagrangian space whereas they make up a rather small part of the volume in Eulerian space \hlt{(and vice versa for voids)}. It can be seen that the GDE and the finite difference distortion tensor are in excellent agreement wherever the stream-density varies slowly - that is in single-stream regions, pancakes and filaments, as we shall see later. However, there are also regions (i.e. haloes) where the stream density varies rapidly. The finite-differencing scheme breaks down here and becomes resolution-dependent, whereas the local distortion tensor of the GDE still remains valid. For infinite resolution the finite-differencing scheme should converge to the GDE result.

\begin{figure}
	\includegraphics[width=\columnwidth]{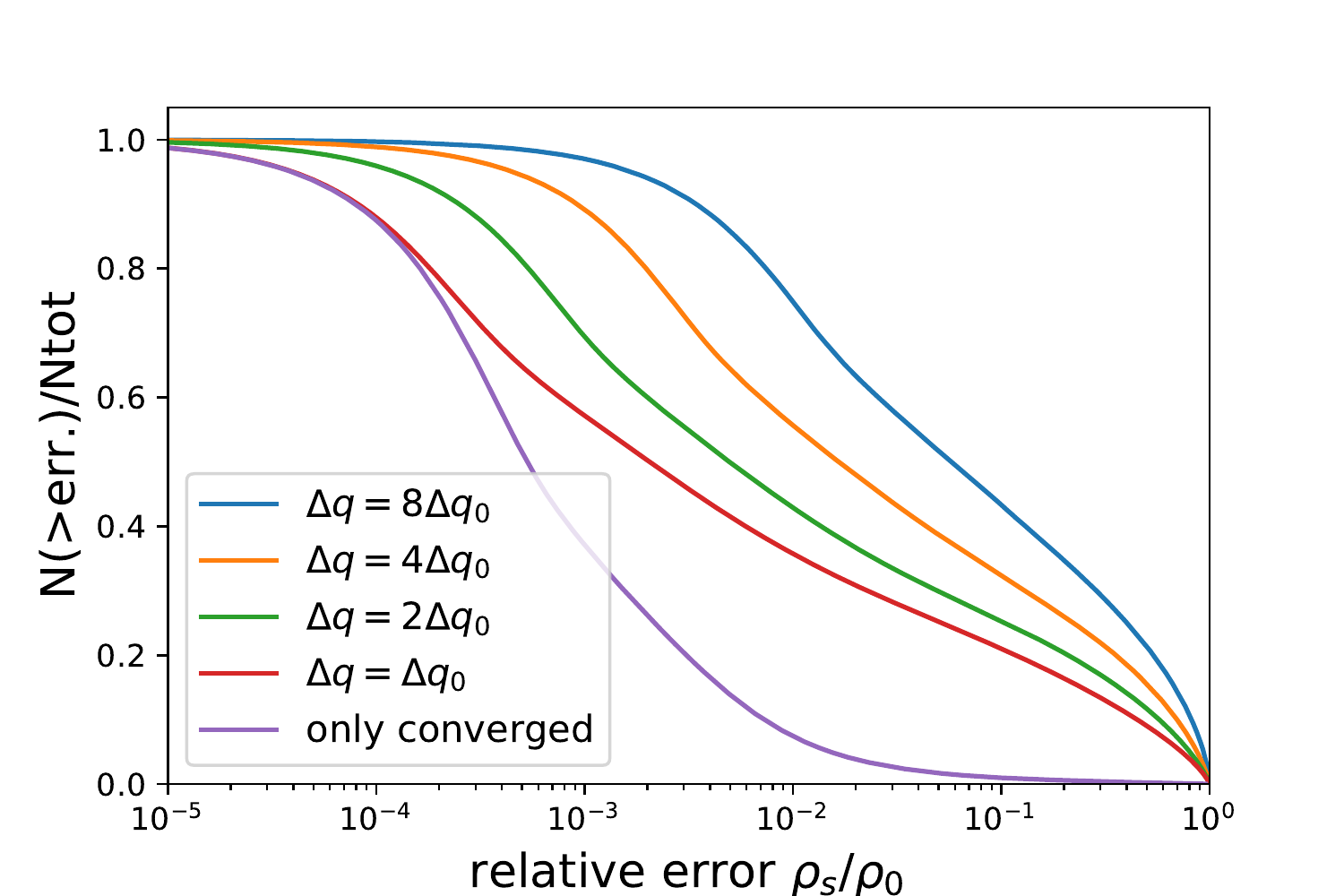}
    \caption{\hlcom{Changed text: <--} Cumulative distribution of differences between the stream densities obtained from a finite-difference version of the Jacobian based on the sheet tessellation, and the Jacobian from the GDE. For the finite differencess, we show results for increasingly coarser sampling (in terms of $\Delta q$). The graph shows how many have an error larger than $\rho_s/\rho_0$. \hlcom{-->} 
    The finite difference distortion tensors converge quickly to the GDE distortion tensor for most of the particles. For roughly $30-40 \%$ of the particles the convergence is rather slow. Particles which have a converged distortion tensor (at the 10\% level) between different resolution levels (purple line) are in excellent agreement with the GDE distortion tensor, showing that the GDE distortion tensor is indeed the limit of the sheet derivative for infinite particle resolution $\Delta q \rightarrow 0$.} 
    \label{fig:streamdensity_convergence}
\end{figure}

We demonstrate this more quantitatively in Figure \ref{fig:streamdensity_convergence} where we plot the \hlt{cumulative histogram of relative differences} between the GDE stream densities $\hlm{\rho_{\text{s,gde}}}$ and the stream densities inferred from finite differences $\hlm{\rho_{\text{s,fd}}}$
\begin{align}
  \epsilon = \left| \hlm{\frac{\rho_{\text{s,gde}} - \rho_{\text{s,fd}}}{\rho_{\text{s,gde}} + \rho_{\text{s,fd}}} } \right|\hlt{\,.}
\end{align}
To test the convergence, we compare the finite differencing for different resolution levels, using all particles, every second particle (per dimension), every 4th and every 8th. It can be clearly seen that the finite difference stream densities converge to the GDE stream density. The GDE provides the limit of the distortion tensor for $\Delta q \rightarrow 0$ -- that is the derivative of the actual dark matter sheet. In contrast the finite differencing provides the derivatives of the interpolated sheet, which are wrong where the interpolation is not converged.

To further emphasize this, we additionally select a subset of particles where the finite differencing is converged, which we define by their stream densities not changing by more than 10 $\%$ when only selecting every second or every fourth particle. For particles where the finite differencing has converged, the agreement with the \hlt{GDE-based density is remarkably good.} We conclude that the comparison of GDE and finite difference distortion tensor can be reliably used as a benchmark for the accuracy of sheet interpolation schemes. 

Looking more closely at Figure \ref{fig:streamdensity_convergence}, it seems puzzling that while for roughly 60$\%$ of the particles the stream densities converge quickly, for the other 40 $\%$ of the particles the stream densities converge rather slowly.  We shall see that those regions where convergence is slow are mostly haloes, and that achieving true convergence here is almost hopeless. Any sheet interpolation scheme will either break down (without using refinement) or become too expensive to be followed (when using refinement) at some point.

\begin{figure}
	\includegraphics[width=\columnwidth]{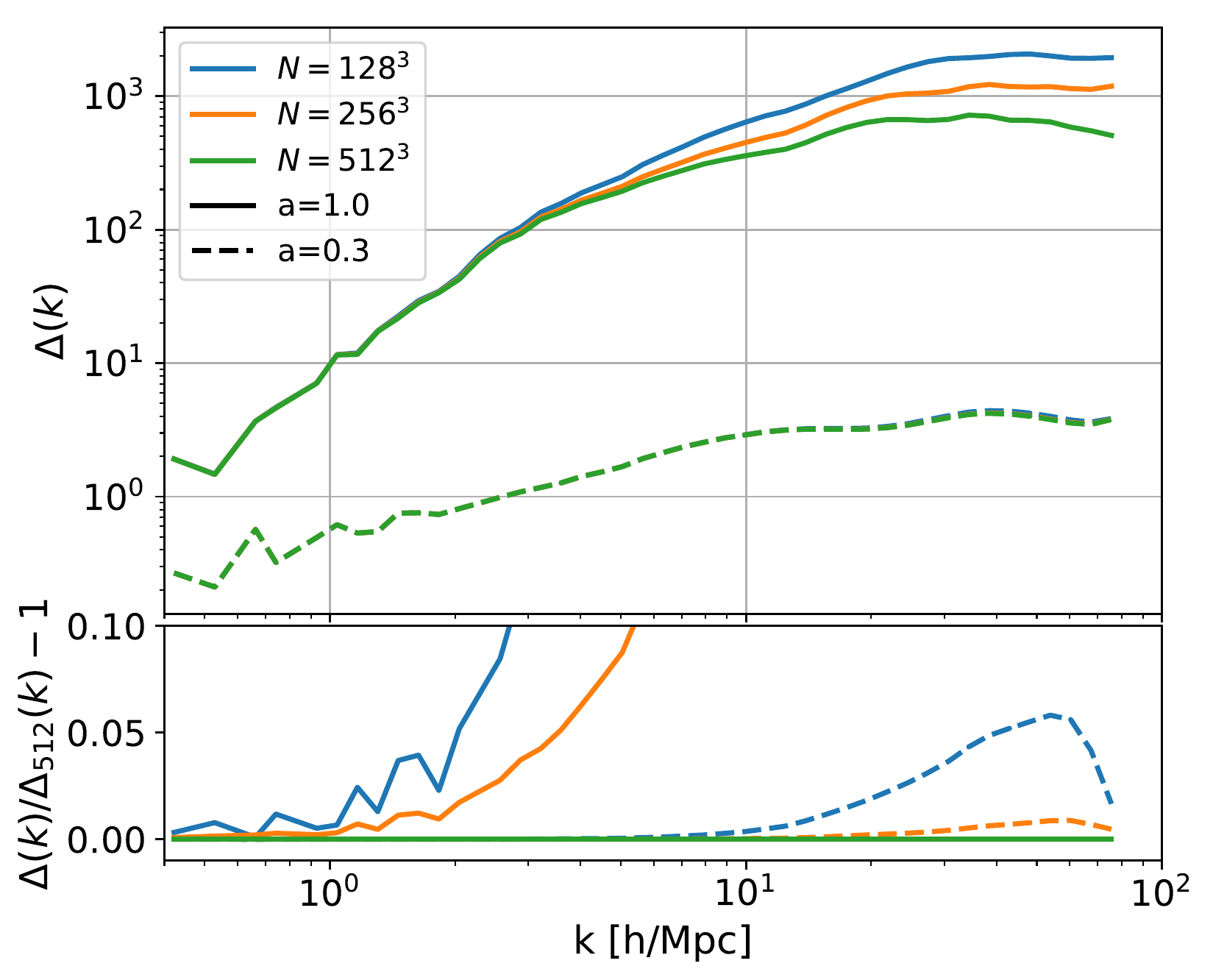}
    \caption{Comparison of dimensionless power-spectra (top) and residuals with respect to the highest resolution (bottom) at two different times. At an early time $a = 0.3$ the power spectra converge well, since the resolution is sufficient for reconstructing the sheet. At late times $a = 1.0$ the power spectra converge only slowly, the sheet is too complex for reconstruction with a fixed finite resolution. }
    \label{fig:power_convergence_sheet}
\end{figure}

To illustrate how important it is that the interpolation is converged, we show in Figure \ref{fig:power_convergence_sheet} power-spectra that have been inferred on sheet-based dark matter simulations (without any refinement) with different resolutions at two different points in time. While it seems that at $a = 0.3$ the power spectra are converged, showing that the sheet-interpolation is working reasonably well at that time, the situation is very different at $a=1$: The complexity of the true dark matter sheet is too high to be captured by the interpolation scheme with a limited number of particles. Therefore the interpolation and subsequently the power spectrum converge only slowly with the particle number. Typically, the densities in the centres of haloes get strongly overestimated by a poorly interpolated reconstruction, as has been already demonstrated by \cite{hahn_angulo_2016}.



\subsection{Structure Classification}
\label{sec:structure_classification}
As we have already mentioned in the introduction, the anisotropic nature of cosmological gravitational collapse together with the absence of thermalisation in collisionless dynamics \citep[cf.][]{buehlmann_hahn_2018} makes anisotropically collapsed structure particularly vulnerable to particle noise. At the same time, the lower dimensional dynamics in those regions restricts the dynamics severely, so that ultimately it would be desirable to disentangle the unproblematic regions where dynamics is close to ergodic in all dimensions (haloes) from the problematic ones where this only true for dynamics in subspaces (filaments, sheets and voids). Therefore we developed a scheme to classify particles into void, pancake, filament or halo particles, purely by their dynamics. 

The distortion tensor describes the three dimensional distortions of an infinitesimal Lagrangian volume element around each particle. These distortions can be a mixture of stretching and rotations (+mirroring). While the stretching happens in all phases of the evolution of the volume elements, rotations do not. The number of rotationally active axes can be used to classify the structure a particle is part of - zero corresponds to a void, one to a pancake, two to a filament and three to a halo. The idea is similar to that of the origami method in \citet{falck_2012}. The origami method uses the number of Lagrangian axes along which a shell crossing has happened to classify the morphology. However, our scheme does not contain any preferred directions along which the shell crossing is tested and is invariant under rotations of the density field. Further it only uses the infinitesimal surroundings of each particle for the classification thus highlighting how the local phase space dynamics is related to the structure a particle is part of.

\begin{figure}
	\includegraphics[width=\columnwidth]{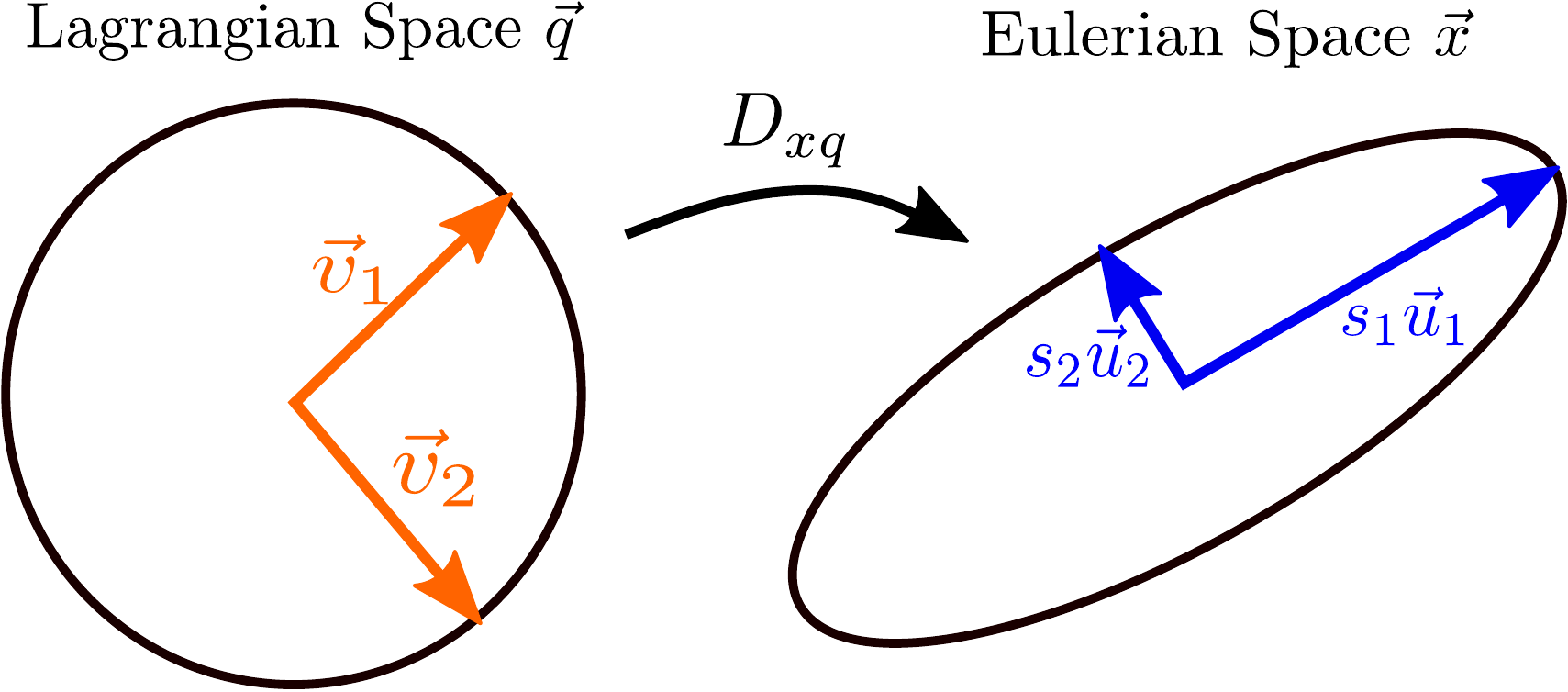}
    \caption{Illustration of the components of the singular value decomposition $\hlm{\Dxq} = U S V^T$. The Matrix $\hlm{\Dxq}$ maps a sphere from Lagrangian space to an ellipsoid in Eulerian space. The column vectors of $U$ give the orientations of the major axes in Eulerian space, the singular values give the relative sizes of the axes, and the column vectors of $V$ give the orientations in Lagrangian space. \hlcom{(Figure changed: arrow v2 color)}}
    \label{fig:svd}
\end{figure}
To disentangle stretching from rotations, we use the singular-value decomposition of the distortion tensor. Any matrix can be decomposed in the form
\begin{align}
  \hlm{\Dxq} = U S V^T
\end{align}
where $U$ and $V$ are orthogonal matrices and $S$ is a diagonal matrix where the diagonal elements are called the singular values $s_i$. If the Matrix $\hlm{\Dxq}$ is symmetric, the singular-value decomposition becomes equivalent to the eigenvalue decomposition: then the singular values are the absolute values of the eigenvalues and $U = V$. However the singular value decomposition also has a simple geometric interpretation in the case of general non-symmetric matrices (like the distortion tensor). We illustrate this in Figure \ref{fig:svd}. The distortion tensor $\hlm{\Dxq}$ maps a unit sphere in Lagrangian space to a distorted ellipsoid in Eulerian space. The column vectors $\myvec{u}_i$ of the matrix $U$ give the orientations of the major axes of the ellipsoid in Eulerian space. The singular values $s_i$ quantify the stretching along the major axes. The column vectors $\myvec{v}_i$ of $V$ give the orientations of the major axes in Lagrangian space. So the general vector $a \myvec{v}_1 + b \myvec{v}_2$ gets mapped by $\hlm{\Dxq}$ as
\begin{align}
  \hlm{\Dxq} \cdot (a \myvec{v}_1 + b \myvec{v}_2) = a s_1 \myvec{u}_1 + b s_2 \myvec{u}_2\hlt{\,.}
\end{align}

  To quantify rotations we define three angles $\alpha_i$ from the singular value decomposition\hlt{,}
\begin{align}
  \alpha_i = \arccos(\myvec{v_i} \cdot \myvec{u_i})\hlt{\,,}
\end{align}
which are the relative angles between the orientation of the major axes in Lagrangian space and in Eulerian space. Note that this choice of angles is independent of the coordinate system (unlike most possible angle definitions). In the example from Figure \ref{fig:svd} the angle $\alpha_1$ would be relatively small whereas the angle $\alpha_2$ would be close to $180^\circ$.

We trace the angles $\alpha_i$ of the distortion tensor of every particle in our simulations and save the maximum value of them for every particle. We classify structures then by the number of angles for which this maximum exceeds $\pi/4$:

\begin{figure}
	\includegraphics[width=\columnwidth]{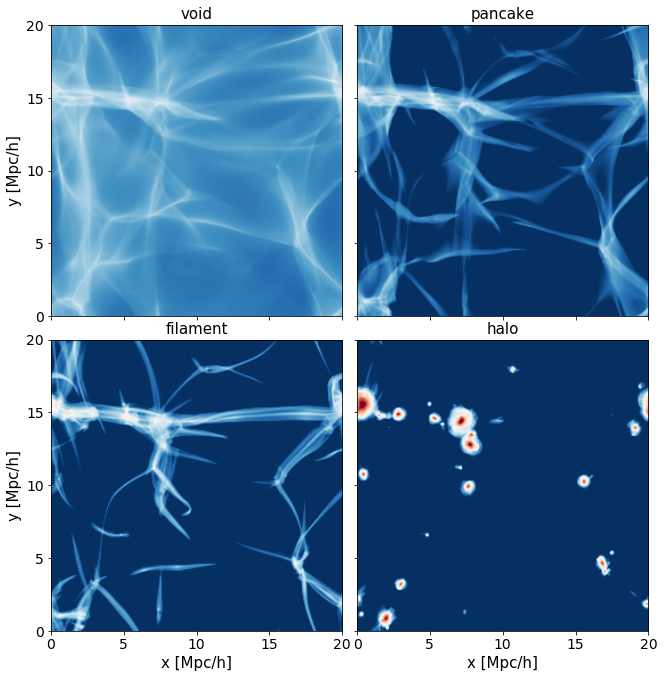}
    \caption{Density projections of a \SI{20}{\mega\parsec\per\h} cosmological warm dark matter box selecting different subgroups of the particles according to their structure class. The structure classification is done by counting the number of axes of the Lagrangian volume elements that are still aligned with their initial orientation: All axes aligned corresponds to a void, one axis misaligned (or flipped) to a pancake, two axis misaligned to a filament, and all axes rotated arbitrarily in comparison to their initial alignment corresponds to a halo.}
    \label{fig:structure-classification}
\end{figure}

In the beginning of the simulations all particles are in a single-stream region (or a void). The Zeldovich approximation \citep{zeldovich_1970} only permits symmetric deformations of the distortion tensor and therefore in the initial conditions all angles are zero by definition. During the evolution in a void a particle mostly undergoes a stretching and/or \hlt{compression} along the three major axes, but almost no rotation. Then when it \hlt{passes} through a first caustic in a pancake, the smallest axis goes through zero and flips its orientation and the associated angle gets close to $\pi$. From that point on, one axis of the volume element is dynamically active whereas the other two axes are still aligned with their initial orientations. Once the particle becomes part of a filament, another axis becomes dynamically active and suddenly rotations in a plane perpendicular to the filament become possible. However, the axis aligned with the filament still roughly maintains its orientation. At this point two angles can be significantly different from zero. The last axis only becomes dynamically active when the particle falls into a halo. 

Therefore, we can classify the particles into structures by counting the number of angles $\alpha_i$ that have deviated substantially from zero. Note that we use the maximum value of each angle along the whole trajectory and not only its current value. This way one can avoid misclassifications in cases where the axes align by chance after having been misaligned for most of the time. In the appendix, in Figure \ref{fig:lagrangian_maps}, we provide a set of Lagrangian maps which show the difference between using the current and the maximum value. This figure also shows how the angles are active in clearly distinct Lagrangian regions.

We show the result of this classification in Figure \ref{fig:structure-classification}. Arguably the classification selects regions in the same way one would intuitively classify them. However, we want to point out that since our classification is based on the dynamical behavior of particles and not purely Eulerian space properties, particles can coexist at the same location but be assigned to different morphological structures.

\begin{figure}
	\includegraphics[width=\columnwidth]{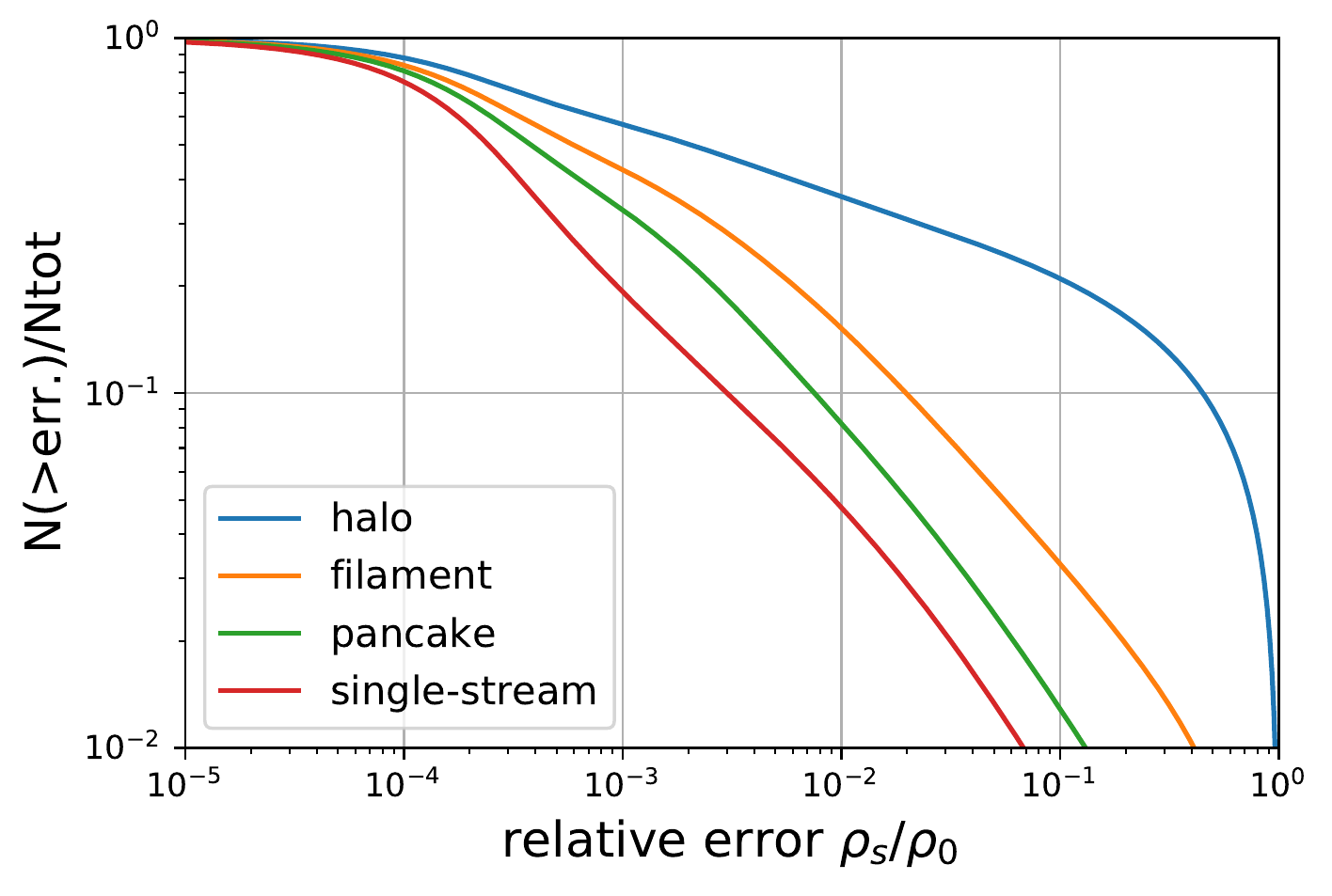}
    \caption{Cumulative histogram of the relative differences of the GDE and finite difference stream densities for different structure types. The dark matter sheet can be reconstructed well outside of haloes \hlt{but only poorly} within haloes.}
    \label{fig:stream_density_error_by_structure}
\end{figure}

We show in Figure \ref{fig:stream_density_error_by_structure} the relative differences between GDE and finite difference stream densities for different structure types. Clearly particles in haloes have significantly larger errors in their stream densities than those in other structures. Simplifying a bit, we can say that the sheet interpolation works well outside of haloes, but tends to break down inside of haloes. This is good news for the possibility of fragmentation-free warm dark matter simulations: It is well known that N-body simulations tend to fragment in filaments. There small discreteness errors in the density estimate can quickly evolve into artificial fragments, since the (coarse-grained) distribution is heated up in two dimensions, but still cold within the third dimension. Errors can easily couple in this situation and amplify. However, in the case of a halo, where the distribution is hot in all dimensions, it is unlikely that small errors cause significant larger scale errors. On the other hand, sheet schemes do not have the problem of fragmentation, because their density estimate is much more accurate and less noisy than the N-body density estimate in low-density regions such as single-stream regions, pancakes and filaments. However, they become intractable inside of haloes because of the rapidly growing complexity.

It is an obvious next step to combine the benefits of the two schemes to achieve fragmentation-free unbiased warm dark-matter simulations. Therefore, we develop a scheme that initially traces all matter by a sheet-interpolation scheme, then detects (Lagrangian) regions where the sheet becomes too complex to be followed accurately by interpolation and switches to an N-body approach for those. We call this switch to an N-body approach ``release''.

\subsection{The Release}

\begin{figure*}
    \includegraphics[width=\textwidth]{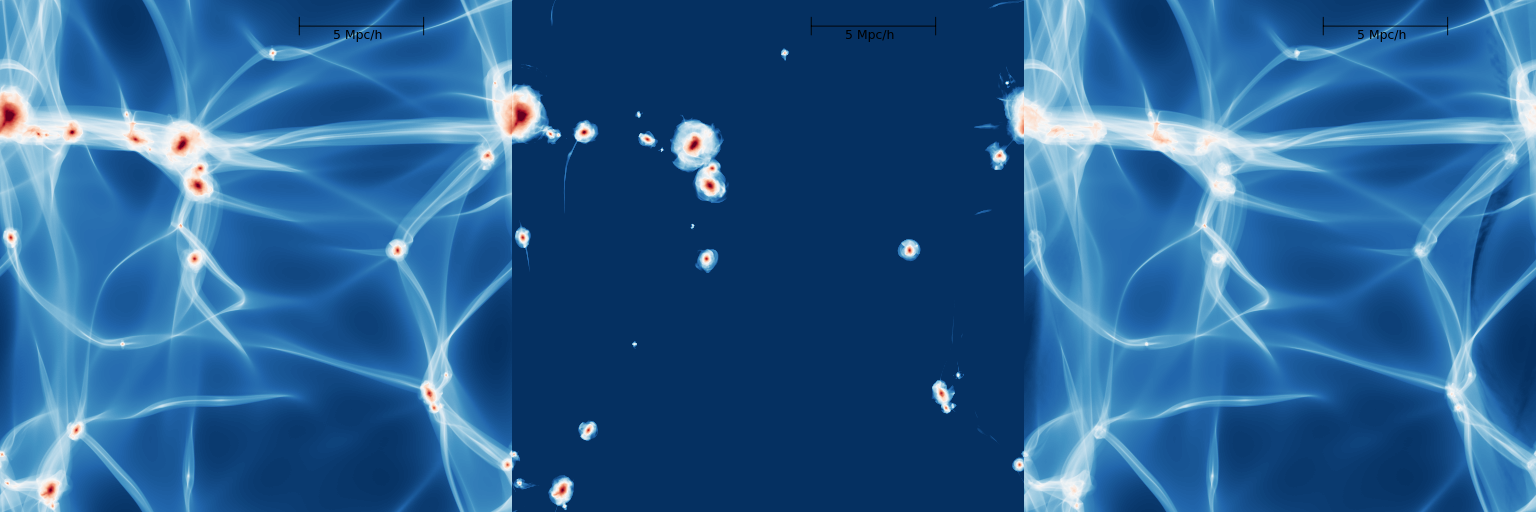}
    \caption{Density projection of a simulation using a hybrid Sheet/N-body scheme. Left: Projection of the total mass. Centre: Projection of the mass that is traced by N-body particles. Right: Mass that is traced by the sheet-interpolation. This combination of schemes produces simulations with high quality density estimates in low-density regions which are simultaneously correct in haloes.}
    \label{fig:released_density}
\end{figure*}

Initially, we trace all mass in the simulation by a sheet interpolation scheme. This means that we follow the dynamics by a set of mass-less particles which we call \emph{flow tracers} which follow normal Newtonian equations of motion. However, forces are not estimated from the usual N-body interactions, but instead by assigning the mass of the interpolated sheet to a mesh (or another Eulerian discretization structure as will be discussed in Section \ref{sec:force}), and solving the Poisson equation for this density field. When we say that we release a Lagrangian volume element, we mean that from that point on its mass is no longer assigned by using the sheet interpolation, but simply by depositing independently traced N-body particles. We illustrate this in the bottom panel of Figure \ref{fig:sheet1d_release}. 

In our code, we have two different methods available for choosing the positions and velocities of the released N-body particles. The first method simply evaluates the phase space coordinates of the interpolated dark matter sheet $(\myvec{x}(\myvec{q}), \myvec{v}(\myvec{q}))$ at a set of Lagrangian coordinates $q_{ijk}$. For example if a cubic Lagrangian volume element starting at $\myvec{q}_0$ with Lagrangian side length $\Delta q$ should be released into $N^3$ N-body particles, their parameters would be chosen the following way:
\begin{align}
  \myvec{q}_{ijk} &= \myvec{q}_0 + \frac{\Delta q}{2 (N+1)}  \left(1+2i,\, 1+2j,\, 1+2k \right)^T \\ 
  \myvec{x}_{ijk} &= \myvec{x} (\myvec{q}_{ijk}) \\
  \myvec{v}_{ijk} &= \myvec{v} (\myvec{q}_{ijk}) \\
  m_{ijk} &= N^{-3} \rho_0 \Delta q^3\hlt{\,,}
\end{align}
where $i$, $j$ and $k$ are going from $0$ to $N-1$. An advantage of this release scheme is, that it is only necessary to trace N-body particles for the released mass, and it is in principle possible to control the mass resolution within released regions (mostly haloes as we shall see) independently. However, while this works fine in principle, it is relatively sensitive to making even a small error in the interpolation at the time of the release. If e.g. the velocity of the newly created N-body particles is off by just a percent it can already affect their future trajectories by a large amount. We found that this can lead to peculiar effects in some cases - for example some N-body particles which are expelled slightly beyond the splashback radius of a halo.

To make sure that there are no artefacts created from small inaccuracies at release, we implemented an additional release method. In this alternate method, we trace an additional set of mass-less particles at the locations $\myvec{q}_{ijk}$ from the beginning of the simulation. We call these particles \emph{silent particles} since they have no impact on the simulation prior to their release. However, when their Lagrangian volume element is released, they are converted to N-body particles with the mass $m_{ijk}$ - thus creating new particles at exactly the correct location with exactly the correct velocity and no interpolation errors. We use this release method for the remainder of this paper.


Note that the release is defined per Lagrangian volume element. Therefore at the same location in Eulerian space released N-body particles and interpolated sheet-elements can coexist. 

\subsection{Release Criterion}

It is important to reliably identify the Lagrangian elements for which the sheet becomes too complex to be traced. We have developed two different criteria to flag elements for the release. The first criterion compares for each particle in a volume element the finite-difference distortion tensor $\hlm{\mathbfss{D}_{\text{xq,fd}}}$ with the GDE distortion tensor $\hlm{\mathbfss{D}_{\text{xq,gde}}}$. If their alignment $a(\hlm{\mathbfss{D}_{\text{xq,fd}}}, \hlm{\mathbfss{D}_{\text{xq,gde}}})$, which we define as
\begin{align}
  a(A, B) = \frac{\sum_{i,j} A_{ij} B_{ij}}{\sqrt{\sum_{ij} A_{ij}^2 \sum_{ij} B_{ij}^2}}, \label{eqn:alignment}
\end{align}
becomes smaller than a threshold value $a_{\text{min}}$ for any particle in a volume element, we flag that element for release. We show density projections of a simulation which uses this release criterion with $a_{\text{min}} = 0.99$ in Figure \ref{fig:released_density}. The benefits of the release technique become clearly evident here. It allows us to get realistic haloes, while at the same time getting the accurate non-fragmenting sheet-density estimate in the low-density regions.

Additionally, we defined an alternative criterion which simply releases all Lagrangian volume elements where at least one particle becomes part of a halo. To detect whether a particle becomes part of a halo, we use the criterion based on the angles of the distortion tensor as described in Section \ref{sec:structure_classification}. 

At first sight, it might seem unnecessary to define an additional criterion here, since the other criterion seems to work well, and has a clearer quantitative justification. However, the benefits of this ``halo-criterion'' are: (1) It can also be used without the need to integrate the GDE for all particles (which can be quite expensive) since also the finite difference distortion tensor can be used for detecting haloes. (2) It can be difficult to get the GDE and finite-difference distortion tensors into exact agreement in the continuum limit. This requires that the tidal tensor $\T$ corresponds exactly to the derivatives of the forces in the sense that $\myvec{F}(\myvec{x} + \Delta\myvec{x}) = \myvec{F}(\myvec{x}) + \T \Delta \myvec{x}$. At first sight this might seem relatively trivial to achieve, but we want to point out here that this is e.g. not the case for a mesh based Poisson solver like that in {\sc Gadget-2} which does not even \emph{exactly} ensure that $\partial_i F_j = \partial_j F_i$.

While we find it possible to bring the two distortion tensors into exact agreement in the continuum limit when using a mesh based scheme, it seems relatively hard when using a tree as we will explain in Section \ref{sec:force}. Therefore, we will use criterion (1) whenever possible, but fall back to the halo-criterion when necessary. However, we will show that the difference is negligible in most cases and the two criteria lead to the same results.

\subsection{Convergence of Power Spectra}
\begin{figure}
    \includegraphics[width=\columnwidth]{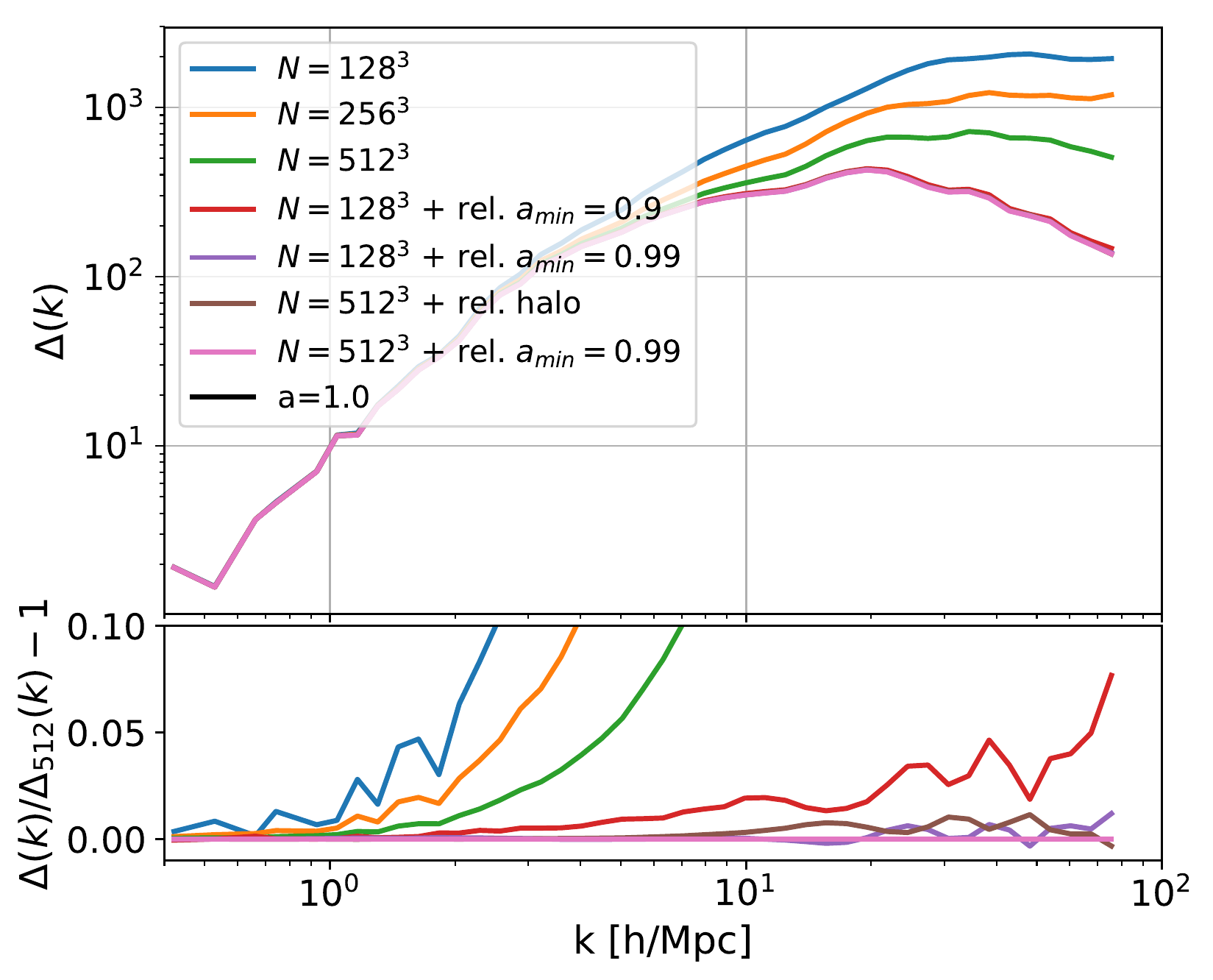}
    \caption{Convergence of the dimensionless power-spectra when using the release. In the top panel the power spectra and in the bottom panel the residuals with respect to the (pink) case with $N=512^3$ and using $a_{\text{min}} = 0.99$ for the release criterion. All simulations that use a release appear to be converged very well relatively independently of the release criterion or resolution. The only release case that disagrees above percent level is the case with $N= 128^3$ and $a_{\text{min}} = 0.9$ showing that the more conservative choice of $a_{\text{min}} = 0.99$ is better.}
    \label{fig:power_release}
\end{figure}

To test the validity and the convergence of the release method, we run a series of sheet-simulations at particle resolutions of $N=128^3$, $N=256^3$ and $N=512^3$ in a $\SI{20}{\mega\parsec\per\h}$ box in an $m_X = \SI{250}{\electronvolt}$ warm dark matter cosmology. The half mode mass of $M_{\text{hm}} = 5.5 \cdot 10^{12} M_\odot$ (according to the formula in \citet{schneider_2012}) is well resolved in all cases with the mass resolutions of $m = 5 \cdot 10^8 M_\odot$, $m = 6 \cdot 10^7 M_\odot$ and $m = 7 \cdot 10^6 M_\odot$ respectively. The number of silent particles which are used for the release are two times as many per dimension in each case. For these simulations we do not use any refinement. We test the different release criteria for these methods; on the one hand the release with the halo criterion and on the other the release with the alignment criterion for different values of $a_{\text{min}}$ as in equation \eqref{eqn:alignment}. We show the power spectra of these runs in Figure \ref{fig:power_release} and also plot the residuals in comparison to the case which we consider most accurate - that is $N=512^3$ with a release criterion of $a_{\text{min}} = 0.99$ which is the same one plotted in Figure \ref{fig:released_density}.

Clearly, the pure sheet simulations (without refinement) have very biased power spectra on small scales and are far from converged as already shown in \citet{hahn_angulo_2016}. Note however, that in most of the volume their density estimate is exactly the same as in the simulations with release (compare right panel of Figure \ref{fig:released_density}). The power spectrum is dominated by the density distribution in haloes. All simulations that use a release agree fairly well and their results seem to be relatively independent of the release criterion or the resolution. In all cases almost no elements outside of haloes are selected for release (since the dark matter sheet is relatively simple outside of haloes) and almost all mass inside of haloes is released. It seems that the distortion tensor based release criterion with $a_{\text{min}} = 0.99$ is a good choice, but also the release criterion based on the halo classification works well.

\begin{figure}
	\includegraphics[width=\columnwidth]{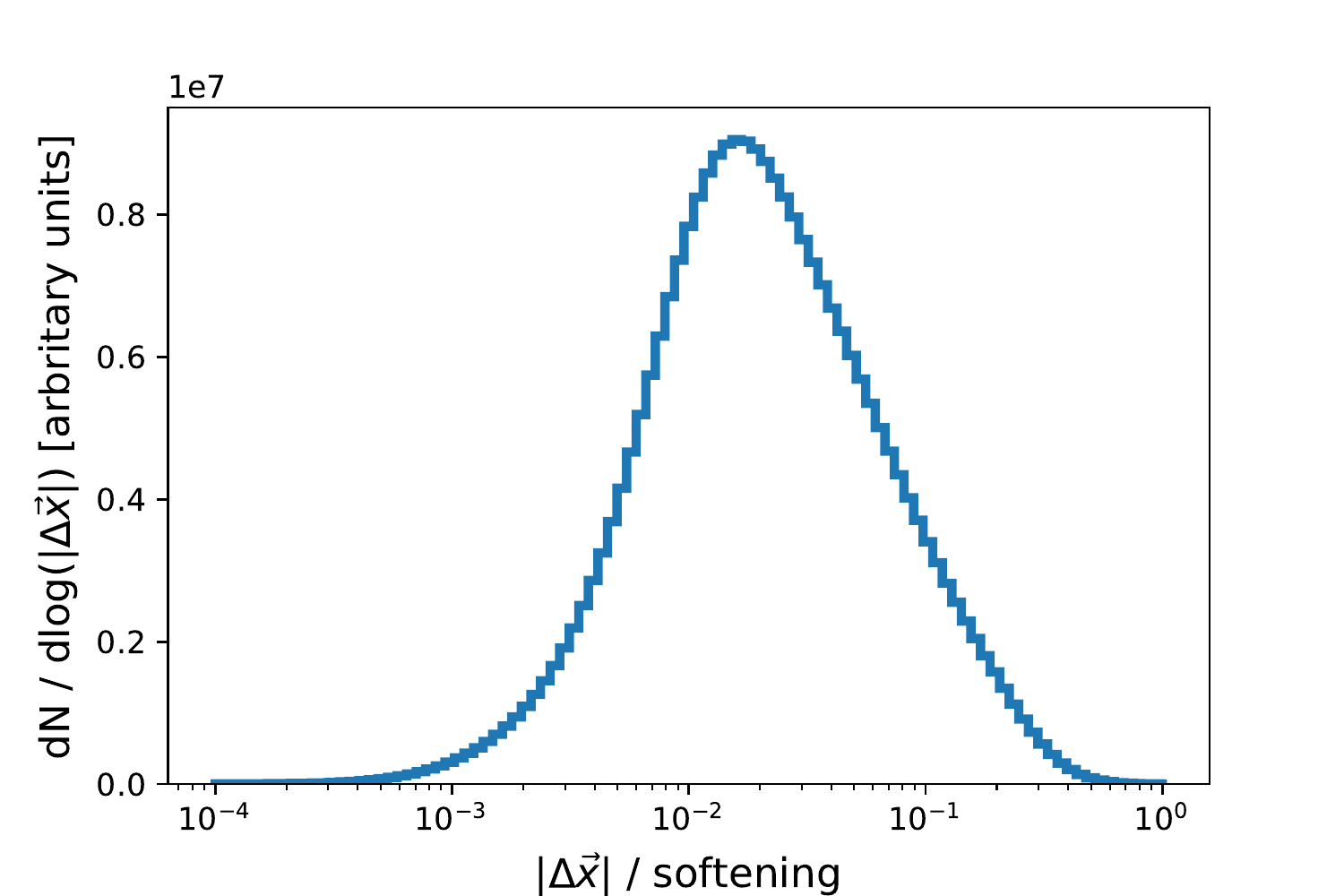}
    \caption{Histogram of the position error of the sheet interpolation at the point of release for the $N=512^3$, $a_{\text{min}} = 0.99$ simulation. This measures the error of the sheet interpolation at the point where it is largest - after this point it is considered unreliable and is no longer used for the corresponding elements. Even at its worst point, the interpolation error is still much smaller than a softening length (the grid spacing) in all cases. This shows that the release is happening early enough to make sure the interpolation is always reliable when used.}
    \label{fig:deltax_error}
\end{figure}

To get a quantitative idea of how large the \hlt{error of the interpolation} is in the worst case, we rerun the simulation with $N=512^3$ and the $a_{\text{min}} = 0.99$ release criterion. We save for every element that is released the difference in position between the silent particles \hlt{(which have no interpolation error)} and the corresponding positions predicted by the interpolated sheet (which can be slightly biased \hlt{through the interpolation}) at the time of release. This measures the absolute error of the interpolation at the point where it is largest and after which our code does not use it anymore for the corresponding mass element. We show those errors in Figure \ref{fig:deltax_error}. Even these worst case errors are much smaller than a softening length (the grid spacing) in all cases. That means that the usage of the interpolation until that point \hlt{indeed gives an accurate representation of the mass-distribution.} \hlcom{old:cannot bias the force calculation significantly.} The release criterion makes sure that mass elements are released early enough.

To highlight the improvement of the sheet simulations with release over pure sheet or pure N-body simulations, we show in Figure \ref{fig:halo_projection} a projection of the density field around the highest mass halo for four different cases: an $N=128^3$ N-body simulation (top left), a high resolution N-body simulation ($N=512^3$) as reference (top right), a pure $N=128^3$ sheet simulation (bottom left) and an $N=128^3$ sheet + release ($a_{\text{min}} = 0.99$) simulation (bottom right). In the pure sheet simulation, the halo appears much rounder than it should. In the sheet + release simulation the halo has the same shape as in the high resolution N-body simulation but there is no artificial clumping in the surrounding structures as in the $N=128^3$ N-body case. Clearly the sheet + release scheme inherits the best of each of the methods. Note that the $N=512^3$ N-body simulation is fragmentation-free, since here we are using a relatively low force resolution. However if the force resolution were higher than the initial separation between particles (as usually assumed) this case would also fragment.

We conclude that realistic, fragmentation-free warm dark matter simulations can be achieved when combining the benefits of sheet and N-body methods into a sheet + release scheme. In the following section we will address how to achieve higher force resolution in the sheet + release scheme.





\section{Towards higher force-resolution} \label{sec:force}

We have shown in section \ref{sec:simulationscheme} how to make unbiased and fragmentation-free warm dark matter simulations with a fixed, relatively poor force resolution and a global time-step. However, the force-resolution that can be achieved with a regular mesh is much below that necessary to get convergence in the centres of haloes. Therefore, we develop here a new scheme to calculate gravitational forces in cosmological  simulations. This allows the usage of adaptive time-steps and much higher force resolutions than in the case of a pure mesh -- which all sheet based methods have used so far.

N-body simulations typically discretize forces as interactions between point-particles. All additional components of the force-calculation like trees,  particle-meshes or adaptive mesh-refinement are just means to speed up the force-calculation. While this approach of pair-wise interactions is relatively simple and works remarkably well, there are also some peculiarities that arise from this approach when compared to the true continuous physical system. One of those is that the ``N-body density-estimate'' is exactly zero in the large majority of the volume: In N-body simulations typically a softening length as small as $1/20$ of the mean particle separation is chosen. The density estimate $\rho = \nabla^2 \phi / 4 \pi G$ is zero outside of a particle's softening radius. Consequently that means that less than $(1/20)^3 \sim 10^{-4}$ of the volume has a non-zero density in such an N-body simulation. However, we know that the true density can be zero nowhere, since at every point in space at least one dark matter stream must be present. Depending on the nature of dark matter, the lowest density regions in the universe may still have densities of $\rho \sim 10^{-1} - 10^{-3} \rho_0$ \citep{stuecker_2018}. Also the fragmentation of N-body simulations is a peculiar effect of the granularity of the N-body density estimate.

\begin{figure}
  \includegraphics[width=\columnwidth]{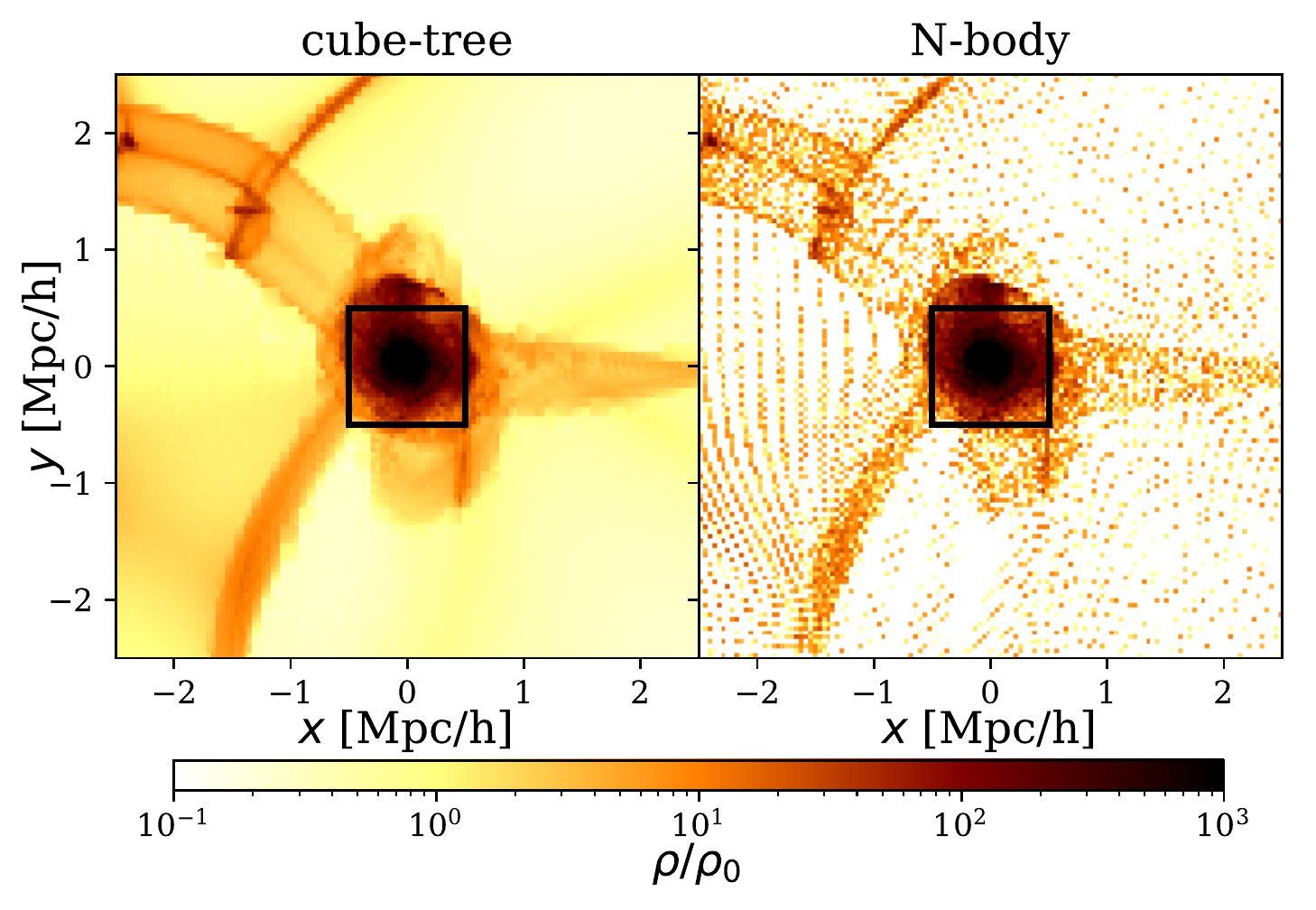}
  \includegraphics[width=\columnwidth]{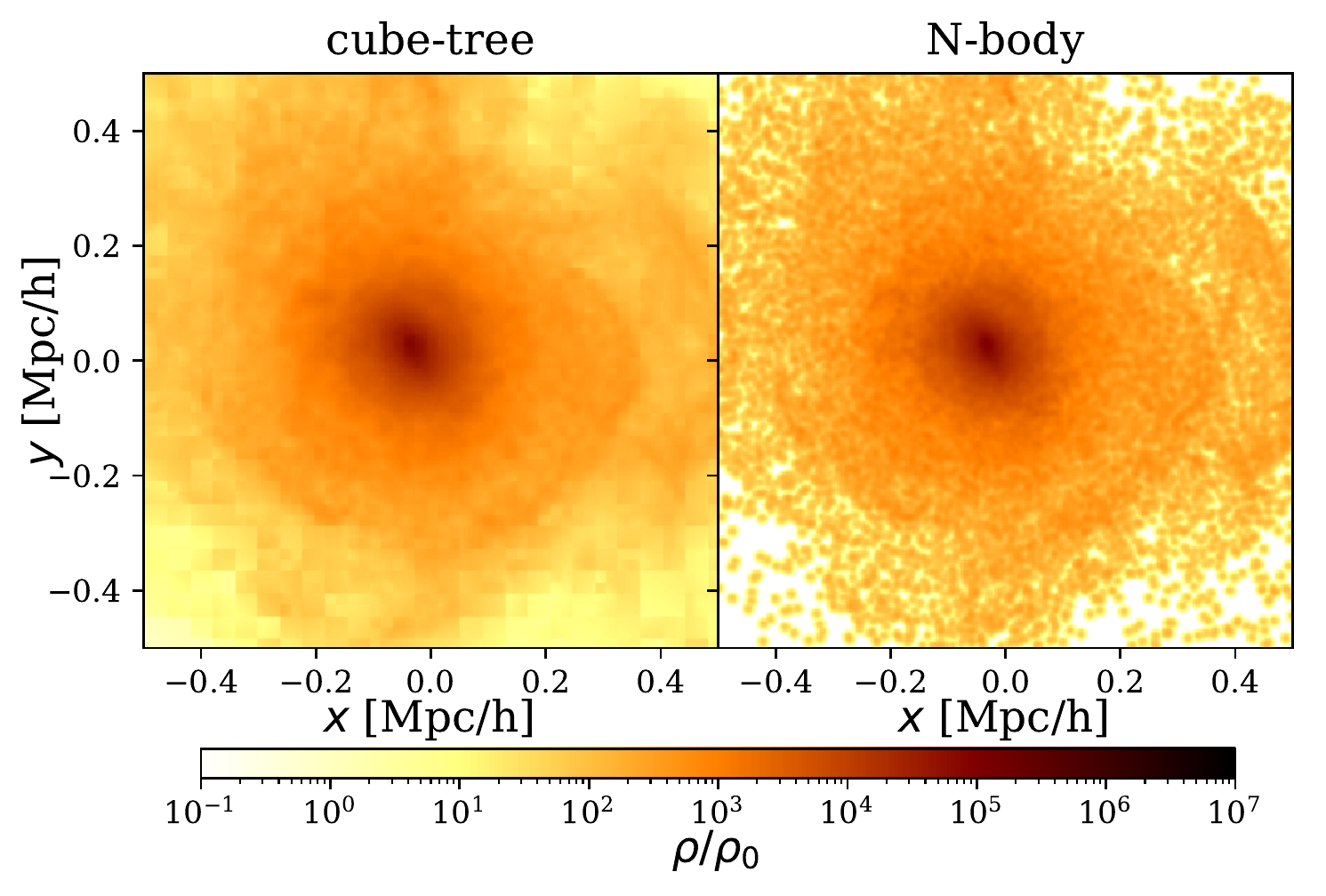}
  \caption{\hlcom{(Increased size of labels, simplified colorbar)}The density field that is represented by an oct-tree of cubes from one of our simulations. Left: using the tree of cubes, right: using an N-body density estimate with fixed softening. Top: \SI{5}{\mega\parsec\per\h} region - the colourmap is clipped at a density of $10^3 \rho_0$ to make the features in the low density regions visible. Bottom: \SI{1}{\mega\parsec\per\h} region showing the internal structure of the halo with an appropriate colourmap range. Inside of the halo mostly ''released'' particles are used for the density estimate so that the quality of the density estimate is similar between the N-body and the sheet + cube-tree case. However, in the outer regions of the halo and in the larger scale structures the sheet density estimate allows for much higher accuracy than the N-body one.  
  }
  \label{fig:tree_density}
\end{figure}

We here propose a different discretization of the density field and the force-calculations that accounts better for the continuous nature of the dark matter sheet. We discretize the density field by a space-filling oct-tree of cubes. The depth of the tree is chosen adaptively and depends (roughly) on the local density. The density of the cubes and their density gradients (assumed uniform) are calculated by sampling them with the N-body tracers (in released regions) and a large number of pseudo-particles which are created from the interpolated sheet (in low-density regions). An example of such a tree can be seen in the left panels of Figure \ref{fig:tree_density} in comparison to an N-body density estimate. The force that a single particle experiences is then the sum of interactions with cubes.

In the following sections we describe in more detail how the tree is constructed, how the densities and density gradients of the tree-nodes are computed, and how the interaction with a cube can be computed. Subsequently we demonstrate for the case of a Hernquist sphere that this discretization of the force field is as good as an N-body representation with the same number of mass-resolution elements, but a much larger number of force-resolution elements. Additionally it makes possible more accurate estimates of the tidal tensor and therefore improves the integration of the Geodesic Deviation Equation (GDE).

\subsection{A tree of cubes}
We discretize the density field in our simulations as an oct-tree of cubes. An oct-tree is a recursive partition of a three dimensional volume into a set of cubes. One starts with a cube which represents the whole simulation box. It is split into eight equal size sub-cubes, each of them representing another oct-tree. Each attached oct-tree can either be a leaf or again be split into eight sub-cubes representing their own subtrees - and so on. It depends on the tree-building procedure which sub-trees are split.

To make minimal changes to the code, we employ the same tree building mechanism as the {\sc Gadget-2} code as presented in \citet{springel_2005}. That is, given a set of particles, the tree is split recursively until each leaf contains either one or zero particles or a minimum node size $\Lmin$ has been reached. That minimum node size is an additional parameter in the code and determines the limit of the force resolution -- similar to a softening parameter. In our simulations the tree structure is not built from all particles, but from a smaller set of distinctly defined ghost particles. 

After the tree has been created it is just considered as a mass-less partition of the volume of the simulation box. Afterwards the mass is assigned to it in an additional independent step. Lagrangian regions that are still described by the sheet interpolation create a large number of mass-carrying pseudo-particles that are deposited into the tree whereas N-body particles are directly deposited into the tree. In \hlt{a} nearest-grid-point (ngp) assignment scheme each mass-carrying particle assigns its mass and its mass weighted position to the tree leaf it belongs to. However, we found that a more accurate density estimate could be inferred by \hlt{a cloud-in-cell} (cic) assignment where the size of the cic-kernel is chosen to be the minimum node-size $L_{\text{min}}$. This way each particle can contribute to a maximum of eight different nodes. Afterwards this information is propagated upwards in the tree so that at the end of the procedure each node has a well defined mass and centre of mass. Typically we choose a much higher number of mass tracing particles than of tree building particles, so that the mass in each tree node is well sampled. For example we have $4^3$ times more N-body particles than particles which define the tree structure, and the number of pseudo particles that are deposited from the sheet interpolation is again much higher than the number of N-body particles. Therefore each tree leaf is sampled by the order of $64$ particles in completely released regions and by many more in regions where the interpolation is active.

Now we interpret this tree as a set of cubic volume elements which have a mean density given by their mass and volume, and which have a density gradient given by the centre of mass. The density distribution within each cube is then approximated as
\begin{align}
  \rho (\myvec{r}) &= \rho_0 + \myvec{g} \cdot \myvec{r} \\
           \rho_0  &= M / L^3 \\
         \myvec{g} &= \frac{12 \rho_0}{L^2} \myvec{r_{\text{cm}}}\hlt{\,,}
\end{align}
where $M$ is the mass of the cube, $L$ is the side-length, $\myvec{r}_{\text{cm}}$ is the offset of the centre of mass from the centre, and $\myvec{r}$ is the offset from the (geometric) centre at which the density is to be evaluated. Note that the gradient is chosen so that the cube with gradient has the same centre of mass as the node.


\begin{figure}
  \includegraphics[width=\columnwidth]{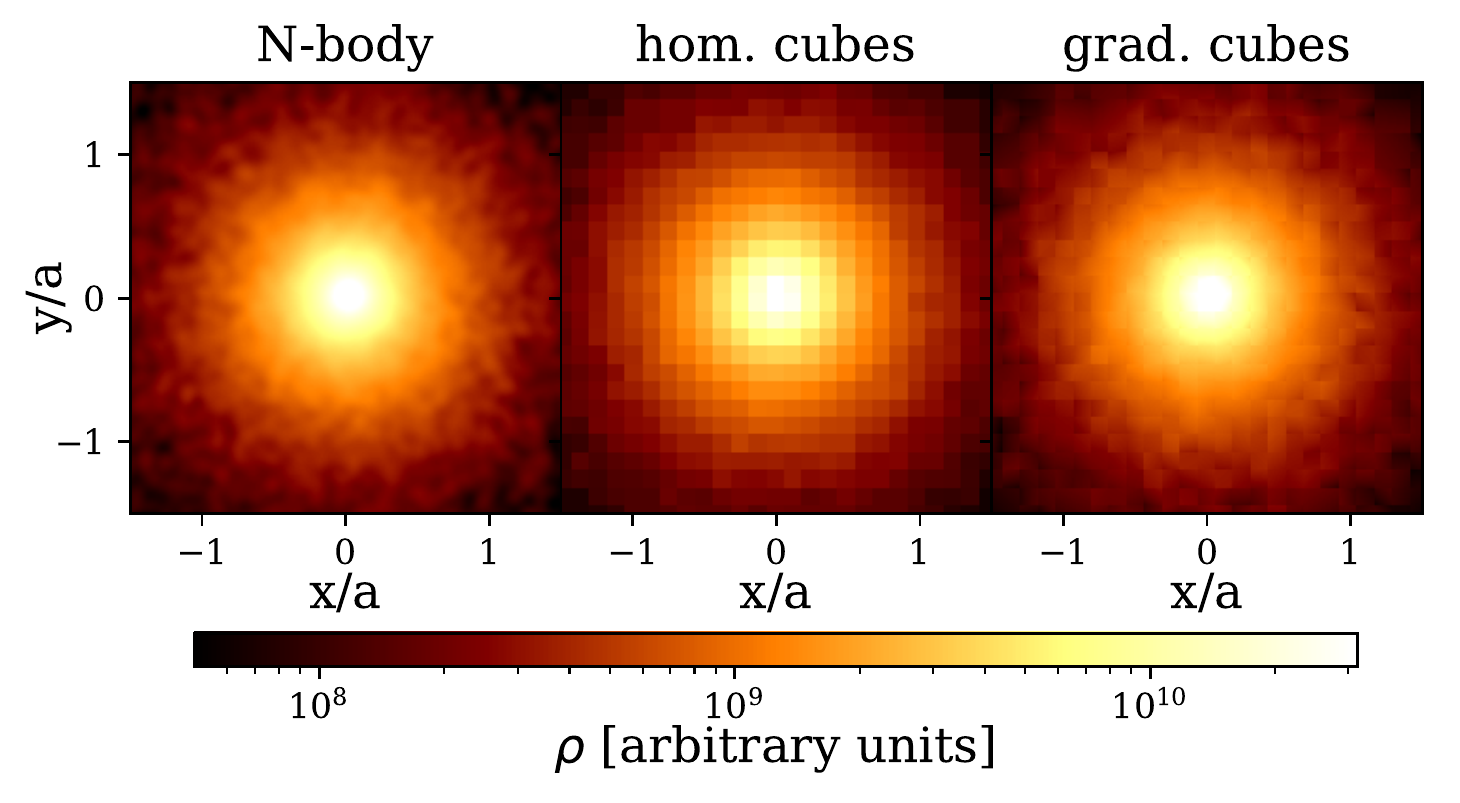}
  \caption{\hlcom{(Increased label size)}Density distribution close to the centre of a Hernquist sphere using different density estimates: N-body mass assignment with softening $\epsilon = 0.05a$ (left), cube-tree mass assignment with $\Lmin \sim 0.1a$ (see text) with zero gradients (centre) and with appropriate gradients (right). The cubes with gradients seem to give a similar good estimate of the density field close to the centre to an N-body density estimate and much higher accuracy than homogeneous cubes. It can represent the spherical geometry reasonably well.}
  \label{fig:density_hernquist_center}
\end{figure}

In Figure \ref{fig:density_hernquist_center} we show a density slice close to the centre of a Hernquist sphere where $N_c = 32^3$ tree-building particles and $N=128^3$ mass-assigning particles have been used which have been each sampled from the Hernquist profile in comparison to an N-body assignment ($M = 128^3$). The cube-trees use a minimal node length of $\Lmin \sim 0.1a$ and the N-body uses a softening of $\epsilon \sim 0.05a$ (which both lead to a similar effective smoothing scale) where $a$ is the characteristic radius of the Hernquist profile \citep{hernquist_1990}. The cubes with gradients with cic-assignment (right panel) seem to give a good representation of the density field even in the centre of a halo-like object. The degree of noise is similar to the N-body density estimate with the same number of mass-assigning particles ($N = 128^3$) though the cube-tree uses a much lower of force resolution elements ($\sim 32^3$ instead of $128^3$). However, recall that the noise should be much smaller in the outer regions of a halo as already seen in Figure \ref{fig:tree_density}. The central panel of Figure \ref{fig:density_hernquist_center} shows how the density estimate would look if the gradients of the cubes would be forced to zero -- this should be similar to the density field that would be represented by a typical adaptive-mesh-refinement (AMR) scheme. Clearly the gradients give a big improvement over this and account better for the spherical geometry. It would be interesting to see whether such a higher order density estimate with gradients could be useful inside of AMR-schemes.

\subsection{The potential of a cube}

While it is relatively simple to define the density field of this oct-tree, calculating its force field is mathematically relatively elaborate. To ease the reading flow we moved the full potential calculation to appendix \ref{app:cube_potential} - \ref{app:multipole}. We just give a brief summary of the necessary steps here.

\begin{figure}
  \includegraphics[width=\columnwidth]{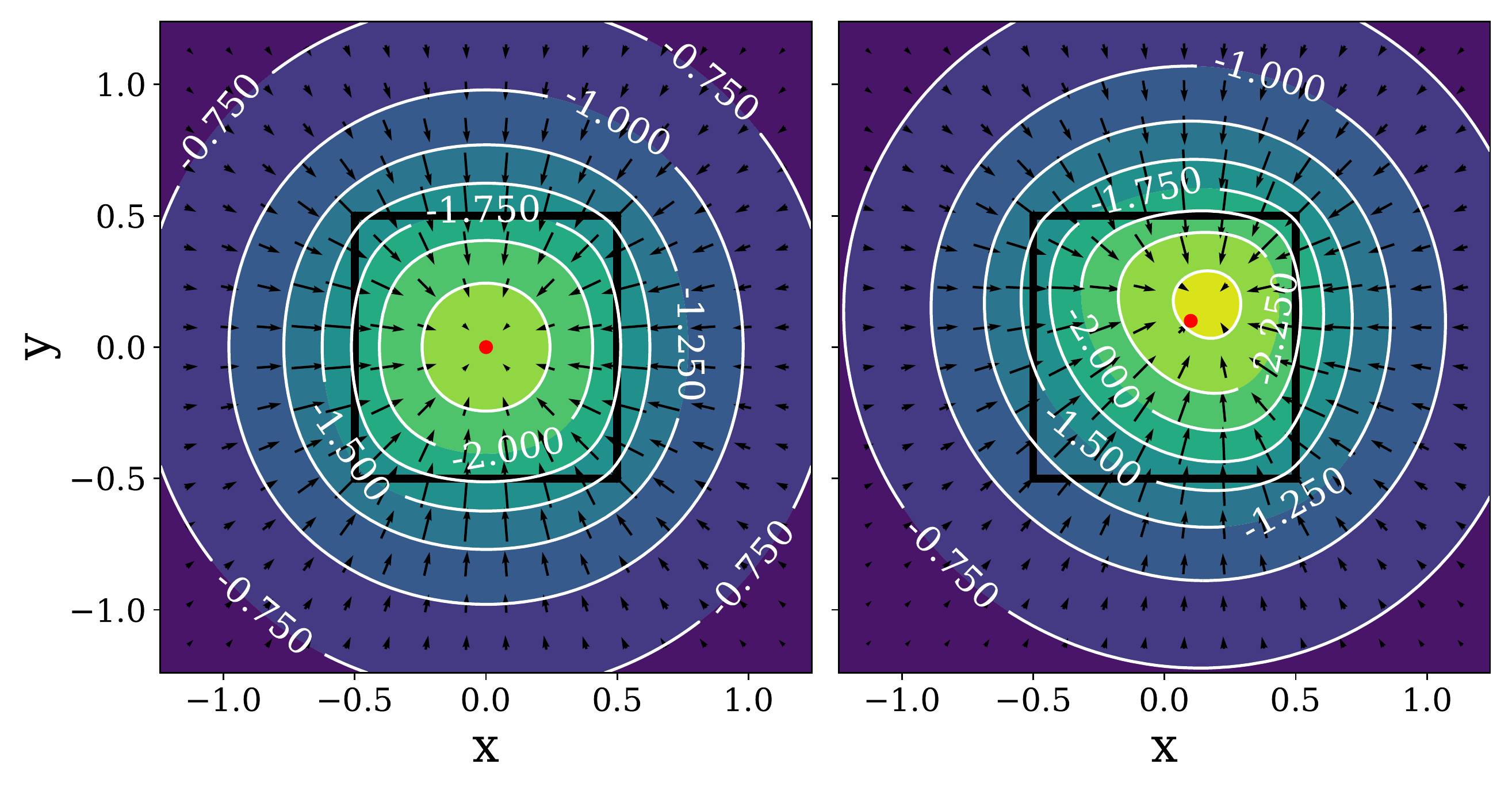}
  \caption{\hlcom{(Increased size of labels)}Potential (contours) and force field (arrows) of a homogeneous cube (left) and a cube with constant gradient (right). The equipotential-lines are closer to a sphere than the cubic mass distribution and farther away from the cube become more and more spherical (centring around the centre of mass -- marked as a red dot). }
  \label{fig:cube_potential}
\end{figure}

The potential of the cubic mass distribution can be calculated as the convolution of its density distribution with the Green's function of the potential\hlt{:}
\begin{align}
    \phi (\myvec{x}) &= \rho \circ G_\phi \\
   \rho(\myvec{x}) &=  
        \begin{cases}
          \rho_0 + \myvec{g} \myvec{x} &\text{ if } -L/2 \leq x_i \leq L/2 \forall i \in \{1,2,3\} \\
          0 & \text{ otherwise \hlcom{(modified)}}
        \end{cases} \\
    G_\phi(\myvec{r}) &= - \frac{G}{\vecnorm{r}} \hlt{\,,}
\end{align}
where $G$ is the gravitational constant. It turns out that the result of this convolution can be expressed in closed form, but yields a rather complicated expression. We present this in detail in appendix \ref{app:cube_potential}. In Figure \ref{fig:cube_potential} we plot the potential in the $z=0$ plane for the example of a homogeneous cube ($\rho_0 = G = 1$, $\myvec{g} = (0,0,0)^T$) and the case of a cube with constant density gradient ($\myvec{g} = (1.2, 1.2, 0)^T$).

However, this is not the kind of interaction that is actually summed over in {\sc Gadget-2}. In {\sc Gadget-2} the potential is split into a long-range part and a short-range part \citep[cf.][]{hockney_1981}. The long-range part $\phil$ is a smoothed version of the potential (described by a convolution with a kernel $f$) and is calculated with Fourier methods on a mesh. The short-range part $\phis$ is then the remaining part of the potential\hlt{:}
\begin{align}
    \phil &= \rho \circ G_{\phi} \circ f \\
    \phis &= \rho \circ G_{\phi} \circ (1 - f)\hlt{\,.}
\end{align}
This is the interaction that needs to be evaluated on the tree for maintaining consistency with the force-split. It turns out that for the Gaussian force-split that is used in {\sc Gadget-2}, this convolution cannot be solved analytically for our mass distribution. Therefore we decided to use a different force-split kernel
\begin{align}
   f(\myvec{r}) &=  
        \begin{cases}
          \frac{3 (a - \vecnorm{r})}{a^4 \pi} &\text{ if } \vecnorm{r} \leq a \\
          0 & \text{ otherwise \hlcom{(modified)}}
        \end{cases} \hlt{\,,} \label{eq:fpk}
\end{align}
for which the potential becomes analytic. As a drawback this piecewise defined kernel creates a large number of different cases. We only calculated the explicit expression for the most typical case that the whole cube is inside the range of the kernel. The full calculation can be found in Appendix \ref{app:treepmsplit}. For all other cases we use numerical approximations by sampling the mass distribution with point-masses or splitting the cube into eight sub-cubes.

Finally, the cost of the calculations can be drastically reduced by using multipole expansions. We give an overview of the approximations that we use in different cases in \ref{app:multipole}. We make sure that the errors in the forces and the tidal tensor due to a single cubic element stay well below 1\% in all cases.

\subsection{Force-field of a Hernquist Sphere}

\begin{figure}
  \includegraphics[width=\columnwidth]{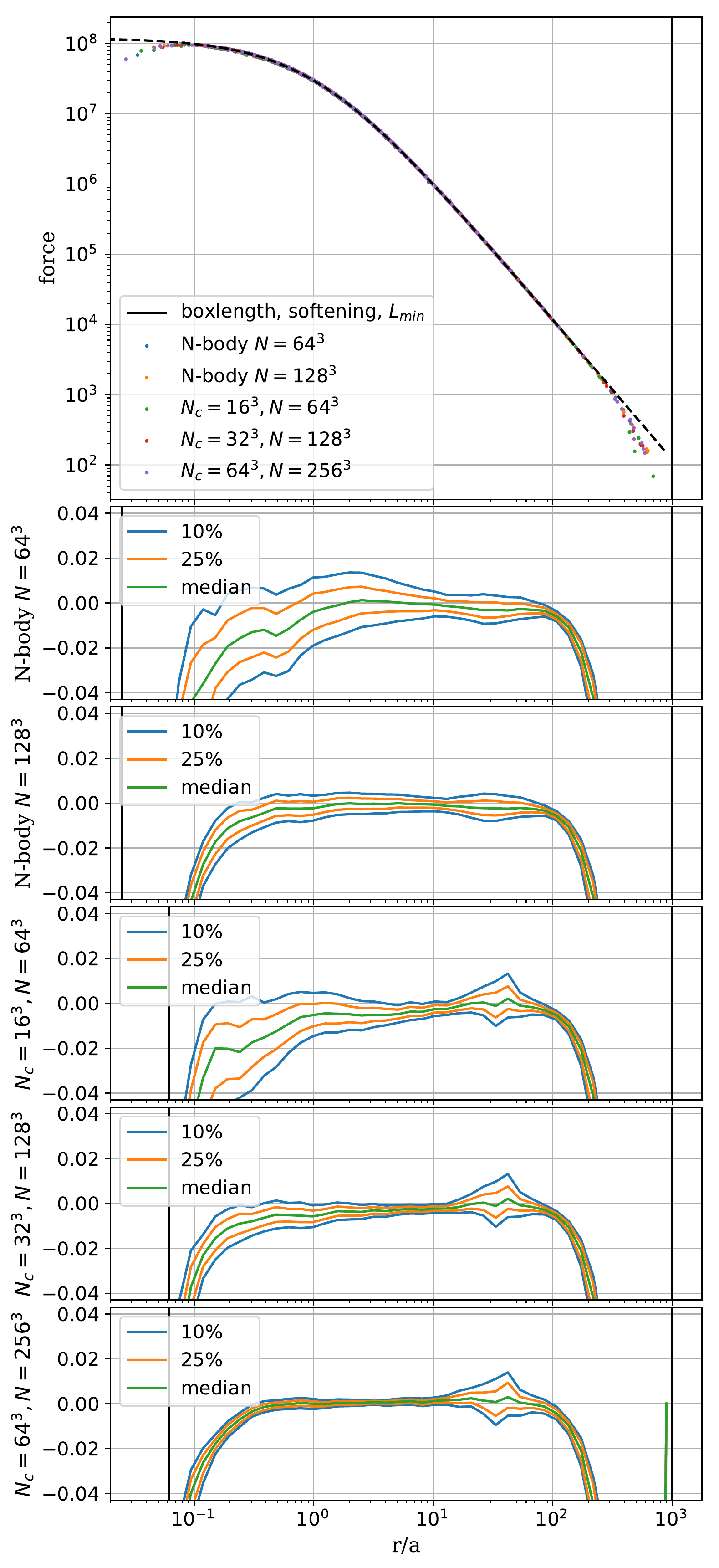}
  \caption{\hlcom{(Increased label size)}Absolute values of the force (top panel) and relative errors (other panels) for the numerically evaluated forces in a Hernquist sphere using different setups. Second and third panel: the quantiles of the force error distributions when using the traditional N-body approach for evaluating the forces with two different particle numbers $N=64^3$ and $N=128^3$. Bottom three panels: Using the cube tree with different number of tree building particles $N_c$ and mass depositing particles $N$. In the cube tree cases the black line indicates the minimal allowed node length which has been chosen to result in a similar effective softening as the N-body cases. The cube tree converges well and gives similar results as the N-body monopole force calculations at the same number of mass depositing particles $(N)$. It requires a significantly lower number of force resolution elements ($\sim N_c$ for the cube-tree).}
  \label{fig:hernquist_force_error}
\end{figure}

To test whether the calculation of the force field by the tree of cubes gives reasonable forces in complex three dimensional scenarios, we set up a Hernquist sphere \citep{hernquist_1990}. To do that we create a set of $N$ particles by Poisson sampling the  3D Hernquist density profile
\begin{align}
 \rho(\myvec{r}) &= \frac{M}{2 \pi} \frac{a}{\vecnorm{r}} \frac{1}{(\vecnorm{r} + a)^3} \hlt{\,,}
\end{align}
where $a$ is the scale radius and $M$ is the mass of the Hernquist sphere. We calculate the forces for that particle realization using the typical N-body softened monopole approach and the cube-tree approach presented in this paper. The structure of the cube-tree is built on a set of $N_c$ particles which are an independent set of Poisson sampled particles and we have typically $N_c \ll N$. The mass and gradients of the nodes are then determined by depositing the $N$ particles into the tree. We show the results of the force computation and the difference with respect to the analytical force\hlt{,}
\begin{align}
 F(r)   &= \frac{G M(r)}{r^2}\\
 M(r)   &= M \frac{r^2}{(r+a)^2}\hlt{\,,}
\end{align}
in Figure \ref{fig:hernquist_force_error}. To calculate the residuals, we use only the radial component of the force:
\begin{align}
   F_{p,r} &= \frac{\myvec{r} \cdot \myvec{F}_p}{\vecnorm{r}} \\
   R_p &= \frac{F_{p,r} - F(r)}{F(r)}\hlt{\,,}
\end{align}
where $\myvec{F}_p$ is the full force vector of a particle and $R_p$ is the relative error of the radial force as plotted in the lower panels of Figure \ref{fig:hernquist_force_error}.

It seems that the cube tree and the monopole interactions give very similar results when compared at the same number of mass-carrying particles $N$. They have a similar spread in the distribution which is mostly caused by shot noise and decreases when increasing the particle number $N$. The error around $r/a \sim 40$ that does not quickly reduce to zero is likely caused by inaccuracies in the Tree-PM force-split. It has a different (smoother) shape in the N-body monopole calculation, since that calculation uses a Gaussian kernel for the force-split. However, the error from the force-split is of the order of $1 \%$, \hlt{similar to the original {\sc Gadget-2} \citep{springel_simulating_2005}}. The force deviates at very large distances, because we use periodic boundary conditions for the force calculation. Further it deviates at very small scales because of the softening. We find that N-body and cube-tree have a similar convergence radius if $\Lmin \sim 2 \epsilon$ where $\epsilon$ is the Plummer-equivalent softening parameter of the N-body calculation and $\Lmin$ is the minimal node-length in the cube-tree.

While the cube-tree has a similar accuracy to the monopole-method in the case where the same number of mass-depositing particles is used, it uses a significantly smaller number of force-resolution elements. In the cube tree this is of the order $N_c$ and in the N-body case this is $N$. For example the cube-tree case with $N_c = 16^3$ and $N=64^3$ has roughly the same accuracy as the monopole case with $N = 64^3$ whereas it has roughly $64$ times fewer force resolution elements. Therefore a cube-tree discretization could also be useful for storing the density field of simulations at any particular point in time. It would only be necessary to write out all the nodes which would require in this case of order $64$ times less memory than storing all the N-body particles. This could be useful for visualizations and for applications where forces and/or the tidal field need to be evaluated in post-processing steps. 

We conclude that the representation of the force field by a tree of cubes will be of similar accuracy to the monopole method inside of haloes where we assume all particles to be released and to roughly follow a Poisson-sampling of the density distribution. However, outside of haloes the cube-tree can provide a much more accurate description of the force-field than the monopole tree, since we can use the sheet-interpolation to sample the tree with many more mass-carrying particles than would be possible in a pure monopole approach. This becomes clear in the top panel of Figure \ref{fig:tree_density}. Therefore the tree of cubes can be used to obtain at the same time a smooth and continuous representation of the density- and force-field outside of haloes and a reasonably good one inside of haloes. 

\subsection{Tidal-field of a Hernquist Sphere}

\begin{figure}
  \includegraphics[width=\columnwidth]{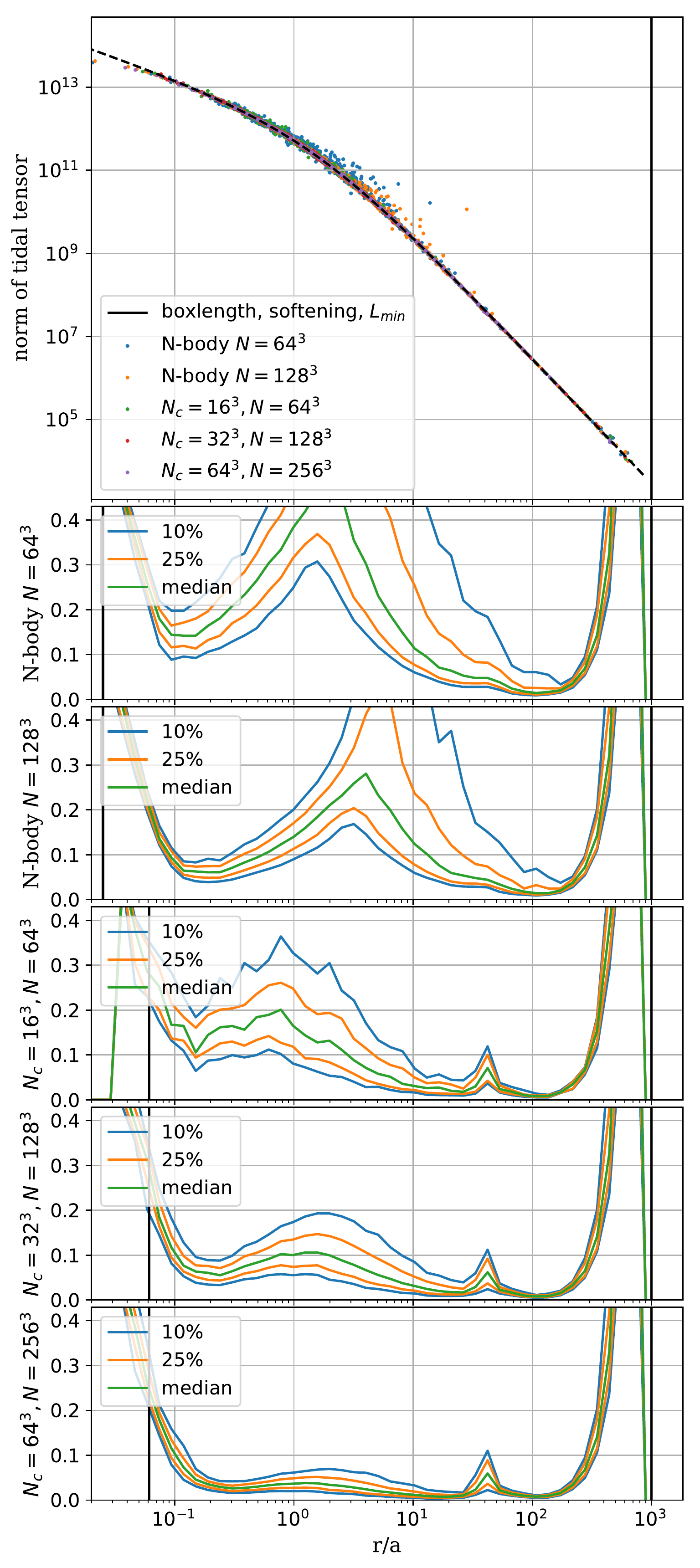}
  \caption{\hlcom{(Increased size of labels.)}L2 norm of the tidal tensor (top panel) and relative errors (other panels) for the numerically evaluated tidal tensors in a Hernquist sphere using different setups. Second and third panel: the quantiles of the error distributions when using the traditional N-body approach for evaluating the tidal tensor with two different particle numbers $N=64^3$ and $N=128^3$. Bottom four panels: Using the cube tree with different number of tree building particles $N_c$ and mass depositing particles $N$. In the cube tree cases the softening (the thick vertical black line) indicates the minimal allowed node length. The cube tree gives a better estimate of the tidal tensor than the N-body calculations. The tidal tensor of the cube-tree converges well with increasing number of particles.}
  \label{fig:cubetidal_error}
\end{figure}

Since our scheme also allows to trace the Geodesic Deviation Equation, it is of further interest for us to check the accuracy of the calculated tidal field in the different approaches.

In Figure \ref{fig:cubetidal_error} we plot the relative errors of the tidal tensor\hlt{,}
\begin{align}
  R &= \sqrt{\frac{\sum_{i,j} (T_{p,ij} - T_{ij}(r))^2}{\sum_{i,j} T^2_{ij}(r)}}\hlt{\,,}
\end{align}
where $\T_p$ is the tidal tensor that has been evaluated and $\T(r)$ is the analytical tidal tensor. It is striking that the errors in the tidal field are much larger than those in the force-field. They easily reach a few tens of percents. The tidal field is a quantity which is much harder to determine from the noisy particle distribution in a simulation. Unlike the force field, it depends crucially on the local density estimate.

The error in the tidal tensor is very large in the N-body cases, and seems to be significantly lower in almost all of the cube-tree cases. The convergence of the tidal-field in the N-body cases seems to be poor at larger radii $r \gtrsim 10 a$. This is likely related to the fact that the density estimate is mostly zero at such radii with a sparse sampling.


With the cube-tree we can significantly reduce the errors in the calculations of the tidal tensor and they seem to converge well with resolution. However, the errors are still of a worrying magnitude. In principle, errors in the evaluation of the tidal-tensor can cause an exponential diffusion in the integration of the distortion tensor. We will investigate this in more detail in section \ref{sec:timevolgde}.


\subsection{Time evolution of a Hernquist sphere}
In our scheme for the force calculation, the force resolution can vary in space and in time. That is similar to the case of an adaptive gravitational softening \citep{price_monaghan_2007, iannuzzi_dolag_2011}. As a consequence the time evolution will not be strictly energy conserving and not formally symplectic. To test whether this causes any major problems for the evolution of systems typical for cosmological simulations, we perform a time evolution of the Hernquist-sphere. For the cube-tree cases that means that both the (massless) tree-structure-defining particles and the mass-carrying particles are integrated along their orbits, and the structure of the tree can change with time. As an additional way of reducing aliasing effects due to the positioning of the oct-tree, we \hlt{randomise} its \hlt{positioning with respect to the simulation box before} each time-step.

\begin{figure}
  \includegraphics[width=\columnwidth]{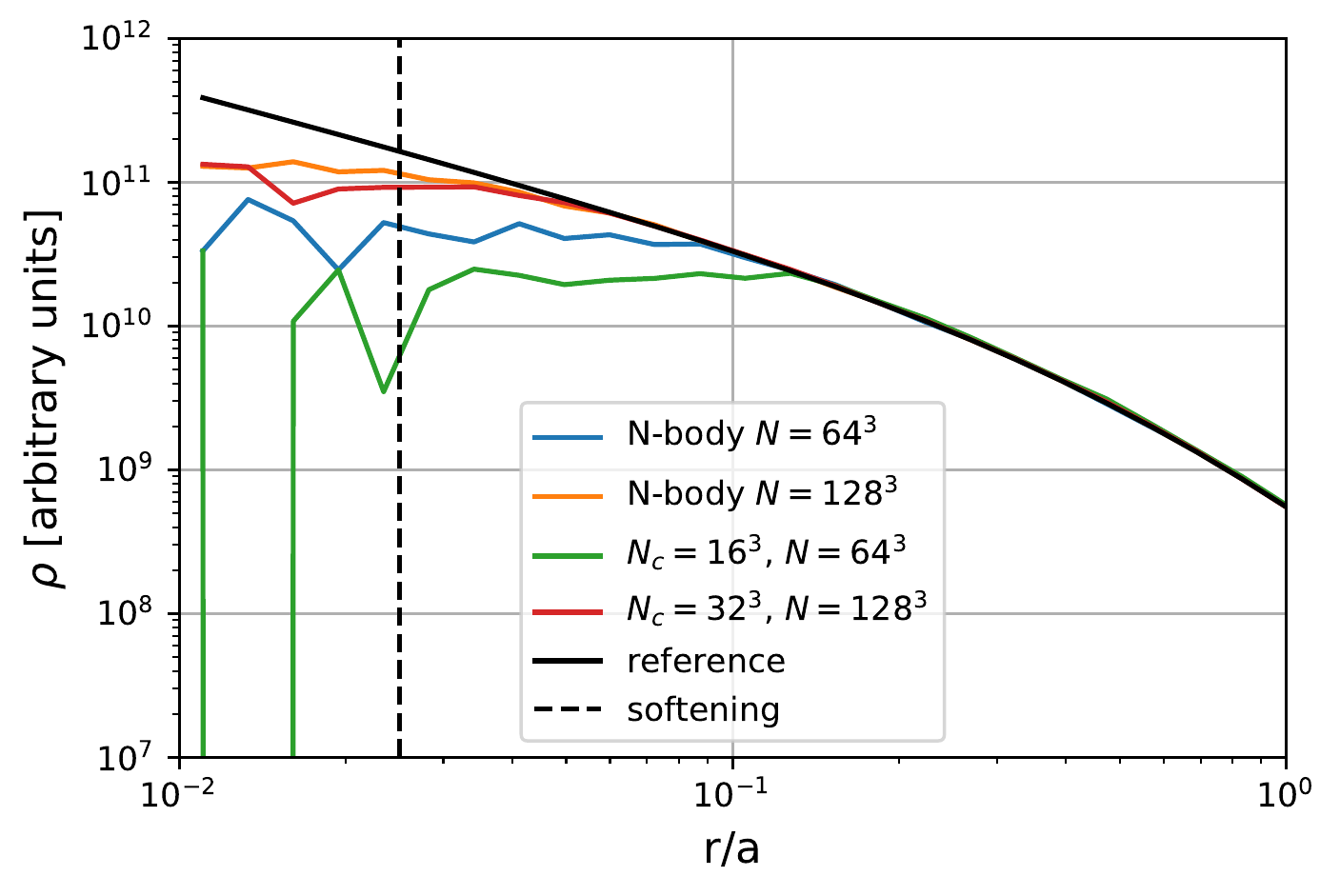}
  \caption{Density profiles of an evolved Hernquist-sphere. The N-body uses a softening of $0.025a$ and the cube trees use $\Lmin = 0.06 a$. At the resolution of $N=64^3$ the cube-tree (green) already shows relaxation at a slightly larger radius than the N-body case (blue). However, this effect seems to converge away so that the profile of the cube-tree method with $N=128^3$ (red) shows an almost identical convergence radius to the N-body scheme with the same mass resolution (orange).}
  \label{fig:hernquist_profile}
\end{figure}

We evolve the Hernquist-sphere for a time span of $t = 110 t_*$ where $t_*$ is the natural time scale of the system
\begin{align}
  t_* = \sqrt{\frac{a^3}{M G}} \text{\,.}
\end{align}
In that time a particle on a circular orbit at a radius of $0.1 a$ will have gone through about 90 periods. If there are any effects beyond the typical two-body relaxation, these should show up by that point.

\begin{figure}
  \includegraphics[width=\columnwidth]{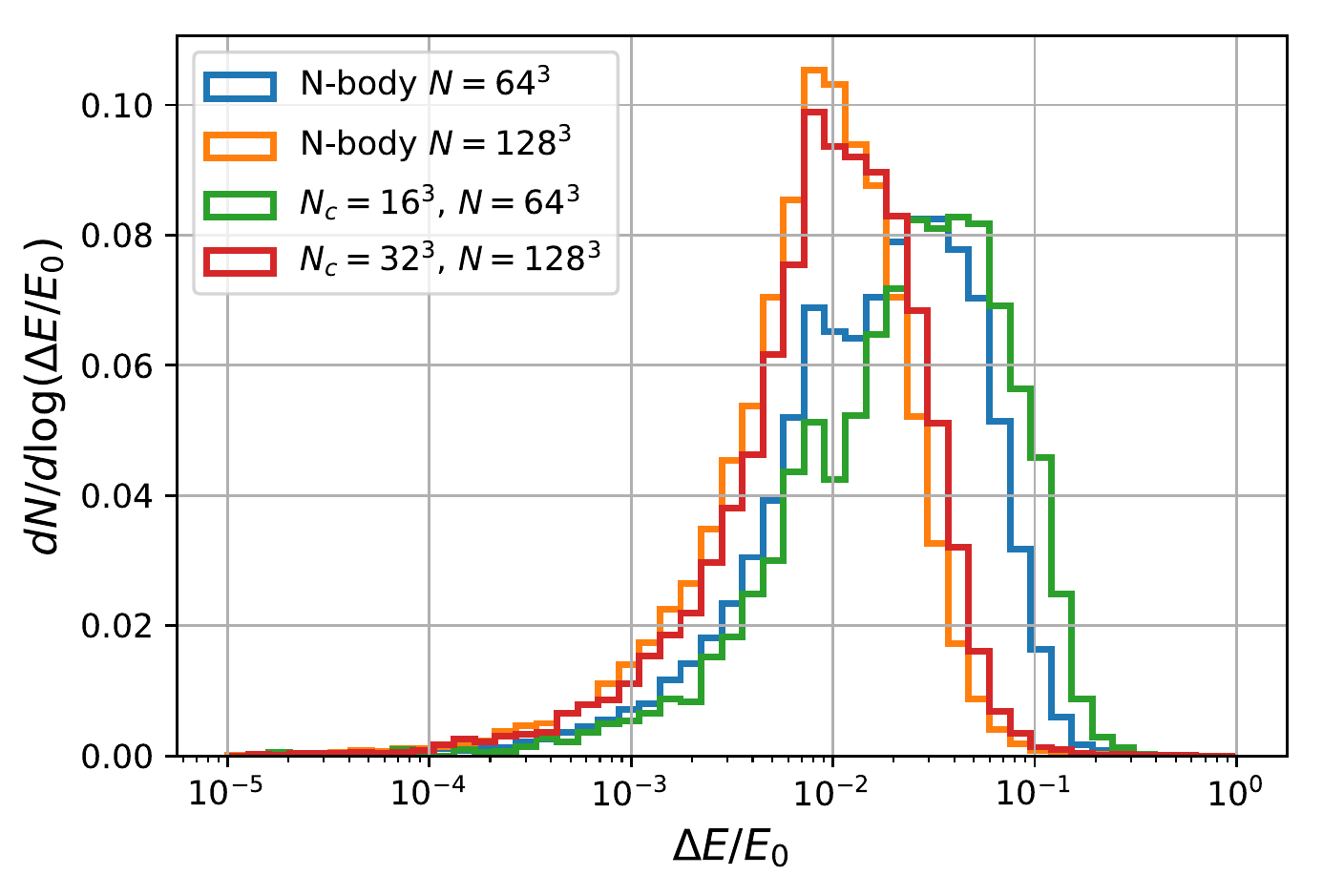}
  \includegraphics[width=\columnwidth]{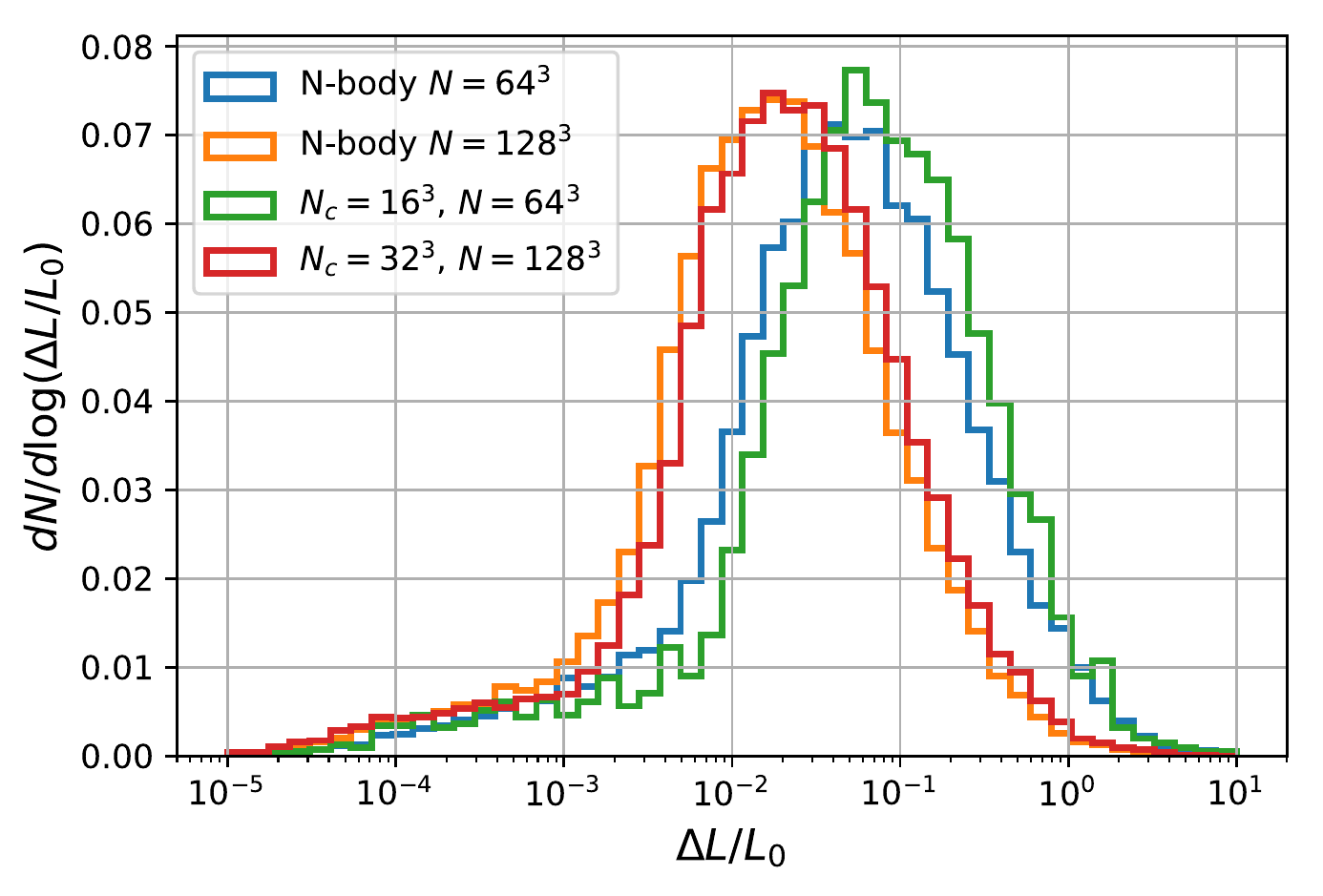}
  \caption{Histogram of the relative change in the energy (top) and angular momentum (bottom) of the particles in the evolved Hernquist-sphere. If there were no numerical errors these should be zero. The numerical errors of the N-body scheme (blue and orange) and of the cube-tree are almost of identical amplitude when compared at the same number of mass resolution elements $N$.}
  \label{fig:hernquist_error}
\end{figure}

In Figure \ref{fig:hernquist_profile} we show the density profile of the evolved Hernquist sphere for different numerical setups and in Figure \ref{fig:hernquist_error} we show the distributions of energy and angular momentum errors. It appears that \hlt{the} N-body approach and \hlt{the} cube-tree approach have almost equivalent accuracy in the case of $N=128^3$ mass tracing particles. The profiles diverge from the theoretical ones at a similar radius and the energy- and angular-momentum errors are of similar amplitude. However, the cube-tree has somewhat higher errors for the case of $N=64^3$ mass-tracing particles ($N_c = 16^3$) and diverges from the theoretical density profile already at a slightly larger radius. However, this effect is not large and it converges away with increasing resolution.

We note that the cube-tree approach uses of order $64$ times fewer force-resolution elements, but achieves similar errors to an N-body simulation on the density-profile, energy-conservation and angular-momentum-conversation. These errors seem to be dominated by two-body effects or shot-noise. We do not find any major additional errors caused by the variable force resolution. 

We conclude that the cube-tree performs roughly as well as a pure N-body scheme when compared at the same mass resolution while requiring significantly less force-resolution elements. It is therefore well suited to be used in a hybrid N-body/sheet scheme, since it can capitalize on the high mass resolution that becomes available through the sheet, without compromising the consistency of the force calculation.

\subsection{Time-integration of the Geodesic Deviation Equation} \label{sec:timevolgde}

Besides typical coarse grained quantities, like for example the density profile, our code also enables us to evaluate fine-grained phase space quantities. In the low-density regions, where the sheet can be interpolated accurately, almost perfect information of the fine-grained phase space distribution is available. However, in released regions the sheet cannot be traced accurately by the interpolation anymore, \hlt{while} it is still possible to trace it statistically in the infinitesimal \hlt{environment} of particles through the Geodesic Deviation Equation (GDE). We have already seen in Figure \ref{fig:lagrangian_stream_densities} that in  low-density regions the GDE can be followed very accurately (thanks to the accuracy of the sheet density estimate). It is not so clear that this is still the case in the dense centres of halos where the noisiness of the density-estimate and the magnitude of the tidal fields could cause problems. In principle the structure of the GDE allows an exponential divergence of the distortion tensor through almost any kind of noise. It is a quantitative question whether this noise can be controlled well enough so that the evolution of the distortion tensor is not dominated by a numerical exponential growth. We will test this here for the case of the Hernquist sphere.

We found that bringing the stream-densities to a reasonable degree of convergence requires the usage of more fine-tuned time-stepping and opening criteria than the fiducial choices in {\sc Gadget-2}. The {\sc Gadget-2} criteria have been optimized to limit the force error and to provide good convergence for density profiles. However, the convergence of stream-densities seems to require more accurate time-stepping and an opening criterion that is more sensitive to the local surrounding of each particle. We have developed new opening and time-stepping criteria to achieve convergence for the stream-densities, but we will discuss these in Appendix \ref{app:streamdensconvergence} to keep the main text concise.

\begin{figure*}
  \includegraphics[width=0.49\textwidth]{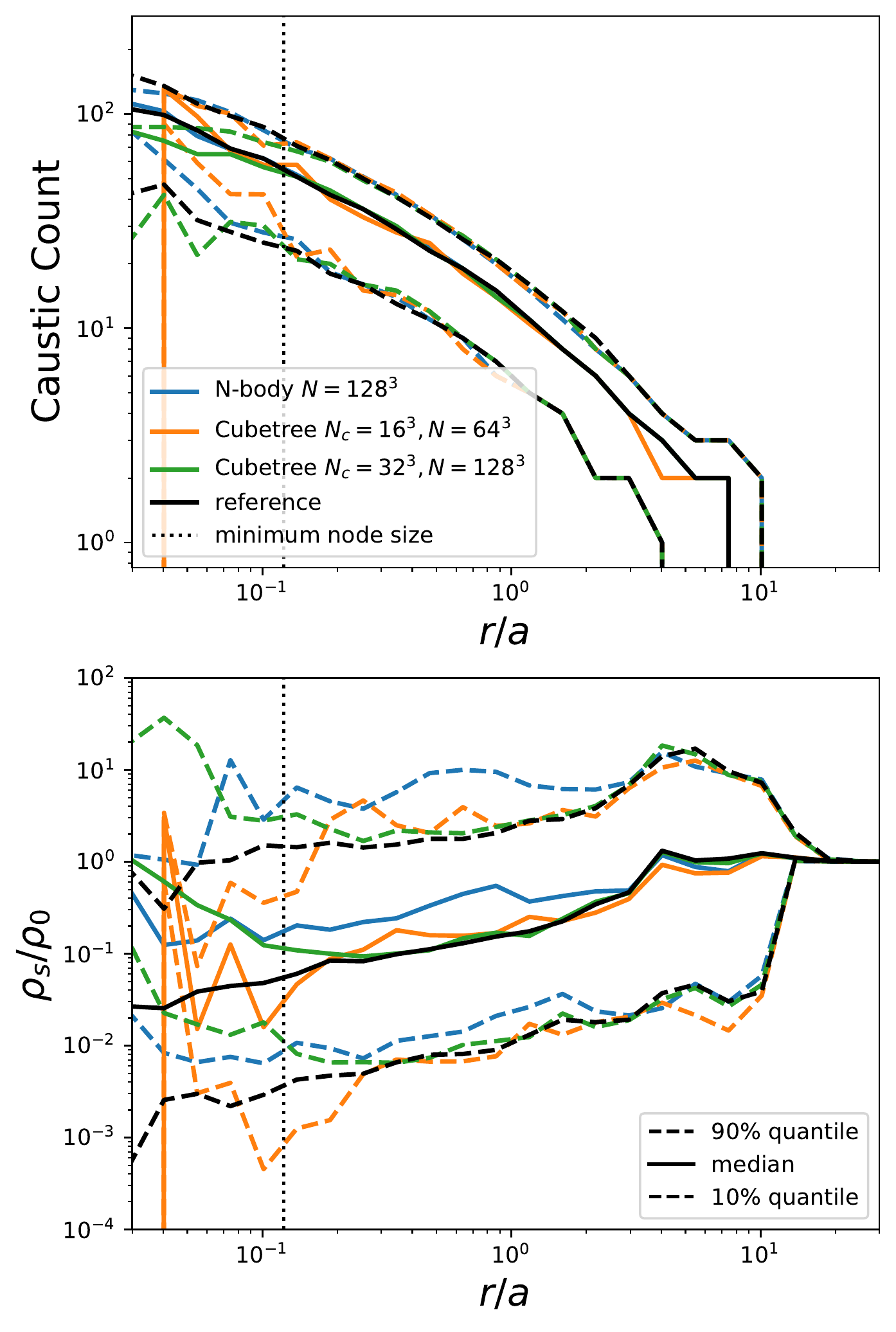}
  \includegraphics[width=0.49\textwidth]{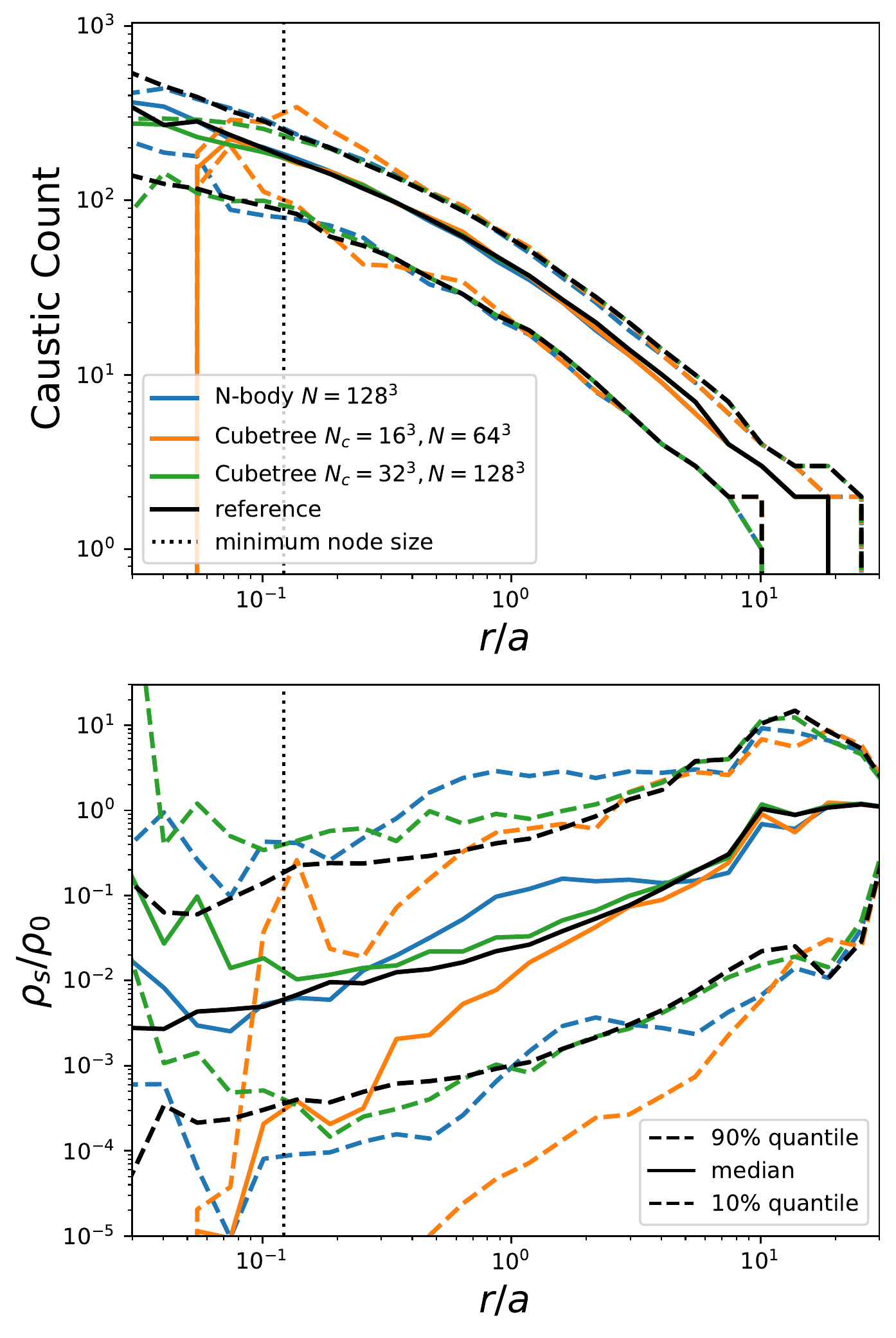}
  \caption{Median- and \hlt{10th and 90th} percentile-profiles for the caustic counts (top) and the stream-densities (bottom) in a Hernquist sphere at two different times $t = 33 t_*$ (left) and $t = 109 t_*$ (right). The caustic counts seem to be well estimated for any setup and any time. The stream-density estimate is more problematic, but it seems to be converged (outside the softening affected region) for the cube-tree case with $N_c = 32^3, N=128^3$ at early and late times.}
  \label{fig:hernquist_strde_profile}
\end{figure*}

In Figure \ref{fig:hernquist_strde_profile} we show the profile of the \hlt{$10$th, $50$th and $90$th percentile} of the caustic counts (top) and the stream-density (bottom). Note that the distortion tensor has been initialized in this simulation as a unit tensor at $t = 0$ so that all stream-densities have started at $\rhos / \rho_0 = 1$. In the left panel we show these at a time of $33 t_*$ and in the right panel at a much later time of $109 t_*$. Particles around $r \sim 0.1 a$ would have gone through order $10^2$ orbits at this time. The N-body case uses a softening of $\epsilon = 0.05a$ and the minimal node size for the cube-tree cases is approximately $\Lmin \sim 0.1 a$. The reference solution is obtained by integrating the orbits and the distortion tensor for a large number of particles in the analytical potential (without softening).

The caustic counts seem to be well converged in all cases. However, for the stream-densities this is not so. At earlier times $t \sim 30 t_*$ the lower and higher resolution cube-tree simulations both seem in good agreement with the reference solution. The N-body case seems to miss part of the dynamics and produces too high stream-densities at this time. At a later time $t \sim 109 t_*$ the $N_c = 16^3, N=64^3$ simulation seems to create far too low stream-densities. The $N_c = 32^3, N=128^3$ case gets still very close to the reference solution. The N-body case still over-estimates the stream-densities at $r \sim a$, but seems to have a dip at smaller radii which might be the onset of an exponential decay of the stream-densities. 

\begin{figure}
  \includegraphics[width=\columnwidth]{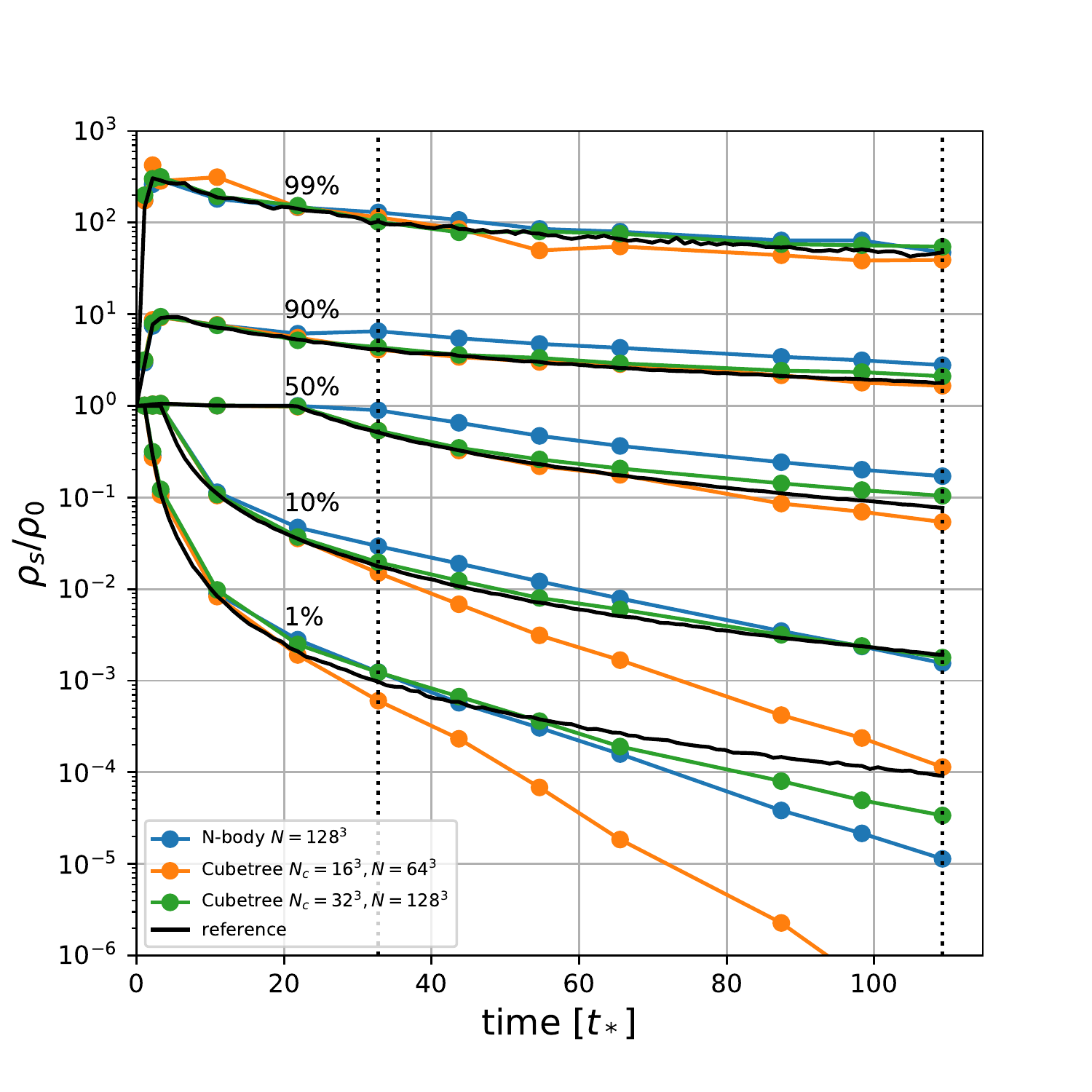}
  \caption{Evolution of the \hlt{$1$st, $10$th, $50$th, $90$th and $99$th percentile} of the stream-density distribution. The dashed lines mark the output times of the plots in Figure \ref{fig:hernquist_strde_profile}. An exponential decay can be seen for the lower quantiles of the stream-density distribution, but the exponential slope is a function of resolution and numerical technique and can therefore be controlled.}
  \label{fig:hernquist_strde_evol}
\end{figure}

In Figure \ref{fig:hernquist_strde_evol} we show the time-evolution of the quantiles of the stream-density distribution (of all particles) for the same three cases. The plot is shown in linear time scale versus logarithmic scale of the stream-densities. Therefore exponential behaviour appears as a straight line. It can be seen that the simulations exhibit an exponential growth $\rhos \propto \exp \left(- \alpha t\right)$ in the lower-quantiles of the stream-density distribution. In the reference case these only decrease according to a power law $\rhos \propto t^{-2}$ (also compare Figure \ref{fig:strde_evolution_timestep}). It seems that errors in the simulations let the stream-densities decay exponentially and the exponential behaviour starts dominating when the ''physical growth'' has a smaller negative slope in logarithmic space than $\alpha$.

Due to the form of the GDE it seems unlikely that one can avoid an exponential growth for an arbitrarily large time, since an exponential with arbitrary small $\alpha$ will dominate over any power-law growth at some time. However, this is also not necessary in the cosmological case. Typical particles will not go through many more orbits than of order $10^2$. Therefore it is only necessary to make sure that this exponential behaviour does not dominate the distribution at the time where the simulation is evaluated. As can be seen in Figure \ref{fig:hernquist_strde_evol}, the exponential index $\alpha$ (of the lower quantiles) is much smaller for the higher resolution case $N_c=32^3, N=128^3$ than for the lower resolution case $N_c=16^3, N=64^3$.  Actually at a time of up to $t \sim 60 t_*$ the exponential growth should be completely invisible in the higher resolution case and even at $t = 109 t_*$ it only affects the $1 \%$ percentile significantly (but still less than an order of magnitude). 

Therefore we conclude, that the exponential decay of stream-densities is present in our simulations, but it is a problem that can be controlled and tested for. Particular care has to be applied when evaluating the distribution at the lowest stream-densities. Caustic counts and \hlt{the central parts of the} \hlt{stream-density distribution (for example the median)} can be evaluated quite robustly.

\section{Conclusions}

Cosmological N-body simulations of warm dark matter suffer from artificial clumping in the sheets and filaments which precede formation of virialised haloes. Simulations based on the dark matter sheet are able to avoid this fragmentation by employing a density estimate that is less noisy and more accurate in low-density and anisotropically collapsed regions. However, they suffer from the intractable complexity of the dark matter sheet in strongly mixing regions like haloes. The two methods are thus optimal in different regions, and we have shown above that their respective strengths can be optimally used in a combined approach: a ``sheet+  release'' scheme that infers densities from the dark matter sheet interpolation wherever the interpolation is valid, and that switches to an N-body approach for mass elements that are too complex to be reconstructed by the interpolation. Through such a combined approach we obtain a fragmentation-free scheme for warm dark matter simulations that converges well inside and outside of haloes at affordable cost.

Further, we have developed a new scheme to calculate the forces in such simulations. The new scheme makes possible for the first time sheet-based simulations with adaptive force-resolution and time-stepping. N-body simulations typically use a multipole expansion of the interactions of point-like particles as the basis for the force calculations. Our scheme instead partitions space into an oct-tree and uses cubic nodes with \hlt{constant} density gradients as the basic force resolution elements. In the regime where we can reconstruct the dark matter sheet, we can determine the masses and gradients of those nodes with very high accuracy. In the regime where the mass is traced by released N-body particles the cubic nodes still lead to an approximation of the force-field that compares favorably with the N-body approach: The accuracy is similar if compared at the same number of mass-resolution elements, but the accuracy is much higher if compared at the same number of force-resolution elements. Although this force calculation scheme is not exactly Hamiltonian, the energy and angular momentum errors are of similar amplitude as in the pure N-body case -- even when integrated over hundreds of orbits. Further, the tree of cubes allows us to infer more accurate estimates of the tidal tensor and can help with following the evolution of fine-grained phase space quantities like the distortion tensor through the Geodesic Deviation Equation.

With these new numerical methods, we are able to carry out reliable non-fragmenting warm dark matter simulations at high force-resolution and with extensive phase space information. We will present such simulations in the sequel paper II \hlt{(St\"ucker et al. 2020, in prep.)}.

\section*{Acknowledgements}

We thank Mark Vogelsberger for providing us with the code for the geodesic deviation equation. Further we want to thank Volker Springel for writing the {\sc Gadget-3} code on top of which we implemented our changes. OH acknowledges funding from the European Research Council
(ERC) under the European Union's Horizon 2020 research and innovation programme (grant agreement No. 679145, project `COSMO-SIMS'). REA acknowledges the support of European Research Council through grant number ERC-StG/716151.




\bibliographystyle{mnras}
\bibliography{bibliography} 




\appendix

\section{Lagrangian Maps}
\begin{figure*}
	\includegraphics[width=\textwidth]{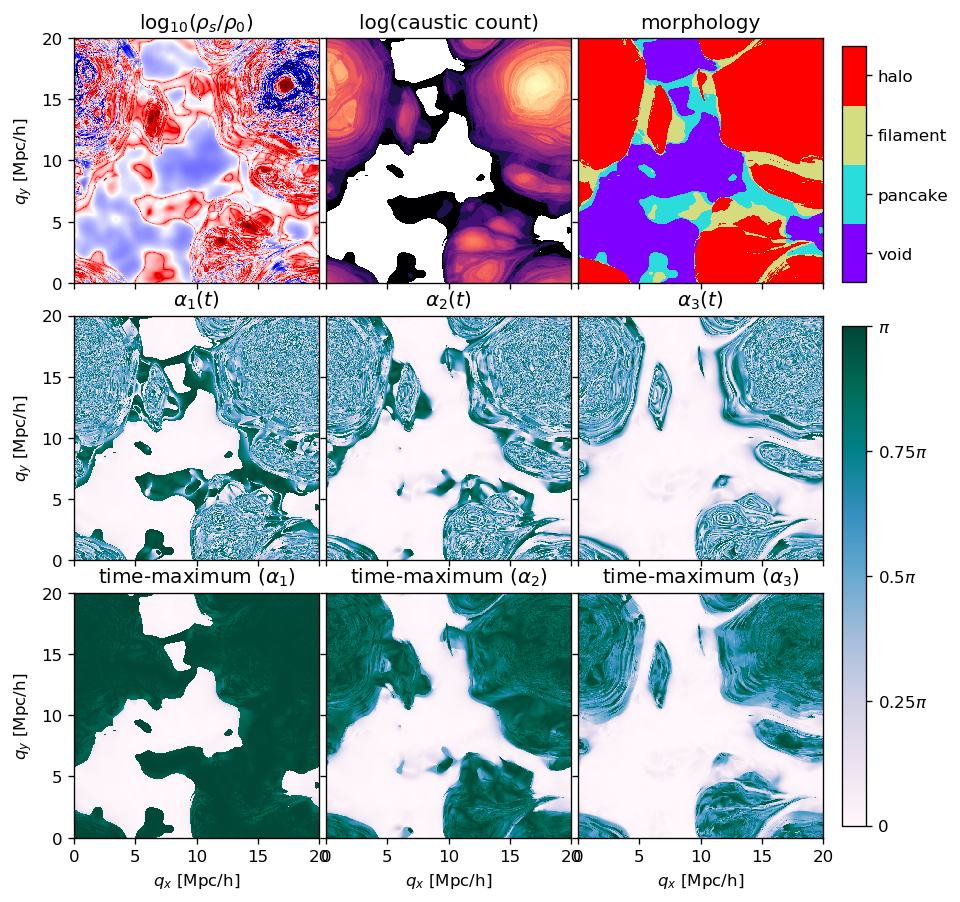}
    \caption{A slice in Lagrangian space through a cosmological warm dark matter simulation with $m_X = \SI{250}{\electronvolt}$. The slices show different quantities that were inferred from the GDE distortion tensor (and could equivalently be inferred from the finite differences distortion tensor). First row: GDE stream density, caustic count (the number of times the sign of the determinant of the distortion tensor of a particle has flipped) and the morphology classification as explained in section \ref{sec:structure_classification}. Second row: angles of the distortion tensor as explained in section \ref{sec:structure_classification}. Third row: maximum of each of the angles over the whole history of a particle. It is striking that the angles are activated in clearly distinct Lagrangian regions. Therefore the morphology classification appears to be very robust. However, it is necessary to take the time-maximum of the angles to avoid misclassification for cases where axes align by chance.}
    \label{fig:lagrangian_maps}
    \vspace{60pt} 
\end{figure*}

Figure \ref{fig:lagrangian_maps} shows a variety of quantities on a thin slice through Lagrangian space similar to that of Fig.4. See the caption for details.

\section{The potential of a cube}

\subsection{The total Potential}
\label{app:cube_potential}
  The gravitational potential $\phi(\myvec{x})$ of a mass distribution $\rho(\myvec{x})$ can be obtained by convolving it with the Green's function $G_\phi(\myvec{x})$ of the gravitational potential\hlt{:}
 \begin{align}
   \phi (\myvec{x}) &= \rho \circ G_\phi \\
        &= \int \int \int \rho(\myvec{r}) G_\phi(\myvec{x} - \myvec{r}) d^3 \myvec{r} \\
   G_\phi(\myvec{r}) &= - \frac{G}{\vecnorm{r}}\hlt{\,,}
 \end{align}
 where $G$ is the gravitational constant. 
 The mass distribution of a homogeneous cube is given by
 \begin{align}
   \rho(\myvec{x}) =  
        \begin{cases}
          \rho_0 &\text{ if } -L/2 \leq x_i \leq L/2 \forall i \in \{1,2,3\} \\
          0 & \text{ otherwise \hlcom{(modified)}}
        \end{cases}\hlt{\,.}
 \end{align}
The potential of a homogeneous cube has already been derived in \citet{macmillan_1958} and can be written as
 \begin{align}
   \phi (\myvec{x}) &= - \int_{-L/2}^{L/2} \int_{-L/2}^{L/2} \int_{-L/2}^{L/2} \frac{G \rho_0}{\norm{\myvec{r} - \myvec{x}}}  d^3 \myvec{r} \\
                    &= - G \rho_0 \int_{-L/2 - x_1}^{L/2-x_1} \int_{-L/2-x_2}^{L/2-x_2} \int_{-L/2-x_3}^{L/2-x_3} \frac{1}{\vecnorm{r}}  dr_3 dr_2 dr_1 \label{eqn:hom_pot_integral} \\
                    &= - G \rho_0 \left[ \left[ \left[ F_{\text{hom}}(\myvec{r} )\right]^{L/2-x_1}_{r_1 = -L/2-x_1} \right]^{L/2-x_1}_{r_2 = -L/2-x_2} \right]^{L/2-x_3}_{r_3 = -L/2-x_3} \\
        F_{\text{hom}}(\myvec{r})  &=  \sum_{i=1}^3 \left( \frac{r_1 r_2 r_3}{r_i} \ln(r_i + \vecnorm{r}) - \frac{r_i}{2} \arctan \left(\frac{r_1 r_2 r_3}{r_i^2 \vecnorm{r} } \right) \right)\hlt{\,,}
 \end{align}
 where we labeled the Cartesian 3d parent function of the integrand $1/r$ by $F_{\text{hom}}(\hlm{\myvec{r}})$ and we used the notation from \citet{chappell_2013}. Note that the type of integral shown in \eqref{eqn:hom_pot_integral} is most easily evaluated in Cartesian coordinates. Transforming to spherical coordinates simplifies the integrand, but makes the integration boundaries very complicated, and is therefore not viable. 
 
We show a slice through the $z=0$ plane of the potential and the force-field of a homogeneous cube with $G = 1$, $\rho_0 = 1$ and $L=1$ in the left panel of Figure \ref{fig:cube_potential}. Close to the centre of the cube forces get close zero. Close to the boundary the forces are largest and the equipotential lines deviate most from spherical ones. Then going farther away from the cube the equipotential lines approach spherical symmetry around the centre-of mass.


The potential of a cube with constant gradient is given by
\begin{align}
  \phi (\myvec{x}) &= - G \int_{-L/2}^{L/2} \int_{-L/2}^{L/2} \int_{-L/2}^{L/2} \frac{\rho_0 + \hlm{\myvec{g}} \cdot \myvec{r}}{|\myvec{r} - \myvec{x}|}  d^3 \myvec{r} \label{eqn:gradcube} \\
                    &= - G \int_{-L/2 - x_1}^{L/2-x_1} \int_{-L/2-x_2}^{L/2-x_2} \int_{-L/2-x_3}^{L/2-x_3} \left(\frac{\rho_0-\myvec{g} \cdot \myvec{x}}{\vecnorm{r}} + \frac{\myvec{g} \cdot \myvec{r}}{\vecnorm{r}}\right)  dr_3 dr_2 dr_1 \label{eqn:grad_pot_integral} \\
                   &= - G \left[ \left[ \left[ (\rho_0 - \myvec{g} \cdot \myvec{x}) F_{\text{hom}}(\myvec{r}) + \sum g_i F_{\text{lin}, i}(\myvec{r} )\right]_{r_1=..} \right]_{r_2=..} \right]_{r_3=..} \hlt{\,.}\label{eqn:grad_pot_parent} 
\end{align}
We already know the first parent function and only need to calculate the second part which is given by
\begin{dmath}
  F_{\text{lin}, i}(\myvec{r}) = \int \int \int \frac{r_i}{\sqrt{r_1^2 + r_2^2 + r_3^2}} dr_1 dr_2 dr_3 \\
  = \frac{1}{6} \left( 2 \frac{r_1 r_2 r_3}{r_i} \vecnorm{r} - 2 r_i^3 \arctan \left(\frac{r_1 r_2 r_3}{r_i^2 \vecnorm{r}} \right) \\+ r_j (3 r_i^2 + r_j^2) \ln(r_k+\vecnorm{r})  + r_k (3 r_i^2 + r_j^2) \ln(r_j+\vecnorm{r}) \right)\hlt{\,,}
\end{dmath}
where 
\begin{align}
j &= i + 1 \mod 3 \label{eq:sum_index_j} \\
k &= i + 2 \mod 3 \label{eq:sum_index_k}
\end{align}
We show the potential of a cube ($\rho_0 = L = G = 1$) with a density gradient $\myvec{g} = (1.2,1.2,0.)$ in the right panel of Figure \ref{fig:cube_potential}. In comparison to the homogeneous cube the centre of mass and the deepest point in the potential shift in the direction of the density gradient. At large radii the equipotential lines approach sphericity around the centre of mass.


\subsection{The Tree PM force-split}
\label{app:treepmsplit}
In {\sc Gadget-2} forces are calculated using a mixture of a short-range tree summation and a long-range force calculation on a periodic particle mesh (PM). Therefore the potential is split\hlt{,}
\begin{align}
  \phi(\hlm{\myvec{r}}) = \phis (\myvec{r}) + \phil (\myvec{r})\hlt{\,,}
\end{align}
where the long-range potential $\phil$ is given by the true potential convolved with a smoothing kernel
\begin{align}
  \phil &= (\rho \circ G_\phi) \circ f\hlt{\,.}
\end{align}
The long-range potential is calculated on a periodic particle mesh. The particles are binned with a \hlt{cloud}-in-cell assignment onto a periodic mesh to get the real-space density field $\rho$. The mesh cannot resolve structures in the density field which have a smaller size than a mesh-cell. However, if the smoothing kernel $f$ is large enough, e.g. the size of a few mesh-cells, the contributions of these small-scale structures to the long-range potential is negligible and the long-range force can be calculated very accurately on the mesh. It is then easy to obtain the long-range potential, simply by Fourier-transforming the density field,
multiplying it by all the convolution components and then transforming the obtained potential back to real space.
\begin{align}
  \phi_{l,k} &= \rho_k G_{\phi,k} f_k\hlt{\,,}
\end{align}
where we denoted 3d-Fourier-transformed functions by a small index $k$. 

The short-range part of the potential cannot be represented accurately on the mesh and is instead calculated by a tree-walk in real-space. It is given by
\begin{align}
 \phis &= (\rho \circ G_\phi) \circ (1 - f) \\
              &= \rho  \circ (G_\phi \circ (1 - f)) \hlm{ = \rho  \circ} \Gs\\\
              \Gs :&= \hlm{G_\phi \circ (1 - f)}\hlt{\,,}
\end{align}
where we defined the Green's function of the short range potential $\Gs$. For the simple case of a point mass the short-range potential is given by $\Gs$. However, in the case of a cubic mass-distribution it is much more complicated to obtain the short-range force, since the integrand in \eqref{eqn:gradcube} must be changed. The default choice of a force-split kernel in {\sc Gadget-2} is a Gaussian kernel
\begin{align}
  \hlm{f_{\text{Gadget}}}(\myvec{r}) &= \frac{1}{8 \pi^3 \rs^3} \exp \left(- \frac{\vecnorm{r}^2}{4 \rs^2} \right) \label{eq:fG2}\\
  \hlm{f_{k, \text{Gadget}}}(\myvec{k}) &= \exp(- \vecnorm{k}^2 \rs^2) \\
  \hlm{G_{\phi,\text{s,Gadget}}}(\myvec{r}) &= - \frac{G}{\vecnorm{r}} \erfc \left( \frac{\vecnorm{r}}{2 \rs} \right)\hlt{\,.}\label{eq:GsG2}
\end{align}
\begin{figure}
  \includegraphics[width=\columnwidth]{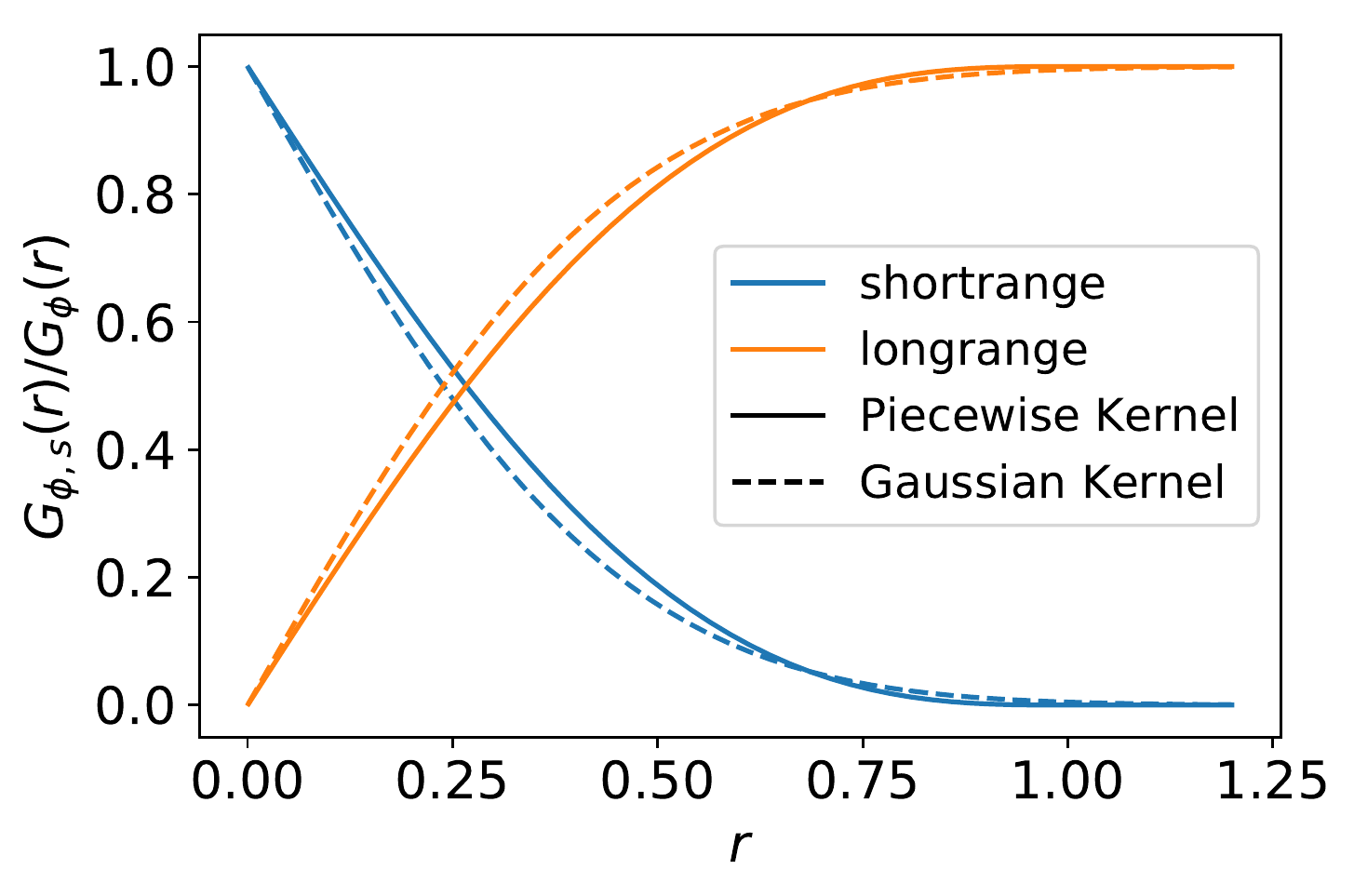}
  \caption{The short-range and long-range parts of the potential, defined as the fractional contribution to the potential at radius $r$ for the two different cases of a piecewise force-split  as in \eqref{eq:fpkap} with $a=1$ and a Gaussian force-split as in \eqref{eq:fG2} with $\rs = 0.25$. The two force-splits have a similar shape for $a \sim 4 \rs$. However, the piecewise kernel has the advantage that the short-range potential becomes exactly zero at the finite radius $a$.}
  \label{fig:force-split}
\end{figure}
\hlcom{(number removed)}However we find that this is not a practical choice in our case, since we cannot find an analytical solution to the convolution of \eqref{eq:GsG2} with the density field of a cube. Instead we choose a different force-split kernel\hlt{\,:}
\begin{align}
   f(\myvec{r}) &=  
        \begin{cases}
          \frac{3 (a - \vecnorm{r})}{a^4 \pi} &\text{ if } \vecnorm{r} \leq a \\
          0 & \text{ otherwise \hlcom{ (modified)}}
        \end{cases} \label{eq:fpkap}\\
   f_k(\myvec{k}) &=  \frac{12 }{a^4 \vecnorm{k}^4} \left( 2 - 2 \cos (a \vecnorm{k}) + a \vecnorm{k} \sin ( a \vecnorm{k}) \right) \\
   \Gs(\myvec{r}) &=  
        \begin{cases}
          - G \frac{(a - \vecnorm{r})^3 (a + \vecnorm{r})}{a^4 \vecnorm{r}} &\text{ if } \vecnorm{r} \leq a \\
          0 &\text{ otherwise } \hlcom{(modified)} 
        \end{cases}
\end{align}
We plot the relative contributions of the long-range and short-range part of the potential in Figure \ref{fig:force-split}. The Gaussian and the piecewise force-split have very similar shapes for $a \sim 4 \rs$ and are both reasonable choices. However, the piecewise split has the advantage that the short-range contribution becomes exactly zero at the finite radius $a$. Therefore it is exactly correct to stop the summation of short-range forces beyond that radius, whereas the Gaussian short-range force never becomes exactly zero and can only be neglected approximately beyond some radius (typically $4.5 - 6 \rs$).

\begin{figure}
  \includegraphics[width=\columnwidth]{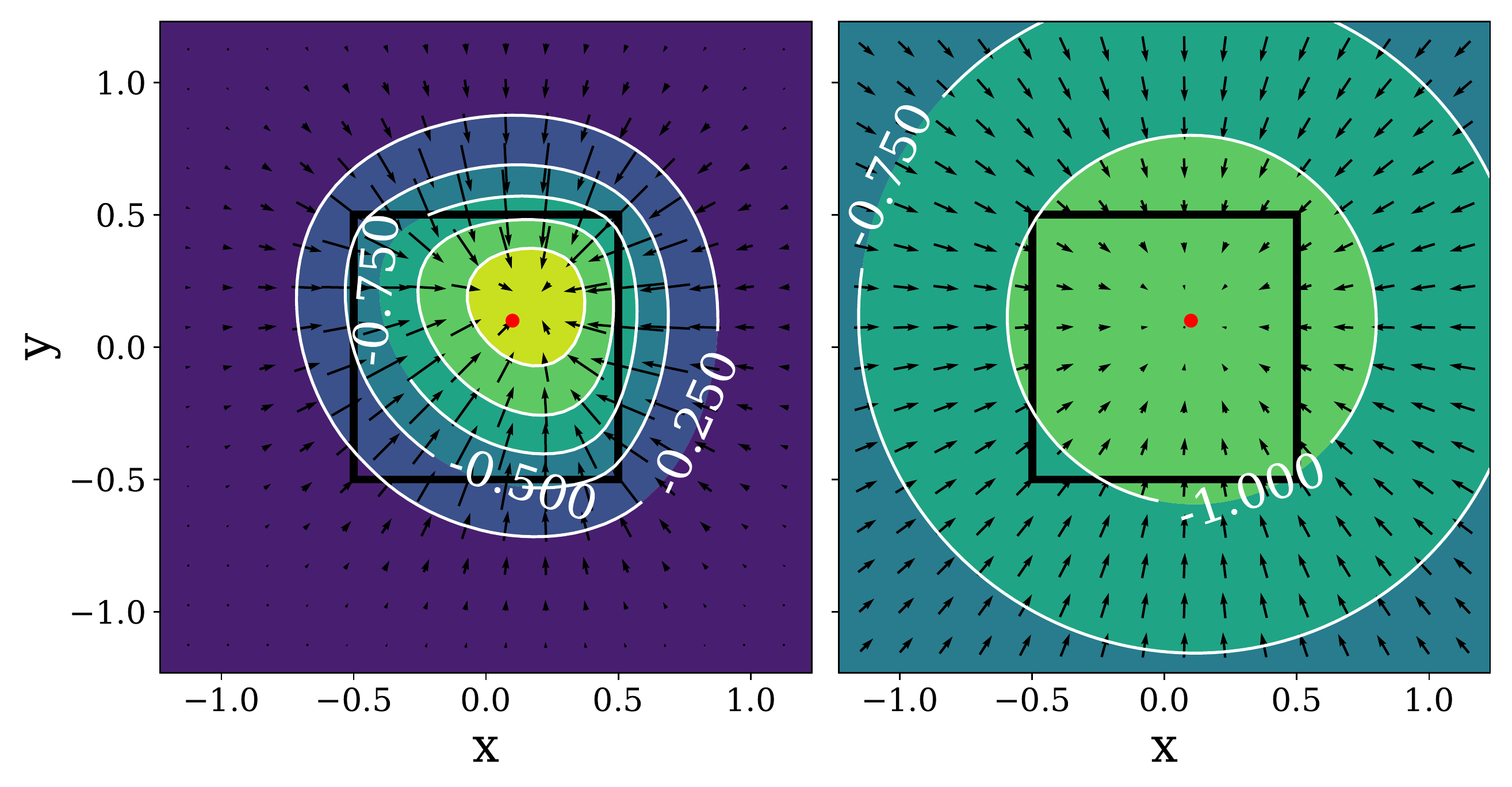}
  \caption{\hlcom{(Increased size of labels)}Potential (contours) and force field (arrows) of the same cube as in the right panel of Figure \ref{fig:cube_potential} with a piecewise force cut with $a = 1.6$. Left: The short-range component and right: the long-range component.}
  \label{fig:cube_potential_force-split}
\end{figure}

To get the short-range potential of a cube with a gradient we have to solve the integral
\begin{align}
   \phis &= \hlm{\int_{-\frac{L}{2} - x_1}^{\frac{L}{2} -x_1} \int_{-\frac{L}{2}-x_2}^{\frac{L}{2}-x_2} \int_{-\frac{L}{2}-x_3}^{\frac{L}{2}-x_3}} (\rho_0-\myvec{g} \cdot \myvec{x} + \myvec{g} \cdot \myvec{r}) \Gs(\hlm{\myvec{r}}) d^3 \myvec{r}\hlt{\,.} \label{eqn:phis}
\end{align}
Note that solving this integral in the general case is very complicated, because the interaction of the integral boundaries with the boundary of the kernel $\Gs$ introduces many different possible cases in the integral. However, in our simulations most of the interactions will be at short distance in comparison to the force-split scale $\vecnorm{x} \ll a$ and $L \ll a$. Therefore we can calculate the analytical solution for these simpler cases, and use a numerical approximation for the other cases. For all cases where the farthest edge is still within the kernel radius $a$\hlt{,}
\begin{align}
  \sum_i \left(|x_i| + \frac{L}{2} \right)^2 \leq a^2\hlt{\,,} \label{eqn:fully_smoothed}
\end{align}
we can simplify
\begin{align}
   \phis = - G  \int_{-L/2 - x_1}^{L/2-x_1} \int_{-L/2-x_2}^{L/2-x_2} \int_{-L/2-x_3}^{L/2-x_3} & (\rho_0-\myvec{g} \cdot \myvec{x} + \myvec{g} \cdot \myvec{r}) \nonumber \hlcom{(number removed)} \\
   & \frac{(a - \vecnorm{r})^3 (a + \vecnorm{r})}{a^4 \vecnorm{r}} d^3 \myvec{r}\hlt{\,.}
\end{align}
We define the 3d parent function of the short-range potential analogous to \eqref{eqn:grad_pot_parent} and find
\begin{dmath}
  F_{\text{hom},s}(\myvec{r}) = \int \int \int \frac{(a - \vecnorm{r})^3 (a + \vecnorm{r})}{a^4 |\vecnorm{r}|} dr_1 dr_2 dr_3 \\
  = \frac{r_1 r_2 r_3}{15} \left(30 a^3 - 10 a \vecnorm{r}^2 + 2 \vecnorm{r}^3 \right) \\
  + \frac{1}{120 a^4} \sum_i \left( r_i^2 (60 a^4 - 4 r_i^4) \arctan \left( \frac{r_1 r_2 r_3}{\vecnorm{r} r_i} \right) \\  +  r_j r_k \left(9 r_j^4 + 10 r_j^2 r_k^2 + 9 r_k^4 - 120 r_k^4 \right) \log (\vecnorm{r} + z) \right)
\end{dmath}
and
\begin{dmath}
  F_{\text{lin},s, i}(\myvec{r}) = \int \int \int r_i \frac{(a - \vecnorm{r})^3 (a + \vecnorm{r})}{a^4 \vecnorm{r}} dr_1 dr_2 dr_3
\\
 = \frac{1}{1680 a^4} \left[ y z \left(41 r^5 + 52 r^3 r_i^2 - 560 a r^2 r_i^2 - 560 a^4 r + 81 r r_i^4 - 30 r y^2 z^2 +1680 a^3 r_i^2 - 280 a r_i^4  \right) \\
  + \arctan \left(\frac{r_j r_k}{r r_i} \right) \left( 560 a^4 r_i^3 - 48 r_i^7 \right)  \\
  + \log(r + r_j) \left(-840 a^4 r_i^2 r_k + 105 r_i^6 r_k - 280 a^4 r_k^3 + 105 r_i^4 r_k^3 + 63 r_i^2 r_k^5 + 15 r_k^7\right) \\
+ \log(r + r_k) \left(-840 a^4 r_i^2 r_j + 105 r_i^6 r_j - 280 a^4 r_j^3 + 105 r_i^4 r_j^3 + 63 r_i^2 r_j^5 + 15 r_j^7\right)  \right]\hlt{\,,}
\end{dmath}
where the indices $j$ and $k$ change with the summation index $i$ as in \eqref{eq:sum_index_j} and \eqref{eq:sum_index_k}. In Figure \ref{fig:cube_potential_force-split} we show the short and- long-range potential and force of the same cube as the right panel of Figure \ref{fig:cube_potential} for a force-cut parameter of $a = 1.6$.

\begin{figure}
  \includegraphics[width=\columnwidth]{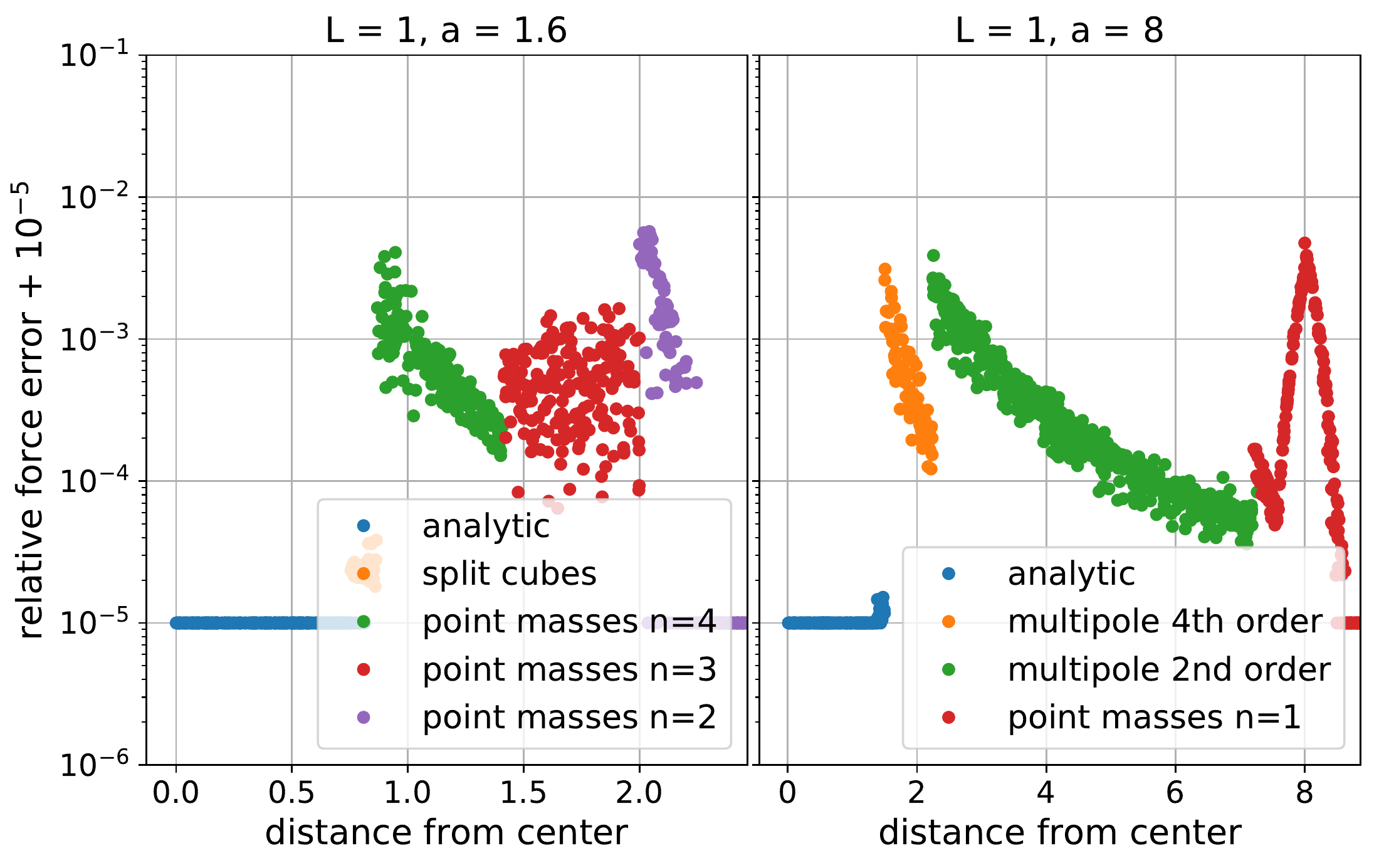}
  \caption{Relative force errors due to different numerical approximations as a function of distance from the centre of a cube with $L = 1$, $G = 1$, $\myvec{g} = (0.5, 0.3, 0.1)^T$. Errors are clipped at $10^{-5}$. Left: with a force-split scale $a = 1.6$ which is similar to the size of the cube. This is a very unusual and challenging case, since our analytical solution is not valid in most of the range. It is however still well approximated (relative error smaller than $10^{-2}$) with the point-mass approximations as explained in the text. Right: for a force-split scale $a = 8$ which is significantly larger than the cube. Not too close to the cube the forces can be very accurately described by multipole expansions which are numerically much cheaper than the analytic solution.}
  \label{fig:cubeforce_error}
\end{figure}

For cases where some parts of the cube are inside and some are outside of the kernel radius $a$ (compare equation \ref{eqn:fully_smoothed}) \hlcom{(changed this sentence)}, we approximate the potential numerically by a sum of point-masses:
\begin{align}
  \phis(\myvec{x}) &\approx \sum_{i=0}^{n-1} \sum_{j=0}^{n-1} \sum_{k=0}^{n-1} \rho(\myvec{r}_{ijk} ) \Delta r^3 \Gs(\myvec{x} - \myvec{r}_{ijk}) \\
  \Delta r &= L / n \\
   \myvec{r}_{ijk} &= \begin{pmatrix} - \hlm{L/2} + (0.5 + i) \Delta r   \\  - \hlm{L/2} + (0.5 + j) \Delta r   \\ - \hlm{L/2} + (0.5 + k) \Delta r  \end{pmatrix}\hlt{\,,}
\end{align}
where $n^3$ is the number of point-masses per dimension. $n$ controls the accuracy of the approximation and we adaptively choose $n$ depending on the distance from the cube. We list the scenarios in which we use different point-mass approximations in the lower part of Table \ref{tab:force_approximations}. In the left panel of Figure \ref{fig:cubeforce_error} we show the relative errors of the point-mass approximations\hlt{,}
\begin{align}
  \epsilon &= \frac{|\myvec{F}_{s,\text{approx}} - \myvec{F}_{s,\text{exact}}|} {|\myvec{F}_{\text{tot}, \text{exact}}|}\hlt{\,,}
\end{align}
where $\myvec{F}_{s,\text{approx}}$ is the calculated force, $\myvec{F}_{s,\text{exact}}$ is a much more accurate reference force (calculated by splitting the cube into $4^3$ sub-cubes and calculating their forces) and $\hlm{\myvec{F}_{\text{tot,exact}}}$ is the total exact force (without force-cut). Our choice of point mass approximations still maintains a relative accuracy better than $10^{-2}$ in the rare but challenging cases of $L \sim a$.

One special case which will almost never happen in a simulation, but which we still include for completeness, is if the whole force-split kernel resides within the cube. In that case the integrand of \eqref{eqn:phis} is only non-zero on the spherical domain $\vecnorm{x} \leq a$:
\begin{align}
   \phis &= \int_0^a 4 \pi r^2 (\rho_0-\myvec{g} \cdot \myvec{x}) \Gs(r) dr \\
          &= - \frac{4 \pi a^2 (\rho_0-\myvec{g} \cdot \myvec{x})}{15}\hlt{\,,}
\end{align}
where we dropped the term $\myvec{g} \cdot \myvec{r}$ from \eqref{eqn:phis}, because it must be zero since it is anti-symmetric with respect to $\hlm{\myvec{r}}$.

\begin{table}
 \caption{Full overview over different cases and the numerical approximations that are used. $r_{\text{max}}$ is the distance to the farthest corner of the cube, $r$ to its centre and $r_{\text{min}}$ to its closest boundary.}
 \label{tab:force_approximations}
 \begin{tabular}{cccc}
  \hline
  \multicolumn{3}{c}{Case}  & Approximation \\
  \hline
  \multicolumn{4}{c}{Fully Smoothed Cases}\\
   $r_{\text{max}} \leq a$ & and & $r \leq 1.5 L$ & analytic \\
   $r_{\text{max}} \leq a$ & and & $1.5 < r/L \leq \sqrt{5}$ & multipole 4th order \\
   $r_{\text{max}} \leq a$ & and & $r/L > \sqrt{5}$ & multipole 2nd order \\
  \hline   
  \multicolumn{4}{c}{Intersecting Cases}\\
   $r_{\text{min}} > a$ & and & $r_i < L/2$ $ \forall i$ & analytic \\
   $r_{\text{max}} > a$ & and & $r/L < \sqrt{3}/2$ & Split into 8 subcubes \\
   $r_{\text{max}} > a$ & and & $\sqrt{3}/2 \leq r/L < \sqrt{2}$ & point masses $n=4$ \\
   $r_{\text{max}} > a$ & and & $\sqrt{2} \leq r/L < 2$ & point masses $n=3$ \\
   $r_{\text{max}} > a$ & and & $2 \leq r/L < 4$ & point masses $n=2$ \\
   $r_{\text{max}} > a$ & and & $r/L > \sqrt{16}$ & point masses $n=1$ \\
  \hline
 \end{tabular}
\end{table}

\subsection{Multipole Expansion of the far-field}
\label{app:multipole}
Since the evaluation of the true parent function of the shortrange potential of the cube is very expensive, we use a multipole expansion to get much cheaper, but still accurate approximations for interactions at distances $\vecnorm{x} > 1.5 L$ from the cube. The multipole expansion is obtained from \eqref{eqn:phis} by expanding the Green's function around a point $\myvec{x}_0$:
\begin{align}
  \phis(\myvec{x}) &= \int \int \int \rho(\myvec{r}) \Gs(\myvec{r} - \myvec{x}) d^3 \hlm{\myvec{r}}  \\
  & \approx \Gs(\myvec{x} - \myvec{x_0}) \int \int \int \rho(\myvec{r}) d^3\myvec{r} \nonumber  \\
   & + \sum_i \left( \partial_i \Gs(\myvec{x} - \myvec{x_0})  \int \int \int \rho(\myvec{r}) (r_i - x_{0,i}) d^3\myvec{r} \right) \nonumber  \\
   & + \frac{1}{2} \sum_{i,j} \left( \partial_i \partial_j \Gs(\myvec{x} - \myvec{x_0})  \int \int \int \rho(\myvec{r}) (r_i - x_{0,i}) (r_j - x_{0,j}) d^3\myvec{r} \right) \nonumber  \\
   & + ...
\end{align}
\hlcom{(Removed some numbers)}In our case it is the most convenient to choose $\myvec{x_0} = 0$ as the expansion point. Since the integration domain is symmetric around this point, many terms drop out. We find the expansion
\begin{align}
  \phis(\hlm{\myvec{x}}) \approx   & -\frac{G L^3 \rho _0 (a-\vecnorm{x})^3 (a+\vecnorm{x})}{a^4 \vecnorm{x}} \nonumber \\
                                       -& \frac{G L^5 \left(a^4-4 a \vecnorm{x}^3+3 \vecnorm{x}^4\right) \sum _i g_i x_i}{12 a^4 \vecnorm{x}^3} \nonumber \\
                                       +& \frac{G L^5 \rho _0 (\vecnorm{x}-a)}{2 a^4} \nonumber \\
                                       -& \frac{G L^7 \left(\left(\vecnorm{x}^4-5 a^4\right) \sum _i g_i x_i^3+\left(3 a^4 \vecnorm{x}^2+7 \vecnorm{x}^6\right)
   \sum _i g_i x_i\right)}{240 a^4 \vecnorm{x}^7} \nonumber \\
                                       -& \frac{G L^7 \rho _0 \left(\left(3 \vecnorm{x}^4-35 a^4\right) \sum _i x_i^4+21 a^4 \vecnorm{x}^4-17
   r^8\right)}{960 a^4 r^9}\hlt{\,,}
\end{align}
where we have written one expansion order per line (from 0 to 4). Note that the dipole moment (order 1) does not vanish, since we expanded around the geometric centre of the cube, not its centre of mass. However, the other terms are simpler than in an expansion around the centre of mass. In the case of a homogeneous cube $\myvec{g} = 0$ in the absence of a force-cut $a \rightarrow \infty$ this simplifies to
\begin{align}
 \phi_{\text{hom}} (\myvec{x}) \approx - \frac{G \rho_0 L^3}{|\myvec{x}|} \left( 1 + \frac{7 L^4}{320 |\myvec{x}|^4} - \frac{7 L^4 \sum x_i^4}{192 |\myvec{x}|^8} \right)\hlt{\,,}
\end{align}
which is in agreement with the multipole expansion in \citet{hummer_1995}. We use the multipole expansion up to fourth order if $1.5 L < r < 2 L$ and only up to second order if $r > 2 L$. We summarize these different cases together with the ones where the point mass approximations have to be used in Table \ref{tab:force_approximations}. In the right panel of Figure \ref{fig:cubeforce_error} we show the errors of the \hlt{multipole} expansion for a typical case $a=8$, $L=1$. In the ranges where we use the multipole expansions they have a relative accuracy better than $10^{-2}$.

\section{Convergence of Stream-Densities} \label{app:streamdensconvergence}

We found that the standard {\sc Gadget-2} opening-criterion \citep{springel_2005} and time-stepping criterion \citep{power_2003} are not \hlt{well suited to achieve convergence in the stream-densities}. We will discuss the modifications that we made to the time-stepping and  opening criteria here.

\subsection{Time-stepping}
The fiducial choice for the time-stepping in the {\sc Gadget-2} code \citep{springel_2005} is
\begin{align}
    \Delta t = \text{min} \left( \Delta t_{\text{max}}, \left( \frac{2 \eta \epsilon}{\vecnorm{a} } \right)^{\hlm{1/2}} \right)\hlt{\,,} \label{eqn:gadget2timestep}
\end{align}
where $\myvec{a}$ is the acceleration of a particle, $\epsilon$ is the softening,  $\eta$ an accuracy parameter and \hlt{$\Delta t_{\text{max}}$ is a maximally allowed time-step}. While this criterion seems to give reasonable results for the density profiles of haloes, we found that it is not very suitable to bring the phase space structure to convergence at affordable cost. We find that it produces unnecessarily small timesteps in the outskirts of haloes and timesteps that are too large in their centres. Further it is not obvious why a decrease in softening should change the timestep globally and not just in the dense centres of haloes (where the softening matters).

Instead we employ the criterion that has been suggested by \citet{dehnen_read_2011}\hlt{,}
\begin{align}
    \Delta t = \text{min} \left( \Delta t_{\text{max}}, \left( \frac{2 \eta}{\vecnorm{T}} \right)^{\hlm{1/2}} \right)\hlt{\,,} \label{eqn:tidalstep}
\end{align}
where $\T$ is the tidal tensor. \citet{dehnen_read_2011} argue that this criterion should naturally work in different scenarios and has as the nice property that it is unaffected by a global uniform acceleration (which does not change the internal dynamics of a system). We can adopt this criterion without additional costs, because we already evaluate the tidal tensor for the particles in the simulation.

\begin{figure}
  \includegraphics[width=\columnwidth]{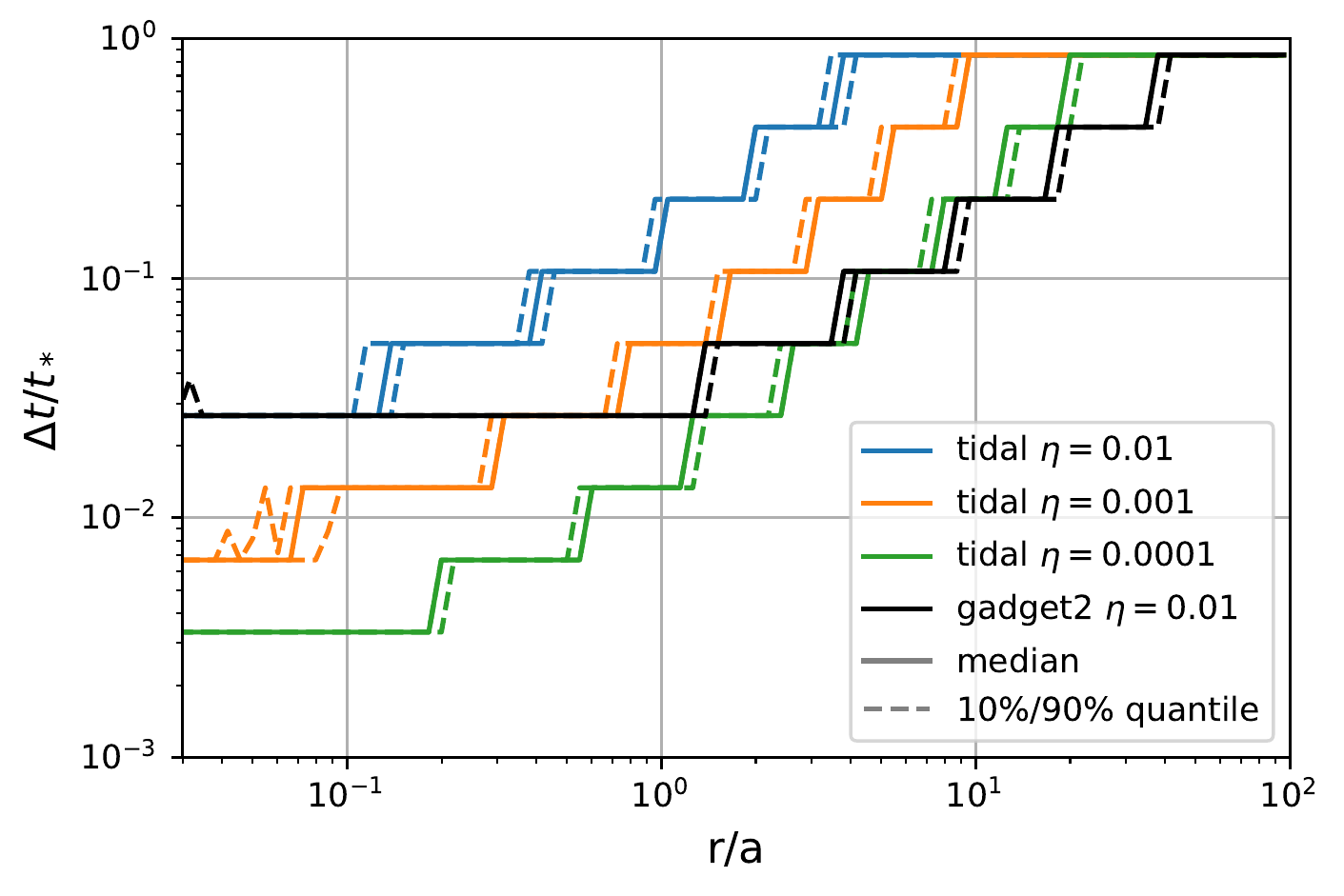}
  \caption{Radial profile of the timesteps in a Hernquist sphere. The solid line denotes the median at that radius and the dashed lines the \hlt{10th and 90th percentiles}. The coloured lines are from the time-step criterion in \ref{eqn:tidalstep} and the black line is the default {\sc Gadget-2} time-step criterion as in \eqref{eqn:gadget2timestep}.}
  \label{fig:timestep_profile}
\end{figure}

In Figure \ref{fig:timestep_profile} we show the radial distribution of timesteps that is produced by this criterion for different accuracy parameters $\eta$ in the Hernquist case. At the same accuracy parameter the tidal criterion produces much larger timesteps in the outer parts of the Hernquist sphere while giving timesteps of the same size in the innermost region that is affected by the softening $r \leq 0.1 a$. This saves a lot of computation time in such regions so that we can employ a factor 10 smaller accuracy parameter at a similar same overall cost (orange line). We only get similar timesteps to the {\sc Gadget-2} criterion in the outskirts if we use a factor 100 smaller accuracy parameter.

\subsection{Opening Criterion}
In {\sc Gadget-2} a relative opening criterion is employed where a node is opened during the tree walk if
\begin{align}
  \frac{G M}{r^2} \left(\frac{L}{r}\right)^2 > \alpha \vecnorm{a} \label{eqn:openingforce}\hlt{\,,}
\end{align}
where $\alpha$ is an accuracy parameter, $r$ is the distance to the interacting particle, $L$ is the size of the considered node and $M$ its mass. This criterion is designed to keep the relative contribution of the (neglected) quadrupole moment of the force small. 

\begin{figure}
  \includegraphics[width=\columnwidth]{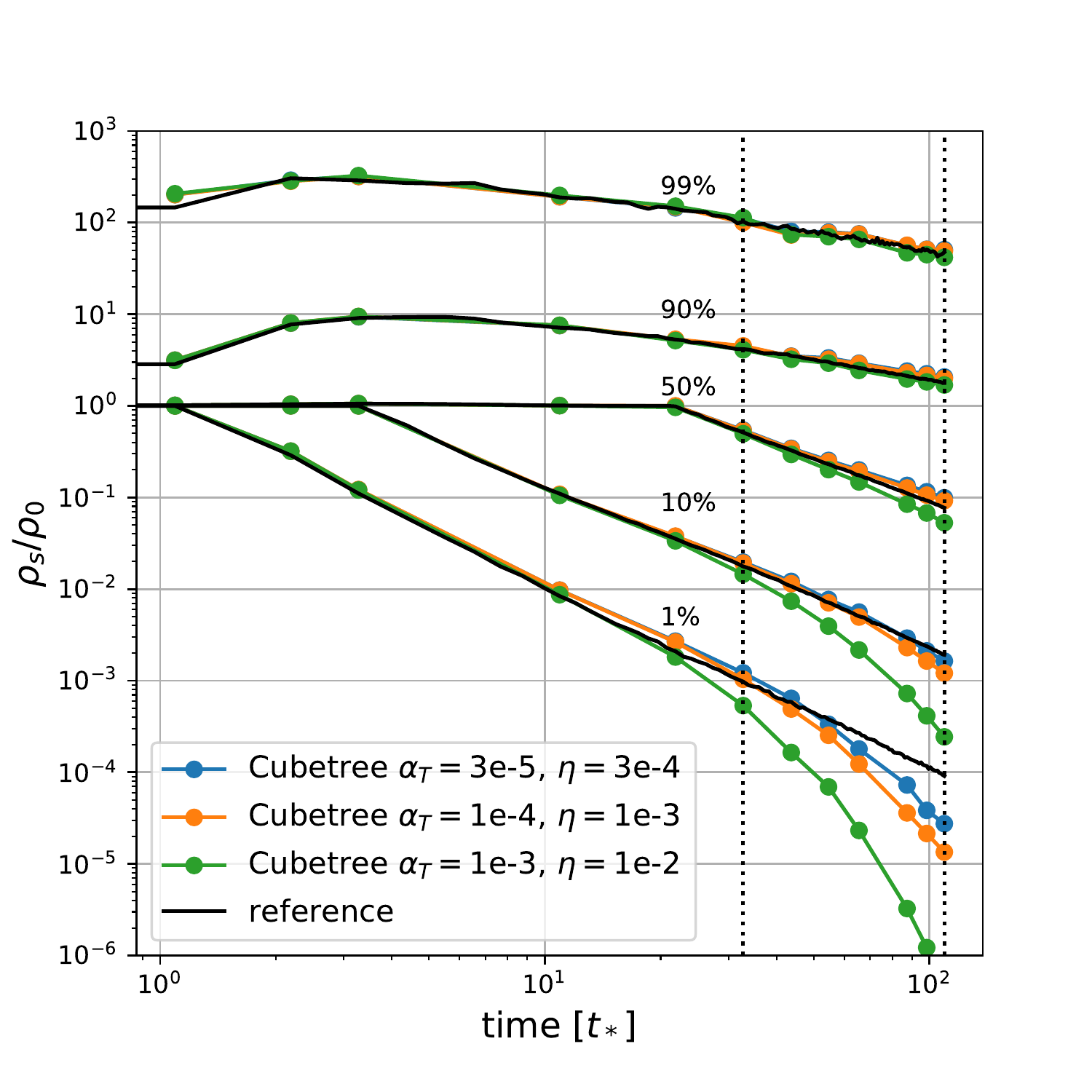}
  \caption{Evolution of the \hlt{1st, 10th, 50th, 90th and 99th percentiles} of the stream-density distribution. The dashed lines mark the output times of the plots in Figure \ref{fig:hernquist_strde_profile}. The lower quantiles of the reference solution show a power law behaviour of the form $\rhos \propto t^{-2}$. The numerical solutions deviate from the power law at late times with an exponential behaviour similar to the one seen in Figure \ref{fig:hernquist_strde_evol}. The opening and time-stepping criterion need to be much smaller for the stream-densities to converge than, for example, for convergence of density profiles.}
  \label{fig:strde_evolution_timestep}
\end{figure}

To further keep the error of the tidal tensor small we additionally employ the opening criterion 
\begin{align}
  \frac{G M}{r^3} \left(\frac{L}{r}\right)^2 \leq \alpha_{T} \vecnorm{T}\hlt{\,,} \label{eqn:openingtidal}
\end{align}
which is designed to keep the relative contribution of the (neglected) quadrupole moment of the tidal tensor small. A tree node is then opened if it satisfies either \ref{eqn:openingforce} or \ref{eqn:openingtidal}. For $\alpha$ we choose $0.005$ (which seems more than good enough to get converged density profiles) and we determine $\alpha_T$ by testing the convergence of the stream-densities. We have found $\alpha_T$ to require significantly smaller choices than is typical for $\alpha$ to have a notable effect. 

In Figure \ref{fig:strde_evolution_timestep} we show a set of simulations where we jointly vary the time-step parameter $\eta$ and the opening parameter $\alpha_T$. We find that the the stream densities do not vary drastically below
\begin{align}
 \alpha_T \sim 10^{-4}, \eta \sim 10^{-3}\hlt{\,.}
\end{align}
Probably at this point further convergence is limited by the level of shot noise and can be improved by using higher resolution. In the plots in section \ref{sec:timevolgde} we used $\alpha_T = 3 \cdot 10^{-5}$ and $\eta = 3 \cdot 10^{-4}$.


\bsp	
\label{lastpage}
\end{document}